\documentclass[twoside,twocolumn,english,superscriptaddres,showpacs,longbibliography,aps,prb]{revtex4-2}

\usepackage[T1]{fontenc}
\usepackage[utf8]{inputenc}
\usepackage[english]{babel}
\usepackage{bm}
\usepackage{amsmath,physics}
\usepackage{amssymb}
\usepackage{graphicx}
\usepackage[unicode=true, pdfusetitle, bookmarks=true,bookmarksnumbered=true,bookmarksopen=true,bookmarksopenlevel=1,
breaklinks=false,pdfborder={0 0 0},pdfborderstyle={},backref=false,color links, citecolor=blue,linkcolor=red,urlcolor=blue] {hyperref}
\usepackage[nameinlink]{cleveref}
\Crefname{figure}{Fig.}{}

\makeatletter
\def\set@firstnote#1{%
 \@ifnum{\firstnote@num=#1\relax}{}{%
  \class@warn@end{Endnote numbers changed: rerun LaTeX}%
 }%
 \immediate\write\@mainaux{%
   \global\mathchardef\string\firstnote@num#1\relax
 }%
}%

\usepackage{braket}
\usepackage{bm}
\usepackage{tikz}
\usepackage{pgfplots}
\usepackage{pgf}
\usepackage{subfigure}
\usepackage{nicematrix}
\usepackage{diagbox}
\usepackage{orcidlink}

\renewcommand{\selectlanguage}[1]{}

\makeatother

\pgfplotsset{compat=1.18}
\begin{document}

\title{Generalizing quantum dimensions: Symmetry-based classification of local pseudo-Hermitian systems and the corresponding domain walls}
\author{Yoshiki Fukusumi}
\affiliation{Physics Division, National Center for Theoretical Sciences, National Taiwan University, Taipei 106319, Taiwan}
\author{Taishi Kawamoto}
\affiliation{Center\! for Gravitational\! Physics\! and Quantum \! Information, Yukawa\! Institute\! for\! Theoretical\! Physics, Kyoto\! University, Kitashirakawa\! Oiwakecho, Sakyo-ku, Kyoto 606-8502, Japan}
\date{\today}
\begin{abstract}
We study conformal field theories (CFTs) and their classifications from a modern perspective based on the abstract algebraic formalism of symmetries or conserved charges, known as symmetry topological field theories (SymTFTs). By studying the algebraic structure of the SymTFTs in detail, we found a natural generalization of the quantum dimensions associated with (pseudo-)Hermitian systems and (non)-unitary CFTs. These generalized data of SymTFTs provide classifications of massless and massive renormalization group flows, which will describe the quantum phase transitions of the corresponding pseudo-Hermitian systems. Moreover, our discussions straightforwardly enable one to relate a general class of coset constructions or level-rank dualities to domain wall problems between topological quantum field theories (or a series of corresponding quantum phase transitions related to the Higgs mechanism). Our work provides a systematic reduction and classification of algebraic data, symmetries, for pseudo-Hermitian systems based on ideas from established mathematical fields, linear algebra and ring theory.

\end{abstract}

\begin{flushright}
YITP-25-174
\\
\end{flushright}
\maketitle

\section{introduction}
\label{introduction}

Non-Hermitian systems and the corresponding nonequilibrium phenomena are among the central focuses in contemporary theoretical and experimental physics. The research is growing very rapidly in these decades, but the characterization of their criticalities and topological orders (TOs) has still been in development. The literature is large, so we only list several reviews in condensed matter \cite{Ashida:2020dkc,Okuma:2022bnb}, a short review on the Lindblad form \cite{Manzano:2020yyw}, and an earlier review\cite{Bender:2007nj}. For readers in high-energy physics theory, we note that many series of non-Hermitian models and corresponding nonunitary conformal field theories (CFTs) have been studied in the communities studying integrable lattice models, with a connection to ordinary differential equations and integrable model (ODE/IM) correspondence\cite{Dorey:2007zx}. Also, nonunitary CFTs appear in the holography of de Sitter spacetime \cite{Strominger:2001pn,Klemm:2001ea}, traversable wormhole \cite{Kawamoto:2025oko}.
We note several earlier works in coset CFTs\cite{Nakanishi:1989cv,Schellekens:1989uf,Mathieu:1991fz} and a work pointing out the connection to the underlying quantum groups\cite{Pasquier:1989kd}. It is remarkable that this earlier research direction has already been commented on in the pioneering work of the Hatano-Nelson model \cite{Hatano_1996}, a landmark non-Hermitian model.

One of the most fundamental ideas in the studies on non-Hermitian systems is the similarity transformation, which maps a pseudo-Hermitian system to the Hermitian system \cite{Mostafazadeh:2001jk,Mostafazadeh:2001nr,Mostafazadeh:2002id}. Especially, when the system has the parity-time (PT) symmetry, this has played an important role\cite{Bender:1998ke}. In quantum field theory, a kind of law of thermodynamics, called Zamolodchikov's $c$-theorem\cite{Zamolodchikov:1986gt} has been generalized to $c_{\text{eff}}$-theorem\cite{Castro-Alvaredo:2017udm} for $PT$ symmetric models where $c$ is the central charge of a CFT and $c_{\text{eff}}$ is the effective central charge. The phenomenology of this theorem is simple: the effective central charge $c_{\text{eff}}$ of a system should not increase during renormalization group (RG) flow under some suitable assumptions, such as $PT$ symmetry. Hence, one may expect a dual relation between the $c_{\text{eff}}$-theorem for non-Hermitian systems and the $c$-theorem for the corresponding Hermitian systems, but the quantum field theoretic (QFT) formulation of the similarity transformation has not been studied widely. Moreover, it is known that the similarity transformation can change the locality of the system, and this is analogous to the gauge transformations\cite{Fernandez:2015aqe}. Hence, at this stage, a local pseudo-Hermitian system is more treatable than the corresponding nonlocal Hermitian system in QFT (For readers interested in the implications of the similarity transformation, we note the complementary work by the authors \cite{Fukusumi:2025fir} and the references therein). In this sense, characterizations of pseudo-Hermitian systems and their quantum phase have still been under development in the QFT language. We note related earlier studies\cite{Guruswamy:1996rk,LeClair:2007iy} and recent one\cite{Hsieh:2022hgi}. It is also worth noting that the corresponding level-rank duality\cite{Kuniba:1990im,Kuniba:1990zh,Nakanishi:1990hj,Altschuler:1990th} has been studied in \cite{Buican:2017rya,Ferrari:2023fez,Creutzig:2024ljv} with a close connection to the $3d-3d$ correspondence\cite{Chung:2014qpa,Chun:2019mal}.

On the other hand, there has been significant progress in the understanding of the classification of CFTs and the corresponding topological quantum field theories (TQFTs) based on their symmetry algebra (or the corresponding category theories). In the previous century, the group symmetries and their classification of various models in physics have been studied widely, and this framework has been generalized to ring (and higher-algebra) structures\cite{Cobanera:2011wn,Cobanera:2012dc,Gaiotto:2014kfa}. Roughly speaking, this paradigm is based on studies on the ring structure corresponding to the fusion rules of anyons in a system, and the corresponding classification has been studied under various names depending on the respective research communities. The renormalization group (RG) domain wall\cite{Brunner:2007ur,Gaiotto:2012np} and the gapped domain wall\cite{Lan:2014uaa} are the corresponding relatively well-established concepts in high-energy physics and condensed matter, respectively. The phenomenology itself is simple. The anyons and the corresponding symmetry operators (or conserved quantities) should form a ring over the complex number field $\mathbb{C}$, and their relation can be transformed (or deformed smoothly) under the ring homomorphism. The expression of the categorical objects as a ring over $\mathbb{C}$, or linear operators in linear algebra, has appeared in the pioneering work \cite{Petkova:2000ip}. In \cite{Petkova:2000ip}, the precise and concrete relationship between the conserved charges and topological defects satisfying the same fusion rule has been clarified. In other words, the fusion categorical structure, which had been studied in (pure) mathematical literature\cite{article,Bockenhauer:1998ef,Bockenhauer:1999ae,Bockenhauer:1999wt}, has become a physical structure by identifying the corresponding objects as defects or conserved charges. To emphasize the role of the set of conserved charges as a ring over $\mathbb{C}$, we denote it as \emph{fusion ring symmetry} rather than fusion category symmetry in the literature\cite{Thorngren:2019iar,Thorngren:2021yso}, because fusion category symmetry sometimes cannot be identified as a ring over $\mathbb{C}$ in the literature.

We stress that the fusion ring, a ring over $\mathbb{C}$, is more general than the fusion category respecting nonnegative integer matrix representation (NIM-rep) at the present stage\footnote{In this sense, a part of category theories appearing in physics is less abstract and less general than linear algebra.}. Hence, as we will demonstrate, there can exist ring homomorphisms that can be treated in linear algebra but cannot be in such a restricted category theory. This aspect has been found in \cite{Zhao:2023wtg} and confirmed more recently in \cite{Fukusumi:2025clr}. The scope of these works is conserved charges of CFTs or the corresponding excitation operators, anyons, of the corresponding topological quantum field theories (TQFTs)\cite{Witten:1988hf}. They are different from defects or boundary conditions respecting \emph{the integer coefficients before objects}, whereas there exists the correspondence between the NIM-rep of \emph{the fusion coefficients} of such operators and that of defects\footnote{We thank Yuji Tachikawa and Jurgen Fuchs for the related discussions and many useful comments.}. To some extent, this undesired elementary fact (or puzzle) is a reason why this simple understanding, classifications of anyons by ring homomorphisms, has not been studied widely. It should be stressed that, whereas the NIM-rep plays a significant role in studying defects, the ring over $\mathbb{C}$ is more elementary, fundamental, and historically established in studying the algebra of conserved charge in quantum Hamiltonian systems in QFTs and lattice models. For readers interested in the RG flows of defects or boundary conditions under bulk perturbations, we note earlier works on the corresponding bulk and boundary RG flows\cite{Dorey:2004xk,Dorey:2005ak,Fredenhagen:2009tn,Dorey:2009vg} and recent works on translational invariant defects \cite{Ambrosino:2025myh,Ambrosino:2025pjj}. They are not in the main scope of the present manuscript, but they contain interesting findings worth for further studies.

The readers who are rather familiar with recent applications of a particular (or too restricted in the above contexts) type of category theories in physics may not be familiar with treating a set of conserved charges as a ring over $\mathbb{C}$. However, this ring-theoretic or linear algebraic formulation of symmetries (or conserved charges) is more standard in other established wider research fields, such as experimental and theoretical physics, especially in quantum physics (also in a wide field of mathematics). One can see many applications of a ring over the number field, or linear algebra, ubiquitously but the class of category theories used in physics cannot capture the properties of such an established structure. Hence, the classification of the rings, such as ring isomorphism, subring, and ring homomorphism, defines the classification of theories more generally. A massive RG corresponds to an operation taking a subring, and a massless RG\cite{Zamolodchikov:1987ti,Zamolodchikov:1987jf,Zamolodchikov:1989hfa} corresponds to a ring homomorphism. The former can be understood by relating the massive RG as the addition of perturbations which break or reduce symmetry, and the latter has been proposed in the studies of RG domain walls\cite{Wan:2016php,Klos:2019axh,Klos:2021gab}. For the massive RGs, to some extent, it has already been established that the ground states or the consequence of the RG can be analyzed by the corresponding (smeared) boundary conformal field theories (BCFTs)\cite{Cardy:2017ufe,Lencses:2018paa,Ares:2020uwy,Kikuchi:2021qxz,Kikuchi:2022gfi,Fukusumi:2024ejk,Choi:2025ebk,Wen:2025xka}. This was first found in the study of massive integrable models\cite{Date:1987zz,Saleur:1988zx,Foda:2017vog} and the corresponding phenomena have been revisited several times\cite{Calabrese:2006rx,Qi_2012,Das:2015oha,Fukusumi:2024ejk} with a connection to the Moore-Seiberg data\cite{Moore:1988qv,Moore:1988ss,Moore:1989vd} or the Li-Haldane conjecture\cite{Li_2008}. It should be pointed out that the topological entanglement entropy\cite{Kitaev:2005dm,Levin:2006zz} and entanglement spectrum have a close connection to such observations, by applying the analysis of entanglement surfaces\cite{Holzhey:1994we} to the TQFTs. 

Even in these elementary algebraic structures, subrings and ring homomorphisms, we demonstrate that the appearance of noninteger coefficients is inevitable by developing the approach in \cite{Fukusumi:2025clr}. However, we strongly believe that it will be possible to formulate the arguments in this work by using some generalized category theories, such as premodular fusion categories with suitable generalizations. This research direction has been still under development, but our arguments will provide clues for constructing such extended categories by requiring their compatibility with abstract algebra or linear algebra. For further studies, we note related recent works investigating related structures \cite{Nakayama:2024msv,Kikuchi:2024cjd,Chen:2025qub,Gaberdiel:2026sfg,Ambrosino:2026umb,Benedetti:2026drn}.

For readers in high-energy physics, we demonstrate that these extended category theories will play a fundamental role in studying the quantum phase transitions related to the Higgs mechanism\cite{Higgs:1964pj} in the recent literature\cite{Cordova:2025eim} (see Sec. \ref{section_massless}). We clarify straightforward, but nontrivial relationships between the tensor structure of CFTs and the homomorphism in a general way. We note that the corresponding phenomena have been studied by several distinct research communities in different names, color confinement\cite{LeClair:2001yp}, $\lambda$-deformations\cite{Sfetsos:2017sep,Georgiou:2018gpe}, and coupled-wire construction\cite{PhysRevB.34.6372,PhysRevLett.88.036401,Teo:2011hq} (or related junction or ladder models) in condensed matter theory\cite{Kimura:2014hva,Lecheminant:2015iga,Kimura:2015nka,Fuji_2017,Quella:2019los}.We also note that a related work with careful historical and terminological explanations on the phenomena appeared very recently\cite{Cheng:2026qax}.

In this work, we study the massless and massive RG flows of two-dimensional CFTs, focusing on the conserved charges forming the fusion rings satisfying the NIM-reps. More phenomenologically, we study the classification of (non)unitary CFTs describing local (pseudo-)Hermitian systems and their quantum phase transition. Most part of the techincal details in this work only require the knowledge of linear algebra, and our method is applicable to higher-dimensional systems in principle. The rest of the manuscript is organized as follows. Sec. \ref{pseudo-Hermitian} is a short introduction to the pseudo-Hermitian systems. We introduce the fundamental properties of (pseudo-)Hermitian systems, which should be correct for the corresponding (non)unitary CFT. We note that this aspect has rarely been discussed in the literature on QFTs. In Sec. \ref{sec_quantum_dimension}, we revisit the fusion ring symmetry formed by the Verlinde line operators of the nonunitary CFTs from the modern view of pseudo-Hermitian systems. The generalization of quantum dimensions, which are less common in the literature, is introduced, and its phenomenologies are studied in this section. We also study the general classification of gapped phase preserving Fibonacci fusion ring symmetry under a few assumptions based on the pioneering works on massive RG\cite{Cardy:2017ufe} and defect and boundary CFTs\cite{Graham:2003nc}. Sec. \ref{section_massless} is the main part of this work, and we introduce a general method for constructing and analyzing the homomorphism between fusion rings. We demonstrate the fundamental importance of introducing the generalization of quantum dimensions. We also clarify the relation between the massless RG and coset construction or level-rank duality, which appeared commonly and has been studied (to some extent heuristically) in the literature. The discussion in this section also provides a straightforward but general understanding of the transitions in \cite{Cordova:2025eim}. In the successive section, Sec. \ref{section_examples}, we study the massless RGs of several known examples. In Sec. \ref{section_spin}, we introduce the anomaly classification for non-group-like objects appearing in the homomorphism by generalizing the notion of the integer spin simple current conditions (or equivalent anomaly-free conditions) in \cite{Furuya:2015coa,Numasawa:2017crf,Fukusumi_2022_c,Kikuchi:2022ipr,Fukusumi:2024ejk} (See also the earlier works \cite{Schellekens:1990ys,Gato-Rivera:1990lxi,Gato-Rivera:1991bqv,Kreuzer:1993tf} and  recent related works \cite{Nakayama:2024msv,Delmastro:2025ksn}). Sec. \ref{section_conclusion} is the concluding remarks of this work. In the Appendix, we comment on related research directions for the application of our methods to more general settings.

\section{Pseudo-Hermitian and linear dual basis}
\label{pseudo-Hermitian}

In this section, we introduce some basic aspects of pseudo-Hermitian systems\cite{Mostafazadeh:2001jk,Mostafazadeh:2001nr,Mostafazadeh:2002id} and several assumptions relevant to the successive sections. For a general review of pseudo-Hermitian systems, we note recent reviews\cite{Ashida:2020dkc,Okuma:2022bnb} in condensed matter and an older review \cite{Dorey:2007zx} which has a close connection to the development of CFTs. We also note that related discussions have been summarized concisely in a recent work \cite{Leng:2025qlu} with applications to the scattering theory.

First, we introduce the following data:
\begin{itemize}
\item{The Hamiltonian $H$ has \emph{real} eigenvalues $\{ E_{\lambda}\}$ labelled by $\{ \lambda \}$.}
\item{Right energy eigenstates are $\{ |\lambda \rangle\}$.}
\item{Linear dual $\{ \widetilde{\langle\lambda |}\}$ of the right energy eigenstates satisfies $ \widetilde{\langle\lambda |}|\eta \rangle=\delta_{\lambda,\eta}$ and forms left energy eigenstates.}
\end{itemize}
where $\delta$ is the Kronecker delta. In other words, just by replacing the complex conjugation with the linear dual, the algebraic description of a general model is the same for Hermitian systems in our settings. To obtain the above properties of the given Hamiltonian, one needs to introduce more respective assumptions, such as $PT$ symmetry\cite{Bender:1998ke,Bender:1998gh}, but the discussions will become more respective to introduce such assumptions. Hence, we assume the above properties\footnote{We thank Naomichi Hatano for clarifying these points.}. We also itemize the aspects that are unusual compared with Hermitian systems.

\begin{itemize}
\item{Hamiltonian $H$ does not satisfy the Hermiticity $H=H^{\dagger}$ in general.}
\item{Linear dual $\{ \widetilde{\langle\alpha |}\}$ does not satisfy the relation $ \widetilde{\langle\alpha |}=(|\alpha\rangle)^{\dagger}$ in general.}
\end{itemize}

In other words, a Hermitian system is a special case of pseudo-Hermitian systems satisfying $H=H^{\dagger}$ and $\widetilde{\langle\alpha |}=(|\alpha\rangle)^{\dagger}$. In the successive sections, we demonstrate their interpretation in two-dimensional unitary and nonunitary CFTs \footnote{Because the nonunitary CFT we study in this section has real eigenenergy, it might be more precise to call them pseudo-unitary CFT or pseudo-Hermitian CFTs.}.

\section{Verlinde line operator and algebraic generalized quantum dimension}
\label{sec_quantum_dimension}

In this section, we revisit the structure of the Verlinde line operators with a formalism applicable to nonunitary conformal field theory corresponding to pseudo-Hermitian systems. We note \cite{DiFrancesco:1997nk,Ginsparg:1988ui} as general references for CFTs, and \cite{Recknagel:2013uja,Northe:2024tnm} as references for BCFTs.

For simplicity, we concentrate our attention on the $A$-type diagonal model. First, we assume the modular invariant as follows,
\begin{equation}
Z=\sum_{\alpha} \chi_{\alpha} (t) \overline{\chi_{\alpha}}(\overline{t})
\end{equation}
where $\alpha$ is the label of primary fields and $t$ ($\overline{t}$) is the chiral (or antichiral) modular parameter and $\chi$($\overline{\chi}$) is the corresponding chiral (antichiral) character\footnote{We have used the symbol $t$ which is less common than $\tau$ to represent the modular parameter for a notational reason. In the successive discussion, we use $\tau$ for the Fibonacci anyon, satisfying the fusion rule $\tau \times \tau =I+\tau$.}. For latter use, we introduce the following modular $T$ and $S$ transformations,
\begin{align}
T:t &\rightarrow t+1, \\
S:t &\rightarrow \frac{-1}{t}.
\end{align}
Phenomenologically, the modular $T$ properties characterize the locality of the systems and the modular $S$ properties characterize the high-low temperature duality of the systems via the open-closed duality\cite{Cardy:1989ir}. 

By assuming the pseudo-Hermiticity of the Hamiltonian, one can express the chiral characters as follows,
\begin{equation}
\chi_{\alpha}(t)= \sum_{M}\widetilde{\langle \alpha,M |}e^{2\pi i t (L_{0}-c/12)}| \alpha, M \rangle
\end{equation}
where $L_{0}$ is the Virasoro generator forming the chiral part of the QFT Hamiltonian and $M$ is the label of descendant fields. In string field theories, the operation mapping a quantum state to the dual vector has been called Belavin-Polyakov-Zamolodchikov (BPZ) conjugate \cite{Belavin:1984vu}(See \cite{Zwiebach:1992ie,Gaberdiel:1997ia,Sen:2016bwe,Erbin:2021smf,Sen:2024nfd}, for example). In the calculations of the chiral characters or modular partition functions in nonunitary CFTs, this structure has been used more or less implicitly in the literature. For the latter use, we introduce the Virasoro algebra $\{ L_{m}\}_{m\in \mathbb{Z}}$,
\begin{equation}
[L_{m},L_{n}]=(m-n)L_{m+n}+\frac{c}{12}(m^{3}-m)\delta_{m,-n}
\end{equation}
where $\mathbb{Z}$ is the set of integer numbers. Corresponding to the eigenvalue of $L_{0}$, and the primary field $\alpha$ has conformal dimension $h_{\alpha}$. By applying the $\{ L\}_{m\le -1}$ recursively to the primary states, one can obtain the corresponding descendant states denoted as $|\alpha, M\rangle$. 

By identifying $\lambda=(\alpha, M)$, the states $\{ |\alpha, M\rangle \}=\{|\lambda\rangle\}$ are in the scope of the general framework of pseudo-Hermitian systems. Hence, corresponding to this representation, we introduce the following projection operators,
\begin{equation}
P_{\alpha, M}=|\alpha,M\rangle \widetilde{\langle \alpha ,M|}
\end{equation}

One can implement corresponding representations for antichiral fields. Then, the following Verlinde line operator\cite{Petkova:2000ip} can be written as follows,
\begin{equation}
\mathcal{Q}_{\alpha}=\sum_{\beta,M, \overline{M}} \frac{S_{\alpha,\beta}}{S_{I,\beta}} P_{\beta,M}\overline{P}_{\beta, \overline{M}},
\label{Verlinde_operator}
\end{equation}
where $S$ is the modular $S$ matrix defined by $\chi_{\alpha}(-1/t )=\sum_{\beta}S_{\alpha,\beta} \chi_{\beta} (t )$ and $I$ is the identity operator corresponding to the vacuum.
One can interprete the Verlinde operator as an exact symmetry of the CFT Hamiltonian because of the representation by projections, i.e., $[\mathcal{Q}_{\alpha},H_{\text{CFT}}]=0$. This is a benefit of this quantum Hamiltonian method, and the set of symmetry operators $\{ \mathcal{Q}_{\alpha} \}$ forms a ring over $\mathbb{C}$ by definition. One can easily detect the noninvertible conserved charges by identifying the zero of the modular-$S$ matrix, and this provides constraints of the phase diagram when studying the massive flows preserving the corresponding symmetries\cite{Fukusumi:2024ejk}. 

For latter use, we introduce the Verlinde formula \cite{Verlinde:1988sn},
\begin{equation}
N^{\gamma}_{\alpha,\beta}=\sum_{\delta}\frac{S_{\alpha,\delta}S_{\beta,\delta}\overline{S_{\delta,\gamma}}}{S_{I,\delta}}
\end{equation}
where $N$ is the fusion matrix defined by the fusion rule $\alpha \times \beta =\sum_{\gamma}N^{\gamma}_{\alpha,\beta}\gamma$. Applying the Verlinde formula, one can obtain the following remarkable relation,
\begin{equation}
\mathcal{Q}_{\alpha}\times \mathcal{Q}_{\beta}=\sum_{\gamma}N^{\gamma}_{\alpha,\beta} \mathcal{Q}_{\gamma}
\label{fusion_ring}
\end{equation}
where $N$ is the nonnegative integer matrix and the symbol $\times$ represents the matrix multiplication.
Hence, one can say that the Verlinde line is a typical object satisfying the algebraic relation between anyons in a CFT, called the fusion ring. This algebraic aspect has captured attentions in the field in the term, symmetry topological field theory (SymTFT)\cite{Apruzzi:2021nmk,Kaidi:2022cpf,Kaidi:2023maf}, by combining these algebraic data and Moore-Seiberg data\cite{Moore:1988qv,Moore:1988ss,Moore:1989vd}. We also note the related earlier work \cite{Fuchs:2002cm} which contains the basic idea of the sandwich construction in the recent terminolgy\cite{kong2015boundarybulkrelationtopologicalorders,Kong:2017etd,Kong:2020cie,Moradi:2022lqp,Bhardwaj:2023bbf,Huang:2023pyk,Wen:2024udn,Fukusumi:2024cnl,Huang:2024ror,Bhardwaj:2024ydc,Fukusumi:2024ejk,Fukusumi:2025ljx}. We also note that the application of SymTFTs to a class of nonequilibrium phenomena, Floquet systems, can be seen in \cite{Motamarri:2023abx,McLauchlan:2025rnw}. We stress that the fusion ring $\{ \mathcal{Q}_{\alpha}\}$ is defined as a ring over $\mathbb{C}$. More phenomenologically, $\{ \mathcal{Q}_{\alpha}\}$ should be treated as a set of conserved charges that has a close connection to topological defects forming a fusion category. Historically, this interpretation of conserved charges has appeared in \cite{Petkova:2000ip}, to demonstrate the physical realization of fusion category structure studied in earlier mathematical literature. In other words, in those days, conserved charges realizing rings over $\mathbb{C}$ were thought of as more physical than category theory. Related to this aspect, it is known that the distinction between conserved charges and topological defects plays a role in studying corresponding lattice realizations\cite{Grimm:2001dr,Belletete:2018eua,Belletete:2020gst,Seiberg:2023cdc,Seiberg:2024gek,Sinha:2025jhh}.  

We also note that the noninvertible symmetry operator defined by linear algebra (or zero of the modular $S$ matrix in our case) can be different from the ``noninvertible symmetry" defined in category theories\footnote{For example, the Fibonacci anyon $\tau$ with the fusion rule $\tau\times \tau=I+\tau$ generates invertible symmetry $\mathcal{Q}_{\tau}$ under the relation $\mathcal{Q}_{\tau}(\mathcal{Q}_{\tau}-\mathcal{Q}_{I})=\mathcal{Q}_{I}$.}. In other words, there can exist categorically noninvertible symmetry operators that are invertible in linear algebra. However, when studying quantum Hamiltonian formalism based on linear algebra, the definition in linear algebra is more dominant when studying the RGs. We believe that there exists some categorical formalisms consistent with the observations in linear algebra, but they seem less common in the physics literature\footnote{A $\mathbb{C}$-linear category will correspond to the linear algebra, but it is less common in the context of fusion rules}. To avoid confusion, we notify the ``noninvertible symmetry" in category theory as categorical noninvertible symmetry or non-group-like symmetry. It also seems reasonable to call the corresponding objects nonabelian symmetry when considering its connection to nonabelian anyons in physics, but the term ``nonabelian'' also appears in group theory and the corresponding group rings in a different meaning.

In this paper, we propose the following quantity as a generalization of the quantum dimension, \emph{algebraic generalized quantum dimension},
\begin{equation}
q_{\alpha,(a)}=\frac{S_{\alpha,a}}{S_{I,a}}
\end{equation}
where $\alpha$ represents type of symmetry or anyons, $a$ represents sector. (We note that our generalization is different from the generalized quantum dimension introduced in the study of representation theory of Grassmannian\cite {onn2006quantumdimensionsnonarchimedeandegenerations}.) By choosing $a$ as the label of the lowest energy state $o$ or the effective vacuum, it is known that the above become positive numbers in unitary CFTs and a wide class of nonunitary CFTs\cite{Gannon:2003de,Beltaos:2010ka}. In a unitary CFT, the relation $o=I$ holds, but in a nonunitary CFT, this does not hold. This swapping of the role of the exact vacuum $I$ and the effective vacuum $o$ is called Galois shuffle\cite{Gannon:2003de} and played a significant role in proving modular noninvariance of gapless fractional quantum Hall states\cite{Milovanovic:1996nj,Ino:1998by,Fukusumi_2022,Fukusumi_2022_c,Fukusumi:2024ejk}. We also note that the quantity $q_{\alpha,(a)}$ has already appeared in a few mathematical studies \cite{Fuchs:1991ci,etingof2015tensor}, or as the coefficients of the Verlinde operators\cite{Petkova:2000ip}, as we revisited Eq. \eqref{Verlinde_operator}. However, its implications for RG flows have not been studied to our knowledge. One can also introduce the antichiral analog of the algebraic generalized quantum dimensions $\overline{q_{\alpha,(a)}}$, and the distinction coming from the chirality plays a fundamental role in studying more general models, such as $Z_{N}$ symmetric models\cite{Kong:2019cuu,Fukusumi:2024ejk,Fukusumi:2025clr,Fukusumi:2025ljx}. However, for simplicity, we mainly focus on the $Z_{2}$ symmetric cases, and focus on the chiral structures in theories. Problems resulting from the chiralities of theories are discussed in Sec. \ref{section_spin} with a possible connection to the Alice ring in the literature\cite{SCHWARZ1982141,SCHWARZ1982427}. 

The appearance of this number can be reproduced straightforwardly, by inserting Eq.\eqref{Verlinde_operator} to Eq. \eqref{fusion_ring} and concentrating attention on the coefficient before the projection labeled by $a$, $P_{a}$. Hence, the following relation holds.
\begin{equation}
q_{\alpha,(a)} \times q_{\beta,(a)}=\sum_{\gamma}N^{\gamma}_{\alpha,\beta}q_{\gamma, (a)}
\end{equation}
In other words, this is an entropic formula representing a conservation of degrees of freedom through the fusions. More mathematically, the quantum dimension can be interpreted as a mapping between the fusion ring $\mathbf{A}$ to $\mathbb{C}$
\begin{equation}
\begin{split}
d_{(a)}: \ \mathbf{A} &\rightarrow \mathbb{C}, \\
d_{(a)}(\alpha)&=q_{\alpha,(a)}.
\end{split}
\end{equation}
As far as we know, this algebraic interpretation of quantum dimension appeared earlier literature\cite{Fredenhagen:1988fj,Fuchs:1989rv,Fuchs:1991ci,Fredenhagen:1992yz} (based on the representation theory), but its phenomenological (or mathematical) implications have not been studied. In the successive discussions, we study their implications for the algebraic structures, modules, and ideals, which provide essential information in classifying the corresponding RGs and TQFTs.

\subsection{(smeared) boundary conformal field theory and entropy}
In this subsection, we propose the form of  boundary conformal field theories (BCFTs) in a way  applicable also to the smeared BCFTs\cite{Cardy:2017ufe}.
The Cardy’s states $|\alpha\rangle$ are given as follows\cite{Cardy:1986gw},
\begin{equation}
|\alpha\rangle =\sum_{\beta}\frac{S_{\alpha,\beta}}{\sqrt{S_{I,\beta}}}|\beta \rangle \rangle
\end{equation}
Where the right-hand side is the corresponding Ishibashi states\cite{Ishibashi:1988kg}. Corresponding to this set of Cardy's states, one can realize the corresponding linear dual of Cardy's states as follows,
\begin{equation}
\widetilde{\langle\alpha|} =\sum_{\beta}\frac{\overline{S_{\alpha,\beta}}}{\sqrt{S_{I,\beta}}}\widetilde{\langle \langle \beta |}
\end{equation}
As can be seen from this expression, Cardy's state is defined by the linear dual, not the complex conjugate. It should be remarked that because we have only modified the definition of the bra and ket to those based on the linear dual, the existing arguments on BCFT is applicable to our case. From this basis, one can obtain the following amplitude of Ishibashi states,
\begin{equation}
 \widetilde{\langle \langle \alpha|}e^{\pi it(L_{0}+\overline{L_{0}}-c/12)}|\beta \rangle \rangle =\delta_{\alpha, \beta} \chi_{\alpha}(t)
\end{equation}

Because of these definitions, one can obtain the following Cardy's condition,
\begin{equation}
\widetilde{\langle\alpha|} e^{\pi it(L_{0}+\overline{L_{0}}-c/12)}|\beta\rangle=\sum_{\gamma} N^{\gamma}_{\alpha,\beta} \chi_{\gamma}(-1/t),
\end{equation}
where we have applied the Verlinde formula for the NIM-rep\cite{Cardy:1989ir}. Hence, the BCFT constructed from our formalism based on the analysis of pseudo-Hermitian systems perfectly produces the existing results on the BCFT spectrum realizable in the corresponding two-dimensional statistical models, such as RSOS models\cite{elDeeb:2015jgf}. The BCFT will also describe the boundary phenomena of the corresponding anyonic chains\cite{Feiguin:2006ydp,Ardonne:2011wxx}, but the corresponding numerical or combinatorial studies are limited. For further studies, we note related works on the defects in the anyonic models\cite{Buican:2017rxc,Belletete:2018eua,Belletete:2020gst,Sinha:2025jhh}.

The boundary entropy is defined by evaluating the dominant contributions of the limit $it \rightarrow -\infty$. From the open string basis, this is a high temperature limit and difficult to evaluate directly, but, from the closed string basis, it can be evaluated easily by observing the coefficient before the lowest energy states. Hence, one can obtain the following dominant asymptotics,
\begin{equation}
\widetilde{\langle\alpha|} e^{\pi it(L_{0}+\overline{L_{0}}-c/12)}|\beta\rangle\sim \frac{S_{\alpha, o}}{\sqrt{S_{I,o}}}\frac{S_{\beta, o}}{\sqrt{S_{I,o}}} e^{-\pi t i c_{\text{eff}}/6 }.
\end{equation}
where $c_{\text{eff}}=c-24h_{o}$ is the effective central charge taking a positive number.
By taking the logarithm, one can obtain the following entropic formula,
\begin{equation}
-\frac{\pi  i tc_{\text{eff}}}{6} +\text{log} g_{\alpha, o} + \text{log} g_{\beta, o}
\end{equation}
where $g_{\alpha, o}=S_{\alpha, o}/\sqrt{S_{I,o}}$ is the so called boundary $g$ entropy\cite{Affleck:1991tk}. We note that these values are positive also in many nonunitary CFTs\cite{Gannon:2003de,Beltaos:2010ka}. Hence, with some appropriate assumptions, such as $PT$-symmetry, we conjecture: 
\begin{equation}
\begin{split}
&\text{The $g$-theorem should be true for nonunitary CFTs}\\ 
&\text{with positive $g$-values.}
\end{split}
\end{equation}
 Compared with the literature introducing negative norm states\cite{Behrend:1999bn}, our discussions on nonunitary BCFTs are simpler. Moreover, one can straightforwardly apply the correspondence between BCFTs and massive RGs in \cite{Cardy:2017ufe}, because we have defined both BCFTs and the symmetry operators in a consistent way, satisfying the following relation,
\begin{equation}
\mathcal{Q}_{\alpha}|\beta\rangle=\sum_{\gamma}N^{\gamma}_{\alpha,\beta} |\gamma\rangle \left(=|\alpha\times\beta\rangle\right)
\label{equation_defect_boundary}
\end{equation}
One can check its consistency only by applying the Verlinde formula\cite{Verlinde:1988sn,Cardy:1989ir}. This relation appeared evidently in \cite{Graham:2003nc} to our knowledge. 

For readers intereted in the implication of these extended boundary states constructed from the application of the defects, Eq. \eqref{equation_defect_boundary} implies the Cardy states with large $g$-value are a consequence of fusing defects to the boundary, and this implements the realization of the general boundary conditions with large $g$-values or boundary degrees of freedom (or qubits)\cite{Fukusumi:2020irh,Okada:2024qmk}. One can interpret these boundary edge modes as a generalization of the Majorana edge modes\cite{Smith:2021luc,Weizmann,Fukusumi:2021zme} or protected edge modes in the symmetry-protected topological phases\cite{Pollmann:2009ryx,Pollmann:2009mhk}. For  readers interested in these edge modes, we note works by the first author\cite{Fukusumi:2023vjm,Fukusumi:2025ljx} and a related work \cite{Wu:2023ezm}. 

\subsection{Examples of our quantum dimension }
In this subsection, we demonstrate some concrete examples of our quantum dimensions in minimal models $M(p,q)$ labelled by integers $(p,q)$. Firstly, we consider the $M(2,5)$ model, which is the simplest non-unitary minimal model.  The $M(2,5)$ model have two chiral primaries $\alpha=\{I,\tau\}$ whose conformal weight are $h_{I}=0,h_{\tau}=-1/5$. Hence, $\tau$ is the lowest energy state $\tau=o$. If we consider the naive quantum dimension \textit{i.e.}, we take $a=I$ for $q_{\alpha,(a)}$, we obtain negative quantum dimension,
\begin{equation}
    q_{I,{(I)}} =1,\; q_{\tau,{(I)}} = \frac{1-\sqrt{5}}{2}<0.
\end{equation}
On the other hand, if we take $a=\tau$, we have a positive quantum dimensions,
\begin{equation}
    q_{I,(\tau)} = 1,\; q_{\tau,{(\tau)}}= \frac{\sqrt{5}+1}{2}>0.
\end{equation}
One can check the consistency of the quantum dimensions by replacing the symbols in the relation $\tau \times \tau=I+\tau$.

Next we consider $M(2,7)$. The $M(2,7)$ model has three chiral primaries $\{I,\Phi,\Psi \}$ whose conformal weights are $h_{I}=0,h_{\Phi}=-2/7,h_{\Psi}=-3/7$. For $a=I$, we again have negative quantum dimensions,
\begin{equation}
    q_{I,(I)}=1,q_{\Phi,(I)}=-2\sin{\qty(\frac{13\pi}{14})},q_{\Psi,(I)}= \frac{\sin{\qty(\frac{\pi}{7})}}{\cos{\qty(\frac{13\pi}{14})}},
\end{equation}
and we see $q_{\Phi,(I)},q_{\Psi,(I)}$ are negative.

On the other hand, when we take $a=\Psi=o$, we have positive quantum dimension,
\begin{equation}
    q_{I,(\Psi)}=1,q_{\Phi,(\Psi)}=\frac{\cos{\qty(\frac{3\pi}{14})}}{\sin{\qty(\frac{\pi}{7})}},q_{\Psi,(\Psi)}= \frac{1}{2\sin{\qty(\frac{\pi}{14})}}.
\end{equation}
It is worth noting that the definition of the quantum dimensions for nonunitary CFTs sometimes differs in the literature, but their mutual implications have not been studied to our knowledge (See \cite{Gannon:2003de} and \cite{Nakayama:2024msv}, for example). We provide them in Sec. \ref{section_massless} and study their implications in concrete exmples in Sec. \ref{section_examples}.

\subsection{Possible gapped phase with Fibonacci fusion ring symmetry}

In this subsection, we demonstrate the benefit of the combination of the smeared BCFT\cite{Cardy:2017ufe} and the Graham-Watts method\cite{Graham:2003nc} for the classification of gapped phases and their quantum states. We concentrate our attention on the Fibonacci fusion symmetry, but the same arguments can be applied to general models. We also note that we provide a complete list of the possible (or realizible) gapped phases, but the realizations in some specific lattice or QFT models need further specific argugements depending on the specific algebraic and analytical data. Our discussion can be regarded as a generalization of the arguments in  \cite{Thorngren:2019iar,Kikuchi:2022ipr,Kikuchi:2023gpj}, by formulating the problem as invariant linear subspace of the modules spanned by smeared BCFTs.

First, let us assume the existence of RG flow from a CFT to a gapped phase while preserving the Fibonacci fusion ring symmetry, $\{ I, \tau\}$, with the following fusion rule,
\begin{equation}
\tau \times \tau =I+\tau
\end{equation}
In Hamiltonian, the RG flow to gapped phase can be expressed as $H_{\text{CFT}}+H_{\text{pert}}$ where $H_{\text{pert}}$ preserve the Fibonacci symmetry $[\mathcal{Q}_{\tau},H_{\text{pert}}]=0$. For simplicity, we concentrate our attention on $\{ I,\tau \}$, but one can consider the RG flow preserving $\{ I,\tau\} \otimes \mathbf{C}'$ where $\mathbf{C}'$ is some arbitrary symmetry operators\footnote{However, to study a larger fusion ring symmetry $\mathbf{C}''(\supset \{ I,\tau\})$, one needs to do the same analysis in the respective situation.}.

Next, we assume that the massive RG flow does not break extended symmetries such as W-symmetry, and the resultant smeared Cardy's states are labelled by the primary fields of the original diagonal CFT. This condition itself is nontrivial and can be broken when the symmetry is anomalous\cite{Affleck:1998nq,Fuchs:1999zi,Quella:2002ct}, but the analysis is nontrivial even restricting our attention to such a well-organized theory. In this setting, the resultant gapped phase should be restricted to the linear subspace of the Hilbert space spanned by the following (smeared) Cardy's states,
\begin{equation} 
\mathcal{H}=\{ x|I\rangle + y|\tau\rangle : x,y\in \mathbb{C} \}
\end{equation}
where $x,y$ are arbitrary complex numbers. By definition, this Hilbert space is invariant under the action of the symmetry $\{ \mathcal{Q}_{\alpha} \}_{\alpha} =\{ \mathcal{Q}_{I}, \mathcal{Q}_{\tau}\}$ implemented as follows,
\begin{equation}
\mathcal{Q}_{\alpha} |\beta\rangle=|\alpha\times \beta\rangle.
\end{equation} 
where $\beta$ is a linear sum of $I$ and $\tau$. 

This action of symmetries on the boundary states has been established in the pioneering work by Graham-Watts\cite{Graham:2003nc}, and its fundamental significance to the classification of gapped states has been studied in the works by the first author\cite{Fukusumi:2024ejk,Fukusumi:2025clr}. It should be stressed that the amplitude of smeared boundary states is defined as the inner product of quantum states. Hence, the amplitude becomes a general complex number, whereas their algebraic structure is determined by the NIM-rep of the symmetry operators. 

As a gapped phase, there exist two choices of one one-dimensional gapped phase,
\begin{align}
\mathcal{H}_{+}=\left\{ x\left(|I\rangle + q_{\tau,+}|\tau\rangle\right) : x\in \mathbb{C} \right\}, \\
\mathcal{H}_{-}=\left\{ x\left(|I\rangle + q_{\tau,-}|\tau\rangle\right) : x\in \mathbb{C} \right\}
\end{align}
where $q_{\tau,\pm}=(1\pm\sqrt{5})/2$.

The above two Hilbert spaces are determined by the following condition,
\begin{equation}
\mathcal{Q}_{\tau} |\alpha \rangle = Q_{\tau, \alpha}| \alpha \rangle
\end{equation}
where $\alpha$ is a linear sum of $I$ and $\tau$ and $Q_{\tau}$ is the corresponding eigenvalue. Hence, the problem is reduced to an eigenvalue problem of the linear transformation. One can obtain the following equations expressing the eigenvalue and eigenvector,

\begin{align}
\mathcal{Q}_{\tau} \left(|I\rangle + q_{\tau,+}|\tau\rangle\right)&=q_{\tau,+}\left(|I\rangle + q_{\tau,+}|\tau\rangle\right)\\
\mathcal{Q}_{\tau} \left(|I\rangle + q_{\tau,-}|\tau\rangle\right)&=q_{\tau,-}\left(|I\rangle + q_{\tau,-}|\tau\rangle\right)
\end{align}
It should be stressed that without changing the normalization, the symmetry operator $\mathcal{Q}_{\tau}$ produces an unusual factor $q_{\tau,\pm}$. By introducing the new normalization as $\mathcal{Q}_{\tau}/q_{\tau,\pm}$, one can obtain unbroken spontaneous symmetry $\{ \mathcal{Q}_{I}, \mathcal{Q}_{\tau}/q_{\tau,\pm}\}$. More generally, the notion of spontaneous symmetry breaking of a state $|\alpha\rangle$ for fusion ring symmetry should be defined up to a constant. 

Interestingly, there exists a correspondence or duality between $\mathcal{H}_{+}$ and $\mathcal{H}_{-}$ by the following $Z_{2}$ operation\cite{Fuchs:1993et},
\begin{align}
I&\rightarrow I, \\
\tau &\rightarrow I-\tau .
\end{align}
In other words, the group automorphism for the fusion ring plays the role of duality $\mathcal{H}_{+}\leftrightarrow \mathcal{H}_{-}$. We note that the above operation cannot be implemented by the multiplication of the linear sum of $I$ and $\tau$ in the original theory. This is different from the usual duality induced from the duality objects, and it seems worth emphasizing. However, we point out that the group automorphisms have played a significant role in studying the construction of extended algebra and associated generalized Verlinde formula in the studies of BCFTs\cite{Affleck:1998nq,Fuchs:1999zi,Quella:2002ct,Ishikawa:2002wx,Ishikawa:2005ea}. Application of the automorphisms for the classification of symmetry-protected topological phases can be seen in \cite{Aksoy:2025rmg}, and a more precise relation between \cite{Aksoy:2025rmg} and our arguments is worth further study. More generally, the objects in an automorphism preserving the unbroken symmetry will exchange the possible gapped phases, and we will study such generalized duality (or $G$-ality) elsewhere.

Consequently, there exist three possible gapped phases, 
\begin{align}
\text{One choice of two-dimensional SSB phase } &\mathcal{H}, \\
\text{Two choices of one-dimensional SUB phase} &\mathcal{H}_{\pm}.
\end{align}
where SSB represents spontaneous-symmetry-breaking and SUB represents symmetry-unbroken. 
By generalizing the notion of the spontaneous symmetry breaking to the fusion ring symmetry, the Fibonacci fusion ring symmetry is spontaneously broken only in $\mathcal{H}$. To our knowledge, this symmetry analysis for the resultant module of the gapped phase is absent in the literature, even when restricting our attention to the simplest model, the Fibonacci fusion ring. We also note that the algebraic objects $I+q_{\tau, \pm}\tau$ form one-dimensional ideals of the original theory, and these ideals induce ring homomorphisms to the one-dimensional ring $\{ I\}$ isomorphic to $\mathbb{C}$ for the latter discussion.

More generally, one can obtain the possible Hilbert space of a gapped phase with symmetry $\mathbf{A}_{\text{ub}}(\subset \mathbf{A})$ by classifying $\mathbf{A}_{\text{ub}}$ invariant modules of $\mathbf{A}$ in this setting. One can also include symmetry-breaking boundary states and their modules (by studying the extended algebra $\mathbf{A}^{\text{ex}}(\supset \mathbf{A})$), and the problem itself can be described in the language of linear algebra. Unfortunately, the category theories in physics are insufficient to solve this problem in linear algebra at this stage. Hence, further studies on more general category theories are fundamental, and we believe that some kind of premodular fusion categories with proper generalizations will solve such classifications in general. We note the works by Kikuchi as related works\cite{Kikuchi:2022ipr,Kikuchi:2023gpj,Kikuchi:2023cgg,Kikuchi:2023eor,Kikuchi:2024hwf,Kikuchi:2024pex} and our discussions provide an elementary method for studying the consequences of the massive RGs only from the established knowledge of linear algebra. It should be noted that the above arguments can provide a series of possible gapped phases, but the problem of detecting which gapped phases are realized in a given model will require more specific arguments. For this purpose, further studies on the eigenvalues $Q_{\alpha_{\text{ub}},\beta}$ of the states $|\beta\rangle$ under the symmetry action $\mathcal{Q}_{\alpha_{\text{ub}}}\in \mathbf{A}_{\text{ub}}$ defined by the relation $\mathcal{Q}_{\alpha_{\text{ub}}}|\beta\rangle=Q_{\alpha_{\text{ub}},\beta}|\beta\rangle$ will be fundamental. Finally, we note several recent works studying or constructing gapped phases with non-group-like (or fusion category) symmetry\cite{Inamura_2021,Inamura:2021szw,Fechisin:2023odt,Seiberg:2024gek,Pace:2024acq,Seifnashri:2024dsd,Cao:2024qjj,Jia:2024bng,Jia:2024zdp,Fukusumi:2024ejk,Cao:2025qhg,Chung:2025ulc}.

\section{Massless flows and matching algebraic generalized quantum dimensions}
\label{section_massless}
The massless RG flow\cite{Zamolodchikov:1987ti,Zamolodchikov:1987jf,Zamolodchikov:1989hfa} (or the RG domain wall\cite{Brunner:2007ur,Gaiotto:2012np}) can be understood as a projection from a ultraviolet (UV) theory to an infrared (IR) theory realizing the ring homomorphism between fusion rings\cite{Wan:2016php,Klos:2019axh,Klos:2021gab}. One of the most surprising facts in \cite{Fukusumi:2025clr} is that the ring homomorphism is \emph{not} unique even when restricting our attention to the simplest flow $M(4,5)\rightarrow M(3,4)$. Hence, it is necessary to study further conditions to detect or characterize each ring homomorphism. In the following, we briefly review the discussions in \cite{Fukusumi:2025clr} expressing the massless RGs or gapped domain walls\cite{Lan:2014uaa} as ring homomorphisms and introduce the benefit of studying the algebraic generalized quantum dimensions in classifying the homomorphisms. The introduction of the algebraic generalized quantum dimension provides systematic data of preserved sectors $\mathbb{S}_{\rho, \times}$ in the following discussions, and this point is a significant progress from the past work \cite{Fukusumi:2025clr}. As a reference for the mathematical discussion in this section, we note an established textbook\cite{atiyah1969introduction}. We also note several recent works in the communities\cite{Kong:2019cuu,Kong:2024ykr,Benedetti:2024utz,Buican:2025zpm,Antinucci:2025fjp}, but the exact method to determine the coefficients has not been studied sufficiently, except for the works on conformal interfaces\cite{Stanishkov:2016pvi,Stanishkov:2016rgv,Poghosyan:2022ecv,Poghosyan:2023brb,Cogburn:2023xzw} \footnote{The literature on conformal interface provides the corresponding analytical data, but they often require extensive analytical calculations}.

First, we introduce the ring homomorphism from a UV fusion ring $\mathbf{A}$ to an IR fusion ring $\mathbf{A}'$. In the subsequent discussions, we distinguish the UV and IR theories by a prime symbol ``\ '". We note that they are equivalent to the transformation law of anyonic objects appearing in the (tensor) functor of category theories or the RG or gapped domain wall in theoretical physics. The ring homomorphism $\rho: \mathbf{A} \rightarrow \mathbf{A}'$ is a mapping satisfying the following conditions,
\begin{align}
\rho(\alpha + \beta)&=\rho(\alpha) + \rho(\beta), \\
\rho(\alpha\times \beta)&=\rho(\alpha) \times \rho(\beta)
\end{align} 
Especially when treating fusion rings with unit, $I$, one can obtain the relation $\rho(I)=I'$. For simplicity, let us assume that the mapping is surjective. From the Noether's second ring isomorphism theorem, one can identify the IR fusion ring $\mathbf{A'}$ as $\mathbf{A'}=\mathbf{A}/ \text{Ker}\rho$, where $\mathbf{I}=\text{Ker} \rho$ is a kernel of $\rho$ satisfying $\rho(\mathbf{I})=\{ 0\}$. The set $\mathbf{I}$ forms an ideal of the UV theory $\mathbf{A}$. Inversely, the classification ideal provides all possible ring homomorphisms. An ideal $\mathbf{I}$ is a set satisfying the relation $\mathbf{A}\times \mathbf{I}=\mathbf{I}$. The remarkable property of the relation $\mathbf{A}\times \mathbf{I}=\mathbf{I}$ is that one can replace the ideal $\mathbf{I}$ to $\{ 0\}$. To our knowledge, the significance of the ideal in fusion rings has been first discussed in \cite{Gepner:1990gr} \footnote{However, it should be kept in mind that this pioneering work appeared much earlier, before the appearance of the fusion category symmetry}.

Corresponding to the fusion rules and algebraic generalized quantum dimensions, we propose the following as a classification criterion,
\begin{equation}  
\begin{split}
&\text{Massless flow $\rho$ preserves some sectors $a\in \mathbb{S}_{\rho}$} \\
&\Rightarrow d_{(a)}(\text{Ker}\rho)=0 \text{ for all $a\in \mathbb{S}_{\rho}$} 
\end{split}
\end{equation}
where $a$ is a label of primary fields and $\mathbb{S}_{\rho}$ is the set specifying the preserved sectors, as we demonstrate in the following discussions. In particular, by choosing $o \in \mathbb{S}$, the flow will correspond to an existing massless integrable RG flow respecting the $c_{\text{eff}}$-theorem.

By generalizing the arguments in \cite{Fukusumi:2025clr}, one can systematically generate the ideal corresponding to $\text{Ker}\rho$ respecting the condition $d_{(a)}(\text{Ker}\rho)=0$. First, let us consider one element $s_{1}=\sum C_{\alpha, 1} \mathcal{Q}_{\alpha}$ with $\sum C_{\alpha,1} q_{\alpha,(a)}=0$ where $\{ C_{\alpha, 1} \}_{\alpha}$ is the set of coefficients. The following will form the ideal preserving sector $a$,
\begin{equation} 
|s_{1}|=\oplus_{\alpha} s_{1}\mathcal{Q}_{\alpha}
\end{equation}
where we have taken the direct sum as a linear space generated by the basis $\{ s_{1}\mathcal{Q}_{\alpha}\}$. Mathematically, this is called the ideal generated by $s_{1}$ (This is often represented as $(s_{1})$ in the literature. We use a less common notation because we apply the homomorphism $d$ to the ideal, $d(|s_{1}|)$, as we show below). Because of the compatibility of the summation and fusion products, one will obtain the relation 
\begin{equation}
d_{(a)} (|s_{1}|)=\{ 0\}. 
\end{equation}
This relation also implies that the objects in $|s_{1}|$ are noninvertible by assuming the matrix representation of $\{ \mathcal{Q} \}$ and their linear algebra. In a similar way, one can construct homomorphisms preserving multiple $(a)$ sectors, and we denote the preserved sector as $\mathbb{S}_{\rho}$. Hence the following defines the $\mathbb{S}_{\rho}$ preserving ideal,
\begin{equation}
d_{(a)} (|s_{1}|)=\{ 0\}, \text{ for all $a\in\mathbb{S}_{\rho}$}. 
\end{equation}
More generally, one can consider the joint or linear sum of the ideals $| s_{i}|$.

Hence one can construct the ring homomorphism preserving the sector $\mathbb{S}_{\rho}$ as follows,
\begin{equation}
\rho_{(a)}: \mathbf{A} \rightarrow \mathbf{A}/\mathbf{I}_{\mathbb{S}_{\rho}}
\end{equation}
where $\mathbf{I}_{\mathbb{S}_{\rho}}$ is an ideal satisfying $d_{a} (\mathbf{I}_{\mathbb{S}_{\rho}})=\{ 0\}$ for all $a\in \mathbb{S}_{\rho}$. In other words, we can obtain the following phenomenologies,
\begin{equation}
\begin{split}
&\text{One can systematically construct IR fusion ring $\mathbf{A}/\mathbf{I}_{\mathbb{S}_{\rho}}$} \\
&\text{by studying the algebraic generalized quantum dimensions.}
\end{split}
\end{equation}
We also note that one can obtain the corresponding symmetry-preserving domain walls\cite{Kaidi:2021gbs} by combining the above discussions with the CFT/TQFT and sandwich construction. 

Interestingly, the algebraic construction implies the existence of the massless flow, which preserves the general sector $(a)$. Even when considering a unitary model, this results in the remarkable consequence that the ground state is not necessarily preserved under the ring homomorphism. In other words, there will exist quantum phase transitions which preserve excited sectors labelled by $(a)$. To some extent, this kind of RGs are conjectural, and we note a few works \cite{Gukov:2015qea,Konechny:2023xvo} in related research directions. However, when interpreting the phenomena as domain walls in $2+1$ dimensional TQFTs, the existence of such unusual homomorphisms is more reasonable (or even trivial).

When assuming the fusion algebraic structure of the UV and IR theories, the above analysis implies the following compatibility of quantum dimensions and homomorphisms for an arbitrary UV object $\alpha$,
\begin{equation}
d_{(a)} (\alpha)=d_{(a')}(\rho(\alpha)) \text{ for all $(a,a')\in\mathbb{S}_{\rho, \times}$}
\end{equation}
where we have used ' to indicate the objects in the IR theories, and $\mathbb{S}_{\rho,\times}$ is the pair of the preserved sector from UV to IR. We also introduce the form of the ring homomorphism as follows,
\begin{equation}
\rho(\alpha)=\sum_{\alpha}A^{\alpha'}_{\alpha} \alpha'
\end{equation}
with 
\begin{equation}
\rho(I)=I'.
\end{equation}

By assuming the compatibility of the quantum dimensions and ring homomorphism, one can obtain the following relations,
\begin{equation}
q_{\alpha, (a)}=\sum_{\alpha'}A^{\alpha'}_{\alpha} q_{\alpha', (a')},
\end{equation}
for the preserved sector $(a,a')\in\mathbb{S}_{\rho, \times}$.
Hence, if we assume that the number of primary fields in the UV and IR theories is $n$ and $n'$ respectively, one will obtain $n-1$ sets of first-order linear equations with $(n-1)\times n'$ variables for each fixed $(a)$. Hence, to obtain a solution uniquely, it is necessary to introduce $\mathbb{S}_{\rho,\times}$ with multiple objects in general. When the homomorphism is not surjective, the solution seems to represent a domain wall between TQFTs rather than a massless RG, whereas we do not provide an explicit proof in this manuscript. We also note that if one takes $(a,a')\notin\mathbb{S}_{\rho,\times}$, the relation $q_{\alpha, (a)}=\sum_{\alpha'}A^{\alpha'}_{\alpha} q_{\alpha', (a')}$ does \emph{not} hold. Hence, corresponding to the choice of algebraic generalized quantum dimension or the preserved reference sectors $\mathbb{S}_{\rho,\times}$ in the RG flows, 
\begin{equation}
\begin{split}
\text{the preservation of quantum dimensions can be broken}. 
\end{split}
\end{equation}
This is a kind of $g$-theorem respecting the quantum symmetry operators and their actions on each quantum state. Phenomenologically, it is fundamental to check that the given homomorphism will preserve the vacuum sectors $(I)$ and $(I')$ or the effective vacuum $(o)$ and $(o')$ with a connection to $c$-theorem\cite{Zamolodchikov:1986gt} or $c_{\text{eff}}$ theorem\cite{Castro-Alvaredo:2017udm} respectively.

\subsection{Duality between massive and massless flows}

In this subsection, we provide a more phenomenological explanation of RG domain walls and gapped domain walls based on the quantum physics of anyons. The discussions here are a variant of those in \cite{Fukusumi:2025clr}, emphasizing the aspects in quantum physics. First, let us consider anyon $\alpha$ in a two-dimensional CFT corresponding to a critical system with fusion symmetry algebra $\mathbf{A}$. In this setting, one can construct a measurement sending an ideal $\mathbf{I}\subset \mathbf{A}$ to zero only from a few calculations in linear algebra (see the corresponding discussions in \cite{Fukusumi:2025clr}). One can obtain the IR theory $\mathbf{A}'=\mathbf{A}/\mathbf{I}$ where ``/'' represents the quotient ring, and one can identify the measurement as application of the ring homomorphism $\rho : \mathbf{A}\rightarrow \mathbf{A}'$. Hence, this measurement induces a quantum phase transition\cite{PhysRevB.98.205136,PhysRevB.100.134306,Skinner:2018tjl} and the anyon $\alpha \in \mathbf{A}$ will split to $\rho(\alpha)=\sum_{\alpha'} A^{\alpha'}_{\alpha}\alpha'$ with $\alpha'\in \mathbf{A}'$. In other words, the massless RG flow is a particular type of measurement-induced quantum phase transition that projects a UV theory to an IR theory by reducing the fusion ring symmetries.

\begin{figure}[htbp]
\begin{center}
\includegraphics[width=0.5\textwidth]{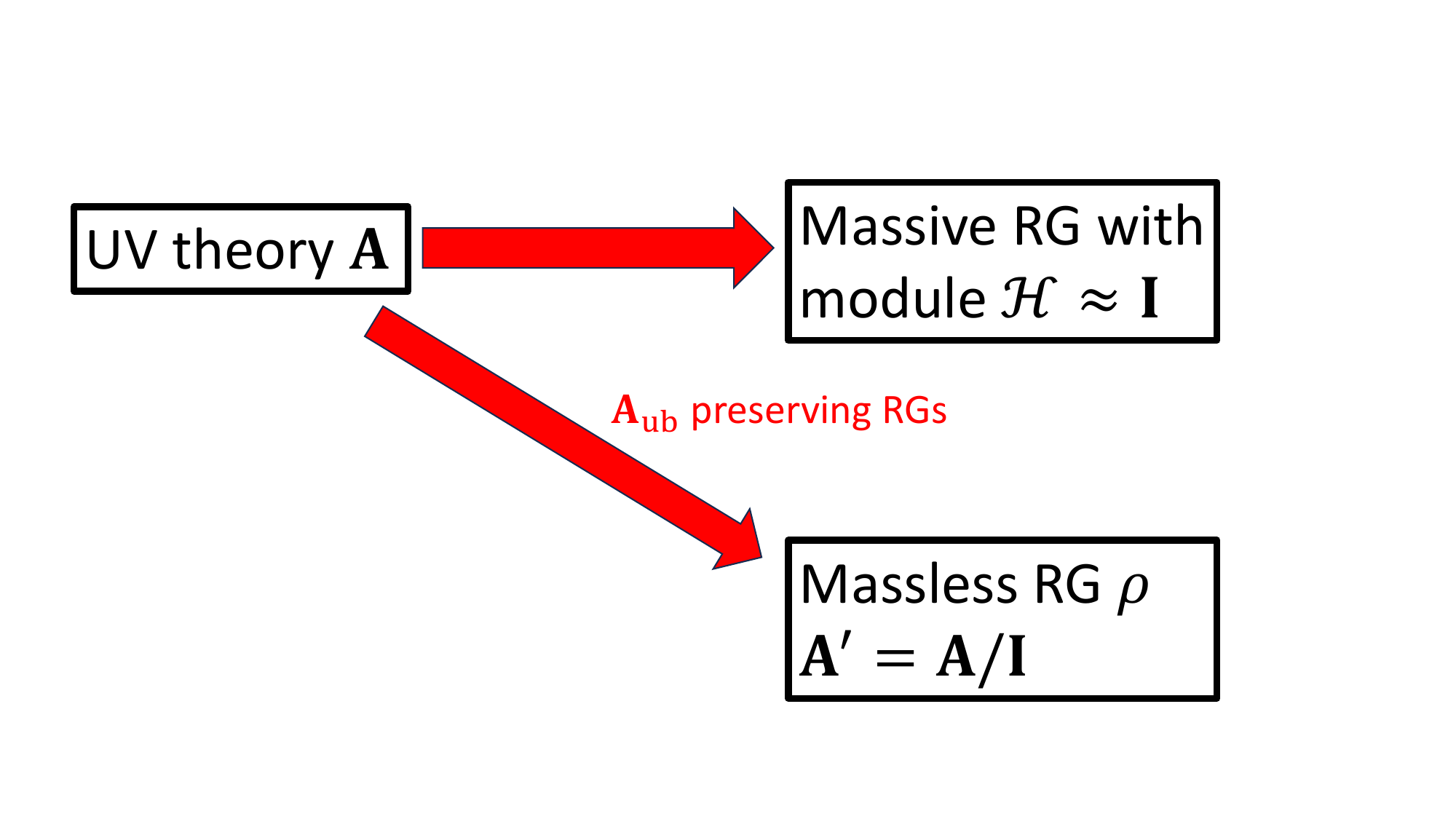}
\caption{ Correspondence between massless and massive RG flows. Existence of an ideal $\mathbf{I}$ preserving $\mathbf{A}_{\text{ub}}$ implies the existence of the corresponding massive RG, by identifying $\mathbf{I}$ as the resultant module $\mathcal{H}$ in the $\mathbf{A}_{\text{ub}}$-preserving massive RG. One can find a similar figure in \cite{Cheng:2026qax}.}
\label{massless_massive}
\end{center}
\end{figure}

In the $2+1$ dimensional TQFTs, this mapping of anyons can be implemented by the application of domain walls\cite{Lan:2014uaa,Hung:2015hfa,Wan:2016php,Zhao:2023wtg}. Interestingly, if the homomorphism $\rho$ preserves a subring $\mathbf{A}_{\text{ub}}\subset \mathbf{A}$ (or satisfying $\mathbf{A}_{\text{ub}} \cap \text{Ker}(\rho)=\{0\}$), the ideal $\mathbf{I}$ can be identified as $\mathbf{A}_{\text{ub}}$ invariant module $\mathcal{H}$ when assuming the Moore-Seiberg data\cite{Moore:1988qv,Moore:1988ss,Moore:1989vd,Fuchs:2002cm} in the original ring $\mathbf{A}$, by identifying the labels of $\mathcal{H}$ to the algebraic objects in $\mathbf{A}$. This mapping from an ideal $\mathbf{I}$ to a module $\mathcal{H}$ is called a fiber functor, and this is a typical example of a forgetful functor in category theory\footnote{A fiber functor should be distinguished from a tensor functor corresponding to a massless RG}. In short, one can express this correspondence as an equivalence $\mathcal{H}=\mathbf{I}$ in the UV theory $\mathbf{A}$. In other words, there exists a general correspondence of condensable block $\mathbf{I}$ in a massless RG with unbroken symmetry $\mathbf{A}_{\text{ub}}$ and some module $\mathcal{H}$ of $\mathbf{A}_{\text{ub}}$ symmetry preserving massive RG flows (Fig.\ref{massless_massive}). We also note that this relationship between a massless RG and the dual massive RG can be seen in a very recent work \cite{Cheng:2026qax}, whereas their methods are different from ours.

In the perturbative QFT formalism, this phenomenology itself can be explained more reasonably. First, let us introduce some perturbation $mH_{\text{pert}}$ preserving $\mathbf{A}_{\text{ub}}$ symmetry where $m$ is a coupling constant. In general, it is known that the sign of the coupling constant determines the property of RG flow, either massless or massive, whereas the unbroken symmetry is fixed to $\mathbf{A}_{\text{ub}}$. Hence, one can expect some kind of dual relation between the corresponding massless and massive RG flows, and the relation $\mathcal{H}= \mathbf{I}$ in the original UV theory $\mathbf{A}$ represents the expected duality. 
However, we also note that there can exist a module $\mathcal{H}$ which cannot be represented as an ideal in the UV theory $\mathbf{A}$. Hence, the existence of the massive RG flow cannot imply the corresponding existence of the massless RG flow in general, whereas the inverse is true for a series of models satisfying the Moore-Seiberg data\cite{Moore:1988qv,Moore:1988ss,Moore:1989vd,Fuchs:2002cm}. The discussions in this subsection seem to be related to gauging operations for non-group-like symmetries\cite{Frohlich:2009gb,Lu:2022ver,Perez-Lona:2023djo,Choi:2023vgk,Diatlyk:2023fwf,Perez-Lona:2024sds}, and the further clarification of their precise relationship is an interesting future problem.

\subsection{Tensor decomposition and gapped domain wall problem: A benefit of introducing algebraic generalized quantum dimension}

In this subsection, we study the implications of the algebraic generalized quantum dimensions to the gapped or RG domain wall problem when the UV CFT $\mathbf{A}$ can be expressed as a tensor product of the domain wall $\mathbf{A''}_{\text{DW}}$ and IR CFT $\mathbf{A'}$. The arguments in this section can be regarded as a generalization of the Higgs transitions involved with coset or level-rank duality structures, and we note several earlier works\cite{Kaplan:1983fs,Kaplan:1983sm,Georgi:1984af,Dugan:1984hq} and a few reviews\cite{Panico:2015jxa,Watanabe:2019xul,Brauner:2024juy}. For readers interested in related discussions on the sigma model or gauge theories, we note a few works with different methods\cite{Fendley:1999gb,Zhao:2025zsb,Apruzzi:2025hvs}. For example, one can starightforwardly applies the argument in this section to the situation where the UV CFT, $G_{K}$ Wess-Zumino-Witten model\cite{Wess:1971yu,Witten:1983ar,Witten:1983tw}, can be decomposed to the $G_{K}/H_{K}$ coset WZW model\cite{Goddard:1984vk,Goddard:1984hg,Goddard:1986ee} and the $H_{K}$ WZW model, by identifying the domain wall as $G_{K}/H_{K}$ and the IR theory as $H_{K}$. We do not assume the chirality of objects and theories, and one can apply the arguments in this section to a more general setting by replacing the chirality of them, such as $\mathbf{A}\leftrightarrow\overline{\mathbf{A}}$. We discuss some involved problems resulted from the chirality in Sec. \ref{section_spin}. We note references on related problems \cite{Kong:2019cuu,Fukusumi:2024ejk,Fukusumi:2025clr,Fukusumi:2025ljx}. 

First, we assume the following decomposition,
\begin{equation}
\begin{split}
\mathbf{A}&=\mathbf{A'} \otimes \mathbf{A''}_{\text{DW}} \\
\alpha &=\sum_{\alpha''_{\text{DW}},\alpha'} B^{\alpha}_{\alpha', \alpha''_{\text{DW}}} \alpha''_{\text{DW}}\otimes \alpha'
\end{split}
\end{equation}
where $B$ is a nonnegative integer matrix realizing the tensor decomposition of the UV theory, $\alpha'$ is an object in the IR theory, and $\alpha''_{\text{DW}}$ is an object in the domain wall theory. For the latter discussion, we stress that this decomposition implements the ring isomorphism $\rho$ between UV theory $\mathbf{A}$ and the tensor product of IR theory $\mathbf{A'}$ by tracing out the domain wall theory $\mathbf{A''}_{\text{DW}}$.

Next, we assume the algebraic generalized quantum dimension is well defined for the domain wall $\mathbf{A''}_{\text{DW}}$ as follows,
\begin{equation}
\begin{split}
d_{(a''_{\text{DW}})}: \ \mathbf{A''}_{\text{DW}}&\rightarrow \mathbb{C} \\
d_{(a''_{\text{DW}})}(\alpha''_{\text{DW}})&=q_{\alpha''_{\text{DW}}, (a''_{\text{DW}})}
\end{split}
\end{equation}

\begin{figure}[htbp]
\begin{center}
\includegraphics[width=0.5\textwidth]{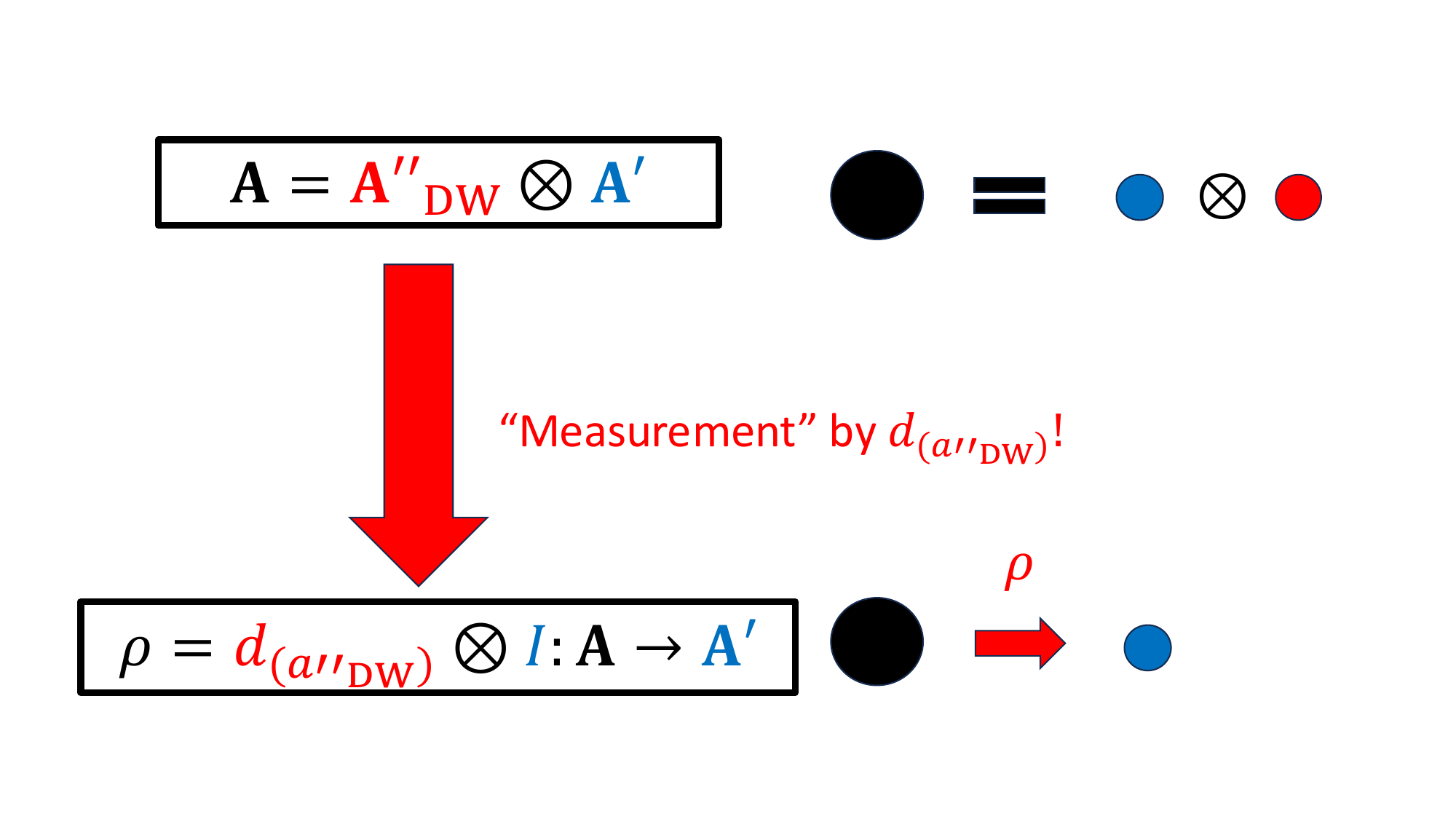}
\caption{ Tensor decomposition of UV CFT and the resultant IR theory. The red color specifies the objects and operations in the theory at the domain wall and the blue color specifies the objects in the IR theory. By tracing out the theory at the domain wall by applying the homomorphism $d_{a''_{\text{DW}}}$ realizing the algebraic generalized quantum dimensions, one can obtain the homomorphism $\rho: \mathbf{A}\rightarrow \mathbf{A'}$. }
\label{domain_wall_projection}
\end{center}
\end{figure}

By applying this homomorphism to the domain wall part of the object in $\mathbf{A'} \otimes \mathbf{A''}_{\text{DW}}$, one can implement the following ring homomorphism straightforwardly,
\begin{align}
\rho_{a''_{\text{DW}}}&=d_{a''_{\text{DW}}}\otimes I': \mathbf{A}\rightarrow \mathbf{A'}, \\
\rho_{a''_{\text{DW}}}(\alpha)&=\sum_{\alpha''_{\text{DW}},\alpha'} B^{\alpha}_{\alpha', \alpha''_{\text{DW}}} q_{\alpha_{\text{DW}},(a''_{\text{DW}})}\otimes \alpha'
\end{align}

Hence, this representation determines the coefficient of the homomorphism $A$ as follows,
\begin{equation}
A^{\alpha}_{\alpha'}=\sum_{\alpha''_{\text{DW}}} B^{\alpha}_{\alpha', \alpha''_{\text{DW}}} q_{\alpha_{\text{DW}},(a''_{\text{DW}})}
\end{equation}
In other words, the multiplicity of the solutions corresponds to the choice of the domain wall particle $a''_{\text{DW}}$. Next section, we study homomorphisms between theories where the domain wall theories $\mathbf{A}''_{\text{DW}}$ are not known, but the existence of the corresponding massless RGs has been established in the literature.

Before moving into the next section, we note that the correspondence between the massless RGs and coset CFTs has already been studied in the literature with a connection to the color confinement \cite{LeClair:2001yp}. There exists more general studies on related integrable deformations, called $\lambda$-deformation or current-current deformation\cite{Sfetsos:2017sep,Georgiou:2018gpe}. We note a recent review for further literature \cite{Borsato:2023dis}. More recently, related domain walls and coset structure have been studied in \cite{Cordova:2023jip,Cordova:2025eim} by applying anyon condensation and a related mathematical approach\cite{Frohlich:2003hm,Bais:2008ni}. For more historical aspects, we note the corresponding remarks in the works by the first author\cite{Fukusumi:2024ejk,Fukusumi:2025clr}.

\section{Examples of ring homomorphism in unitary and nonunitary models}
\label{section_examples}

Before moving into the detailed studies, we note the ring isomorphism of the Fibonacci fusion rule. For readers interested in existing RG flows of nonunitary CFTs, we note several earlier references\cite{Berkovich:1997ht,Dorey:2000zb,Dunning:2002cu} and recent ones\cite{Blondeau-Fournier:2017otv,Nakayama:2024msv}. Related structures have also been observed in the hierarchical structure of fractional quantum Hall states\cite{PhysRevLett.60.956,PhysRevB.75.075317,Davenport:2012fcs,Bernevig_2008,Yuzhu_2023,Bourgine:2024ycr}. More recently, there exist several developments on the unbroken symmetries $\mathbf{A}_{\text{ub}}$ in some series of massless RG flows\cite{Nakayama:2024msv,Kikuchi:2024cjd,Chen:2025qub,Gaberdiel:2026sfg,Ambrosino:2026umb,Benedetti:2026drn}. In other words, even the studies on the unbroken symmetry have not been studied sufficiently. We also remind that the unbroken symmetries in a massless RG are fundamental, but they are insufficient to characterize or determine the corresponding homomorphism uniquely\cite{Fukusumi:2025clr}. Hence, in this section, we show some detailed calculations determining the homomorphisms with suitable assumptions for further studies. Our method is straightforward just determining the transformation laws by the relation $\rho(\alpha\times \beta)=\rho(\alpha)\times \rho(\beta)$. We intend to provide concrete algebraic data of the possible set of homomorphisms in a few series of models. Detection of homomorphisms in more physical or realistic settings require more respective methods, such as the truncated conformal space approach\cite{Yurov:1989yu,Yurov:1991my}, and this problem is out of the scope of the present manuscript.

For later use, we note the analysis of the automorphisms between Fibonacci fusion rules. The Fibonacci fusion rule is formed by the two objects $\{ I, \tau \}$ satisfying the following relations,
\begin{equation}
\tau \times \tau =I +\tau,
\end{equation}
where $I$ is the identity operator again. The ring isomorphism is necessary to satisfy the following relations, 
\begin{align} 
\rho(I)&=I, \\
\rho (\tau) &= A_{\tau}^{I} I + A_{\tau}^{\tau}\tau
\end{align}
where $A_{\tau}^{I}, A_{\tau}^{\tau}$ is constants satisying the requirement of ring isomorphism,
\begin{equation}
\rho(\tau \times \tau)=\rho(\tau) \times \rho(\tau)
\end{equation}
We also assume $A_{\tau}\neq 0$ to obtain the surjective homomorphism. Then one obtains the following relations,
\begin{align}
A_{0}^{2}+A_{\tau}^{2}&=A_{0}+1, \\
2A_{0}A_{\tau}+A_{\tau}^{2}&=A_{\tau}
\end{align}
Hence, one can obtain the following two solutions,
\begin{equation}
(A_{0}, A_{\tau})=(0,1), (1,-1)
\end{equation}
The solution $(A_{0}, A_{\tau})=(0,1)$ corresponds to the trivial ring isomorphism and $(A_{0}, A_{\tau})=(1,-1)$ is the nontrivial one. As far as we know, the ring isomorphism corresponding to the latter first appeared in \cite{Fuchs:1993et}. Remarkably, this ring isomorphism combined with the quantum dimension satisfies the following relation,
\begin{equation}
d_{(0)} (\rho(\tau)) =1-q_{\tau, (\tau)} 
\end{equation}
or equivalently,
\begin{equation}
d_{(\tau)} (\rho (\tau))=1-q_{\tau, (0)}.
\end{equation}
One can confirm these relations by studying the $M(2,5)$ minimal model, known as Yang-Lee CFT. More phenomenologically, this changes the quantum dimensions to its dual of the characterstic polynomial equation $x^{2}=x+1$ obtained from the fusion rule $\tau \times \tau =I+\tau$. Because the corresponding ideal is trivial, only such an exchange is permitted. We also note that there exist two non-surjective ring homomorphisms,
\begin{equation}
(A_{0}, A_{\tau})=\left(\frac{1\pm\sqrt{5}}{2},0 \right)
\end{equation}
for the latter use.

\subsection{$M(4,5)$ to $M(3,4)$}

First, we revisit the ring homomorphism in unitary CFTs from $M(4,5)$ to $M(3,4)$ with emphasis on the quantum dimensions\cite{Fukusumi:2025clr}. The $M(4,5)$ model or tricritical Ising CFT can be represented as a tensor product of Ising and Fibonacci fusion ring, $\{ I, \tau \} \otimes \{ I, \psi , \sigma \}$. Following the analysis in \cite{Fukusumi:2025clr}, one can represent the homomorphism as follows,
\begin{align}
\rho(\{ I, \psi, \sigma\})&=\{ I', \psi', \sigma' \}, \\
\rho(\tau)&=A_{I'} I' +A_{\psi'} \psi'
\end{align}
where we have assumed that the homomorphism $\rho$ acts on the Ising fusion ring part trivially, and the $Z_{2}$ anomaly is preserved. One can obtain the four solutions satisfying the above conditions,
\begin{equation}
(A_{I'}, A_{\psi'})= \left(\frac{1\pm\sqrt{5}}{2},0\right), \left(\frac{1}{2}, \pm \frac{\sqrt{5}}{2}\right).
\end{equation}
Corresponding to the sign $\pm$, the resultant quantum dimensions $d_{(0')}\left(\rho\left(\tau\right)\right)$ takes the form $q_{\tau, (0)}=\frac{1+\sqrt{5}}{2}$ or $q_{\tau, (\tau)}=\frac{1-\sqrt{5}}{2}$. Hence, the preserved sector, $I$ or $\tau$, can be different depending on the choice of homomorphism. We also remind that the conditions $A_{\psi'}=0$ or $A_{\psi'}\neq 0$ determine whether the fermion $Z_{2}$ parity is conserved or not. Hence, we conclude that the massless RG from $M(4,5)$ to $M(3,4)$ or the embedding of the theory $M(3,4)$ to $M(4,5)$ is determined by the shuffle structure of $\{ I, \tau \}$ and fermion parity conditions.

\subsection{$M(3,4)$ to $M(2,5)$}
It is widely known that there exists a massless  integrable flow between the Ising CFT and the Yang-Lee CFT, denoted as $M(3,4) \rightarrow M(2,5)$. The corresponding ring homomorphism has not been studied to our knowledge, whereas the related conformal interfaces have been studied in \cite{Quella:2006de}. Hence, we provide the ring homomorphism that corresponds to the RG flow.

First, we fix the theories with nonnegative integer fusion coefficients for simplicity. The Ising CFT has the three chiral primary fields $\{ I, \psi, \sigma\}$ with the following relations,
\begin{align} 
\psi \times \psi &=I, \\
\psi \times \sigma &=\sigma, \\
\sigma \times \sigma&=I+\psi.
\end{align}
where $\times$ represents the fusion product and $I$ is the identitiy operator. The operator $\psi$ is called  the (chiral) Majorana fermion field and generates $Z_{2}$ symmetry, and $\sigma$ is the (chiral) order field or spin field. In the literature on TQFT or TO, $\sigma$ is called an Ising anyon.

The $M(2,5)$ model has two objects $\{ I, \tau \}$ that satisfy the Fibonacci fusion rule. To distinguish the UV and IR theory, we denote the IR objects as $\{ I', \tau' \}$.

As one can see, the Ising CFT has three objects and the Yang-Lee CFT has two objects. Moreover, the Ising model has $Z_{2}$ group ring structure $\{ I, \psi\}$. It is known that the $Z_{2}$ symmetry should be violated under the massless RG. Hence, the following is a candidate for the corresponding homomorphism,
\begin{align} 
\rho(I)&=I', \\
\rho(\psi) &= A^{I'}_{\psi} I' +A^{\tau'}_{\psi} \tau',\\
\rho (\sigma) &= A^{I'}_{\sigma} I' + A^{\tau'}_{\sigma}\tau',
\end{align}
where $A_{0}, A_{\tau}$ is constants satisying the requirement of ring homomorphism. For this case, the following are the sufficient conditions
\begin{align}
\rho(\psi\times \psi)= \rho(\psi) \times \rho(\psi), \\
\rho(\psi\times \sigma)= \rho(\psi) \times \rho(\sigma), \\
\rho(\sigma\times \sigma)= \rho(\sigma) \times \rho(\sigma), 
\end{align}

The first relation results in the following,
\begin{align}
(A_{\psi}^{I'})^{2}+(A_{\psi}^{\tau'})^{2}&=1, \\
2A_{\psi}^{I'}A_{\psi}^{\tau'} +(A_{\psi}^{\tau'})^{2} &=0, 
\end{align}
Hence, the solution is,
\begin{equation}
(A_{\psi}^{I'},A_{\psi}^{\tau'})=\pm (\frac{1}{\sqrt{5}}, -\frac{2}{\sqrt{5}}), (\pm 1,0)
\end{equation}

The second condition results in the following relation,
\begin{align}
A_{\psi}^{I'}A_{\sigma}^{I'} +A_{\psi}^{\tau'}A_{\sigma}^{\tau'} &=A_{\sigma}^{I'}, \\
A_{\psi}^{I'}A_{\sigma}^{\tau'} +A_{\psi}^{\tau'}A_{\sigma}^{I'} +A_{\psi}^{\tau'}A_{\sigma}^{\tau'} &=A_{\sigma}^{\tau'}, 
\end{align}

The third condition results in the following equations,
\begin{align}
(A_{\sigma}^{I'})^{2}+(A_{\sigma}^{\tau'})^{2}&=1+A_{\psi}^{I'}, \\
2A_{\sigma}^{I'}A_{\sigma}^{\tau} +(A_{\sigma}^{\tau'})^{2} &=A_{\psi}^{\tau'}, 
\end{align}

For the solution $(A^{I'}_{\psi},A^{\tau'}_{\psi})=(0,-1)$, the second equation results in $(A^{I'}_{\sigma},A^{\tau'}_{\sigma})=(0,0)$. Hence, the solution does not provide a surjective homomorphism. For $(A^{I'}_{\psi},A^{\tau'}_{\psi})=(0,1)$, there exist two solutions,
\begin{equation}
(A_{\sigma}^{I'},A_{\sigma}^{\tau'})=\pm (\frac{\sqrt{2}}{\sqrt{5}}, -\frac{2\sqrt{2}}{\sqrt{5}})
\end{equation}

By applying the same analysis to , one can obtain,
\begin{equation}
(A_{\sigma}^{I'},A_{\sigma}^{\tau'})=\pm (\frac{\sqrt{5}+1}{\sqrt{10}}, -\frac{\sqrt{2}}{\sqrt{5}})
\end{equation}
for the case with $(A^{I'}_{\psi},A^{\tau'}_{\psi})=(1/\sqrt{5},-2/\sqrt{5})$, and
\begin{equation}
(A_{\sigma}^{I'},A_{\sigma}^{\tau'})=\pm (\frac{\sqrt{5}-1}{\sqrt{10}}, \frac{\sqrt{2}}{\sqrt{5}})
\end{equation}
fo the case with $(A^{I'}_{\psi},A^{\tau'}_{\psi})=(-1/\sqrt{5},2/\sqrt{5})$.

Hence, there exist six solutions by assigning the condition of ring homomorphisms, and it is necessary to interpret the solutions. For this purpose, we observe the algebraic generalized quantum dimensions and their compatibility under the homomorphism $\rho$ for the solution with $(A^{I'}_{\psi},A^{\tau'}_{\psi})=(0,1)$ for simplicity.

First, the sector $\rho(I)=\rho(\psi)= I'$ does not produce nontrivial relations of quantum dimensions. Hence, we concentrate our attention on the relation $\rho (\sigma) = A I' + B\tau'$.

In general, one can calculate the quantum dimensions of $\rho(\sigma)$ as follows,
\begin{equation}
d_{(a')} (\rho(\sigma)) = A + Bq_{\tau', (a')}
\end{equation}
When the UV vaccum flows to the sector $a'$, the above implies the relation $A+Bq_{\tau',(a)}=\sqrt{2}$.

Hence, one can interprete the solution $(A,B)=(\frac{\sqrt{2}}{\sqrt{5}}, -\frac{2\sqrt{2}}{\sqrt{5}})$ as the vacuum, $(a,a')=(I,I')\in \mathbb{S}_{\rho,\times}$, preserving flow and the solution $(A,B)=(-\frac{\sqrt{2}}{\sqrt{5}}, \frac{2\sqrt{2}}{\sqrt{5}})$ corresponds to the effective vacuum $(a,a')=(I,\tau')\in \mathbb{S}_{\rho,\times}$ preserving domain wall. Reflecing the choices of the preserved sector, the structure $\mathbb{S}_{\rho, \times}$ becomes $\mathbb{S}_{\rho,\times}=\{ (I,I')\}$ or $\mathbb{S}_{\rho,\times}=\{ (I,\tau')\}$ accordingly. Because of the $c_{\text{eff}}$-theorem for the $PT$-symmetric systems\cite{Castro-Alvaredo:2017udm}, the latter solution corresponds to the existing massless integrable flows. One can apply the same analysis to the other four solutions and their relation to the results in \cite{Quella:2006de} is an interesting problem.

\subsection{$M(3,5)$ to $M(2,5)$}
In this section, we study the flow from the $M(3,5)$ minimal model to the $M(2,5)$ minimal model. First, the fusion rule of the $M(3,5)$ minimal model can be represented as $\{ I, J \} \otimes\{ I, \tau\}=\{ I, J, \tau, J\tau\}$, where $J$ is the anomalous $Z_{2}$ simple current and $\tau$ is the Fibonacci anyon again. Then, by specifying the action of the ring homomorphism to the tensor decomposition, one can obtain the following homomorphism from the part $\{ I, J\}$,
\begin{align}
\rho(I)&=I' \\
\rho(J)&= A^{I'}_{J}I'+A^{\tau'}_{J}\tau'
\end{align}
This results in the following relations,
\begin{align}
(A^{I'}_{J})^{2}+(A^{\tau'}_{J})^{2}&=1, \\
2A^{I'}_{J}A^{\tau'}_{J}+(A^{\tau'}_{J})^{2}&=0, \\
\end{align}
This results in the following surjective solutions, with $\rho:\{ I, J\}\rightarrow \{ I', \tau'\}$,
\begin{equation}
(A^{I'}_{J}, A^{\tau'}_{J})= \pm (\frac{1}{\sqrt{5}}, -\frac{2}{\sqrt{5}})
\end{equation}
and nonsurjective ones, $\rho:\{ I, J\}\rightarrow \{ I'\}$, 
\begin{equation}
(A^{I'}_{J}, A^{\tau'}_{J})=(\pm 1,0), 
\end{equation}
where the former two relations are nontrivial ring homomorphisms, and the rest correspond to the group homomorphisms $Z_{2}\rightarrow \{ I\}$. 

For the solution $(A^{I'}_{J}, A^{\tau'}_{J})= \pm (\frac{1}{\sqrt{5}}, -\frac{2}{\sqrt{5}})$, one can choose the action of homomorphism to $\{ I, \tau\}$ as ring isomorphisms $\{I,\tau\}\rightarrow \{ I,\tau'\}$ or ring homomorphisms $\{ I,\tau\}\rightarrow\{ I'\}$. Hence, there exists $2\times 4=8$ ring homomorphisms.

For the solution $(A^{I'}_{J}, A^{\tau'}_{J})= \pm (1, 0)$, it is necessary to introduce the ring isomorphism $\{ I, \tau\}\rightarrow \{ I,\tau'\}$. Hence, there exist $2\times 2=4$ solutions. Totally, one can obtain $8+4=12$ homomorphisms. One of the remarkable points of the above homomorphism is the breaking of the $Z_{2}$ structure and the correspondence of the Fibonacci fusion ring sector. Interestingly, there exist solutions which break the $\{ I, \tau \}$ but produce $\{ I',\tau'\}$ from the sector $\{ I,J\}$ for the case with $(A^{I'}_{J}, A^{\tau'}_{J})= \pm (\frac{1}{\sqrt{5}}, -\frac{2}{\sqrt{5}})$. In other words, the symmetry $\{ I, \tau\}$ and the emergent symmetry $\{ I', \tau' \}$ match accidentally or by coincidence in these homomorphisms\footnote {The term, accidental, is different from that used to mean the small (or irrelevant) breaking of symmetry.}. Further research on such accidental matching or coincidence of the symmetry subring in UV and the emergent symmetry subring at IR is worth further study.

\subsection{$M(2,7)$ to $M(2,5)$}

In this subsection, we consider the ring homomorphism from $M(2,7)$ to $M(2,5)$ as a more complicated example. The $M(2,7)$ minimal model has three chiral primaries $\{I, \Phi, \Psi\}$ with the fusion rule,
\begin{align}
    \Phi \times \Psi &= \Phi + \Psi,\\
    \Phi \times \Phi &= I + \Psi,\\
    \Psi \times \Psi &= I + \Phi + \Psi.
\end{align}
Following the analysis so far, we consider a ring homomorphism defined by
\begin{equation}
    \begin{split}
        \rho(I) &= I',\\
        \rho(\Phi) &= A_{\Phi}^{I'} I' + A_{\Phi}^{\tau'} \tau',\\
        \rho(\Psi) &= A_{\Psi}^{I'} I' + A_{\Psi}^{\tau'} \tau'.
    \end{split}
\end{equation}
To preserve the fusion ring, we obtain six equations:
\begin{align}
     A_{\Phi}^{I'} A_{\Psi}^{I'} + A_{\Phi}^{\tau'} A_{\Psi}^{\tau'} &= A_{\Phi}^{I'} + A_{\Psi}^{I'},\\
     A_{\Phi}^{I'} A_{\Psi}^{\tau'} + A_{\Phi}^{\tau'} A_{\Psi}^{I'} + A_{\Phi}^{\tau'} A_{\Psi}^{\tau'} &= A_{\Phi}^{\tau'} + A_{\Psi}^{\tau'},\\
     (A_{\Phi}^{I'})^2 + (A_{\Phi}^{\tau'})^2 &= 1 + A_{\Psi}^{I'}, \label{eq:A_Psi^I}\\
     2A_{\Phi}^{I'} A_{\Phi}^{\tau'} + (A_{\Phi}^{\tau'})^2 &= A_{\Psi}^{\tau'}, \label{eq:A_Psi^tau}\\
     (A_{\Psi}^{I'})^2 + (A_{\Psi}^{\tau'})^2 &= 1 + A_{\Phi}^{I'} + A_{\Psi}^{I'},\\
     2A_{\Psi}^{I'} A_{\Psi}^{\tau'} + (A_{\Psi}^{\tau'})^2 &= A_{\Phi}^{\tau'} + A_{\Psi}^{\tau'}.
\end{align}

Since $\rho$ is surjective, we require $A_{\Phi}^{\tau'} \neq 0$ and $A_{\Psi}^{\tau'} \neq 0$. Using these conditions, the six equations can be reduced to a single polynomial equation for $A_{\Phi}^{I'}$:
\begin{equation}\label{eq:eq_A_phi^I}
\begin{split}
     &\quad 125 (A_{\Phi}^{I'})^6 - 250 (A_{\Phi}^{I'})^5 - 25 (A_{\Phi}^{I'})^4 
     \\&+ 225 (A_{\Phi}^{I'})^3 - 30 (A_{\Phi}^{I'})^2 - 45 A_{\Phi}^{I'} - 1 = 0.
\end{split}
\end{equation}

We confirm that this equation has six real solutions. For each solution, $A_{\Phi}^{\tau'}$ is determined by
\begin{equation}
    A_{\Phi}^{\tau'} = \frac{5 (A_{\Phi}^{I'})^3 - 5 (A_{\Phi}^{I'})^2 + 4 A_{\Phi}^{I'} - 2}
    {-15 (A_{\Phi}^{I'})^3 + 10 A_{\Phi}^{I'} + 3}.
\end{equation}
Similarly, $A_{\Psi}^{I'}$ and $A_{\Psi}^{\tau'}$ are determined by Eq. \eqref{eq:A_Psi^I} and Eq. \eqref{eq:A_Psi^tau}, respectively. Numerically, we obtain the following solutions:
\begin{equation}
    \begin{split}
        &(A_{\Phi}^{I'}, A_{\Phi}^{\tau'}, A_{\Psi}^{I'}, A_{\Psi}^{\tau'}) \\
        =& (0.820079, 0.606822, 0.0407623, 1.36352), \\
        & (1.4269, -0.606822, 1.40428, -1.36352), \\
        & (-0.0226214, -0.756695, -0.426901, 0.606822), \\
        & (-0.40428, 1.36352, 1.02262, 0.756695), \\
        & (-0.779316, 0.756695, 0.179921, -0.606822), \\
        & (0.959238, -1.36352, 1.77932, -0.756695).
    \end{split}
\end{equation}

In conclusion, we find six solutions satisfying the homomorphism conditions. Interestingly, there exists only one positive solution $(A_{\Phi}^{I'}, A_{\Phi}^{\tau'}, A_{\Psi}^{I'}, A_{\Psi}^{\tau'})= (0.820079, 0.606822, 0.0407623, 1.36352)$, and the resultant homomorphism has the preserving sectors $\mathbb{S}_{\rho,\times}=\{(I,I'),(o,o') \}$. We conjecture that this positive solution corresponds to the existing massless RG flow, and further studies based on other methods are necessary for the confirmation of our prediction. Moreover, there exists a controversy between the perturbative expressions on the flow $M(2,7)\rightarrow M(2,5)$ in \cite{Lencses:2023evr,Lencses:2024wib} and those in \cite{Katsevich:2025ojk}, and this controversy will be explained by the nonuniqueness of the homomorphism in this subsection\footnote{We thank Andrei Katsevich for notifying us the subtelties of $M(2,7)\rightarrow M(2,5)$ in the literature.}. From the analysis so far, we observe that these conditions lead to algebraic equations such as Eq. \eqref{eq:eq_A_phi^I}. It should be noted that the identification of the massless RGs as projections has been proposed in the study of $M(2,2k+1)$ models \cite{Smirnov:1990vm}. 

\section{Spin-classification for non-group-like objects: (Half) integer spin conditions}
\label{section_spin}

Interestingly, there exists a nontrivial structure when restricting our attention to the chiral and antichiral property of the flow $M(4,5)\rightarrow M(3,4)$ preserving the Ising fusion ring, $\{ I, \psi, \sigma\} \sim\{ I', \psi', \sigma'\}$. Assuming the $Z_{2}$ symmetry preservation, one can consider RG flows $\{ I, \psi\}\rightarrow \{ I,\psi'\}$ and $\{ I, \psi\}\rightarrow \{ \overline{I}',\overline{\psi}'\}$ because these the $Z_{2}$ simple currents satisfying the $Z_{2}$ anomaly condition $h_{\psi}- h_{\psi'}=0 \ (\text{mod}. 1)$ or $h_{\psi}+\overline{h}_{\overline{\psi}'}=0 \ (\text{mod}. 1)$. This anomaly matching condition, a kind of 't Hooft anomaly matching codition\cite{tHooft:1979rat}, is nothing but Lieb-Schultz-Mattis anomaly matching condition\cite{Schultz:1964fv,Cho:2017fgz,Lieb:1961fr} with a close connection to the Haldane conjecture\cite{Haldane:1981zza,Haldane:1982rj,Haldane:1983ru,haldane2016groundstatepropertiesantiferromagnetic,Wamer:2019oge}, and this condition results in the appearance of the general electron operators constructing the singlevalued wavefunction of the TQFT $\mathbf{A} \otimes \mathbf{A}'$ or $\mathbf{A} \otimes \overline{\mathbf{A}}'$ \cite{Kaidi:2021gbs}(or its $Z_{2}$ extension more precisely\cite{moore_nonabelions_1991,Cappelli:1996np,Frohlich:2000qs,Schoutens:2015uia,Fukusumi:2022xxe}). Hence, one will expect the same condition for the preserved structure $\{ I, \psi, \sigma\} \sim\{ I', \psi', \sigma'\}$, but $\sigma$ and $\sigma'$ only satisfies the relation $h_{\sigma}+h_{\sigma'}=0$. Hence, the flow $\{ I, \psi, \sigma\} \rightarrow\{ I', \psi', \sigma'\}$ only preserves $Z_{2}$ structure when applying the anomaly classification to the non-group like symmetry whereas the flow $\{ I, \psi, \sigma\} \rightarrow\{ \overline{I}', \overline{\psi}', \overline{\sigma}'\}$ is an anomaly preserving flow. More categorically, these two flows correspond to the choice of the common connected etale algebra $\{ I, \psi\}=\{ I', \psi'\}$ or $\{ I, \psi, \sigma\} =\{ \overline{I}', \overline{\psi}', \overline{\sigma}'\}$. Consequently, if one believes that the massless flow $M(4,5)\rightarrow M(3,4)$ preserves the Ising fusion ring structure in the anomaly-free way, the application of the corresponding domain wall should change the chirality of objects. Phenomenologically, the domain wall changing the chiralities seems to correspond to the Alice ring in the literature\cite{SCHWARZ1982141,SCHWARZ1982427}. 

In other words, when introducing the notion of braiding and chiralities, the classification of the RGs and the corresponding domain walls in TQFTs becomes more nontrivial. However, one can still expect that the (half-)integer spin conditions, which appeared in the studies of the discrete torsion models \cite{Vafa:1986wx,Dixon:1986jc,Hamidi:1986vh,Schellekens:1990ys,Gato-Rivera:1990lxi,Gato-Rivera:1991bqv,Kreuzer:1993tf}, should govern the properties of the systems. The term is relatively less common, but this condition should be called (half-)integer spin non-simple current condition\cite{Frohlich:2003hm}. This condition has played a role in the studies of exceptional modular invariants, related heterotic string theories, and maverick coset CFTs\cite{Dunbar:1993hr,Gannon:1992np,Gannon:1998rw,Pedrini:1999iy}. We note recent works\cite{Nakayama:2024msv,Delmastro:2025ksn} investigating the corresponding structure and a related work\cite{Kikuchi:2024cjd}.

For the flow $M(3,5) \rightarrow M(2,5)$, for example, the same phenomena will appear. The nontrivial point of the representation of the $M(3,5)=\{ I, \psi\} \otimes \{ I, \tau\}$ is that one can choose $(h_{\tau}, h_{J\tau})=(1/5, -1/20)$ or $(h_{\tau}, h_{J\tau})=(-1/20, 1/5)$. Reflecting the conformal dimensions of the IR theories, $(h_{I'}, h_{\tau'})=(0,-1/5)$, only the flow $\overline{M(3,5)}\rightarrow M(2,5)$ with $(h_{\tau}, h_{J\tau})=(1/5, -1/20)$ satisfies the integer spin nonsimple current condtion $h_{\tau}+h_{\tau'}=0$. The condition $h_{\tau}+h_{\tau'}=0$ is the generalized version of the anomaly cancellation condition of nongroup-like structures in the UV and IR theories. More generally, one can express the anomaly-free conditions for the unbroken algebraic structures $a_{\text{ub}}\in \mathbf{A}_{\text{ub}}$, where an unbroken structure satisfies the relation $\rho(\mathbf{A}_{\text{ub}})=\mathbf{A}_{\text{ub}}$, as follows,
\begin{itemize}
\item{The anomaly matching flow, $\mathbf{A}\rightarrow \mathbf{A'}$, satisfies $h_{a_{\text{ub}}}-h'_{a'_{\text{ub}}}\in \mathbb{Z}/2$ for all $a_{\text{ub}}\in \mathbf{A}_{\text{ub}}$.}
\item{The anomaly cancellation flow, $\overline{\mathbf{A}}\rightarrow \mathbf{A'}$ satisfies $h_{a_{\text{ub}}}+h'_{a'_{\text{ub}}}\in \mathbb{Z}/2$ for all $a_{\text{ub}}\in \mathbf{A}_{\text{ub}}$.}
\end{itemize}
where we have distinguished the IR objects by adding ``'", as in the previous sections and $\mathbb{Z}/2$ represents the set of integers or half-integers. One can expect appearance of ``electron operators" with the fusion rule $\mathbf{A}_{\text{ub}}$ which construct single-valued wavefunctions of the corresponding coupled TQFTs, $\mathbf{A}\otimes \mathbf{A'}$ or $\overline{\mathbf{A}}\otimes \mathbf{A'}$, but this research direction has been studied only in several cases, as we have remarked \cite{Dunbar:1993hr,Gannon:1992np,Gannon:1998rw,Pedrini:1999iy} \footnote{Corresponding to this anomaly-free condition of the non-group like objects, there will exist intrinsically nonabelian anyonic TOs, or $\mathbf{A}_{\text{ub}}$-symmetry enriched TOs where the fundamental ``electron operators'' constructing the wavefunctions are nonabelian anyon belonging to $\mathbf{A}_{\text{ub}}\boxtimes \mathbf{A'}_{\text{ub}}$ or $\overline{\mathbf{A}_{\text{ub}}}\boxtimes \mathbf{A'}_{\text{ub}}$.}. 

Before moving into the concluding remark, we note the unusual matching of the symmetry ring appearing in the flow $M(3,5) \rightarrow M(2,5)$. As we have demonstrated, there exist homomorphisms which break UV symmetry $\{ I, \tau\}$ but (emergently) produce the IR symmetry $\{ I', \tau'\}$ accidentally or by coincidence from the UV $Z_{2}$ group ring $\{ I, J\}$.
For the readers interested in such unusual emergent symmetry, we formulate the definition of the coincident emergent symmetry or coincidence of the UV-IR symmetries. First, let us assume subring of the UV fusion ring, $\mathbf{A}_{\text{coi}} \subset \mathbf{A}$ and there exists the IR subring $\mathbf{A'}_{\text{coi}}\subset \mathbf{A'}$ which are ring isomorphic to $\mathbf{A'}$, i.e. $\mathbf{A}_{\text{coi}}=\mathbf{A'}_{\text{coi}}$. One can identifies $\mathbf{A'}_{\text{coi}}$ as coincident emergent symmetry when $\rho(\mathbf{A}_{\text{coi}})\neq \mathbf{A'}_{\text{coi}}$ under a homomorphism $\rho$. In other words, the same structure in the UV and IR theories is produced under the action of $\rho$ to other structures $\mathbf{A}_{\text{ot}}\neq \mathbf{A}_{\text{coi}}$. There exist related discussions in \cite{Nakayama:2024msv}, but the exact implications of this matching have not been studied to our knowledge. By combining the analysis on the anomaly-free condition in this section, one can find more nontrivial examples.

\section{Conclusion}
\label{section_conclusion}

In this work, we have studied fusion ring symmetry of pseudo-Hermitian systems and their RG flows in a rigorous abstract algebraic formalism. The algebraic interpretation of the generalization of quantum dimensions plays the most fundamental role in our classifications, and these structures result in the appearance of (unusual) non-integer coefficients in the classifications. The notion of the spontaneous symmetry breaking of generalized symmetry for the gapped phase has been introduced in a linear algebraic way applicable to more general systems. Moreover, we provide a series of algebraic data of RG domain walls in CFTs or gapped domains in the corresponding TQFTs. We have explicitly demonstrated that the appearance of non-integer coefficients is inevitable when assuming the ring homomorphism structures, which are also fundamental for the categorical formulations of such domain walls. In other words, if one believes the RG domain walls or gapped domain walls are compatible with categorical formulations, it is necessary to introduce generalized category theories applicable to linear algebra (or intrinsically quantum systems). If one believes that the existing category theory without noninteger coefficients is sufficient, one cannot apply the notion of the (tensor) functor and subcategory to the analysis of massless and massive RG flows. We note that this view has validity when studying defect and boundary RG flows, and this research direction is worth further study, whereas they are outside of the main scope of the present manuscript. Finally, we note that our method is applicable to higher-dimensional systems in principle and cite some recent references \cite{Johnson-Freyd:2020usu,Zeev:2022cnv,Kong:2024ykr,Antinucci:2025fjp}.

\section{Acknowledgement}
  
The authors thank Po-Yao Chang, Naomichi Hatano, and Hosho Katsura for sharing their knowledge on non-Hermitian systems. YF thanks Yuma Furuta and Shinichiro Yahagi for related collaborations. YF also thanks Sylvain Ribault and Ingo Runkel for the discussions on fusion rules and Jurgen Fuch and Yuji Tachikawa for the helpful discussions on the distinction between fusion categories and fusion rings. We thank Hosho Katsura, Ingo Runkel and Gerard Watts for helpful comments and for checking some technical calculations in the manuscript. Also, we thank Tadashi Takayanagi for commenting on topological entanglement entropy and its importance, and Andrei Katsevich notifying us controvercies in the flow $M(2,7)\rightarrow M(2,5)$. YF thanks the support from NCTS. TK is supported by Grant-in-Aid for JSPS Fellows No. 23KJ1315.

\appendix

\section{Comments on conjectural structures in pseudo-Hermitian models}

\subsection{Comment on noncommutative fusion ring and magma}

In this section, we comment on the noncommutative fusion ring and magma, outside of the scope of the main text. For a noncommutative fusion ring, there can exist matrix representations, but the underlying operations can be intertwining operations. This intertwining operation comes from degeneracies of the spectrum between different primary fields\cite{Petkova:2000ip}. Hence, for example, a $Z_{N}$ symmetric model will naturally have this structure. Because of the noncommutativity, one cannot map the fusion ring to $\mathbb{C}$. Hence, quantum dimension will be defined in a different way as a ring homomorphism to some noncommutative number fields.

Next, we discuss the magma appearing in recent literature\cite{Nivesvivat:2025odb}\footnote{More precisely, linear magma satisfying the pentagon relation.}. In such models, the fusion rule can be nonassociative. Hence, there exists no matrix representation by identifying the fusion as matrix multiplication. However, by changing the definition of multiplication, it is possible to obtain a magma from the representation of matrices or linear operators. For example, commutator $[A,B]=AB-BA$ and anticommutator $\{A,B\}=AB+BA$ are naive candidates which implement the fusion $\times$, by replacing ``$\ \times \ $" to $[ \ ,\ ]$ or $\{ \ . \ \}$. Even in this magma, the bilinearity of objects will survive, and this will simplify the calculations. We note a historical review for this research direction\cite{liebmann2019nonassociativealgebrasquantumphysics}. In this setting, the quantum dimension will be generalized to the mapping, magma homomorphism, to nonassociative number fields. 

\subsection{Phenomenology on entanglement cut: Naturality of complex spectrum}

In the main texts, we studied two types of boundary phenomena: smeared BCFT appearing in a gapped phase or BCFT appearing in physical boundaries. In this subsection, we discuss some phenomenology on the entanglement surface\cite{Holzhey:1994we,Lauchli:2013jga,Ohmori:2014eia}.

First thing to note is that the entanglement spectrum in a pseudo-Hermitian system can be complex\cite{Fossati:2023zyz,Rottoli:2024tvr,Shimizu:2025kse,Chou:2025awd}. However, following the discussion in the previous sections, this phenomenon seems unusual because the complex energy value cannot appear in the previous discussion. Moreover, there exist established studies on the universal behaviors of real entanglement entropy in the nonunitary CFTs by analyzing the corresponding two dimensional statistical models and their quantum group structure\cite{Bianchini:2014uta,Couvreur:2016mbr}, combined with the twist field method\cite{Knizhnik:1987xp,Calabrese:2004eu,Calabrese:2009qy}. We also note \cite{Lu:2025myv} as a reference summarizing a variant of related entropies, such as those in \cite{Tu:2021xje,Yang:2024ebm} (for a recent analysis on symmetry resolved entanglement entropy\cite{Goldstein:2017bua,Xavier:2018kqb,Kusuki:2023bsp,Saura-Bastida:2024yye,Das:2024qdx,Choi:2024tri,Heymann:2024vvf,Choi:2024wfm,Castro-Alvaredo:2024azg} in nonunitary CFTs, see \cite{Bhattacharyya:2025tmg}).

The most fundamental difference between our discussions and the phenomenology of the entanglement cut in recent studies is that the basis of the entanglement cut is not a linear dual basis. When keeping in mind this difference, the appearance of a complex spectrum is reasonable. Moreover, when assuming such an unusual basis, some condition of charge conservation will be broken, because we have chosen an unusual basis from the entanglement cut. Hence, some charged character will appear, and one of the most natural candidates is the \emph{nonspecialized character} appearing in the corner-transfer-matrix (CTM) method\cite{BAXTER198118,Baxter_2007}. It is known that there exists a correspondence between CTM and the entanglement Hamiltonian, and the complex-valued CTM for a series of two-dimensional statistical models have been studied in \cite{Chui:2001kw,Chui:2002bp}. Hence, as in the Hermitian systems, the following correspondence should be true for the non-Hermitian systems,
\begin{equation}
\begin{split}
 &\text{\{Non-specialized character in a CFT\}} \\
=&\{ \text{CTM of a statistical mechanical model}\} \\
\sim &\{ \text{Entanglement spectrum of a quantum model}\}
\end{split}
\end{equation}
For readers interested in the phenomenological understanding of the nonspecialized character, we note related works on the charged Cardy formula\cite{Breckenridge:1996is,Maldacena:1997de,Dijkgraaf:2000fq,Hosseini:2020vgl}. It is worth mentioning that the nontriviality of the nonspecialized characters has already been remarked in the earlier works in defect and boundary CFTs\cite{Petkova:2000dv,Petkova:2000ip}. 

We also note that the appearance of the complex entanglement spectrum can be explained by using pseudo-entropy theories \cite{Nakata:2020luh,Akal:2021dqt,Caputa:2024gve,Kawamoto:2025oko} which implement the natural entropy under the final state projection.  Moreover, the complex entanglement spectrum appears from the "double cone" regularization, which is one way of regulating UV divergence of QFT by introducing complex deformations \cite{Kawamoto:2023ade,Saad:2018bqo,Chen:2023hra}. In this setting, the entanglement Hamiltonian is non-Hermitian. 

We remind that even in Hermitian systems, it has been reported that the Cardy state with the highest boundary entropy appears at the entanglement surfaces\cite{Roy:2025hew}. This seems unusual from the $g$-theorem\cite{Affleck:1991tk}, but one can understand this as a consequence of high-low temperature duality phenomenologically. The open-closed duality is a relation between high-temperature phenomena and low-temperature phenomena. In high-energy physics, this has been used to evaluate or estimate high-temperature or UV phenomena in those of IR. However, when considering the low-temperature physics realized in the ground states of a quantum lattice model, the opposite is used. Hence, more high-temperature (or UV-like) theories will appear at the entanglement surface when studying more low-temperature (or IR-like) theories. By combining this understanding with symmetry analysis of boundaries, one can expect the appearance of boundary states with large $g$-values (including symmetry-breaking branes\cite{Affleck:1998nq,Fuchs:1999zi,Quella:2002ct} and the Graham-Watts(GW) states\cite{Graham:2003nc}) at the entanglement surfaces. The same reasoning will be true in non-Hermitian systems, but one needs to study the complex conjugate operation or linear dual operation more carefully.

\bibliographystyle{ytphys}
\bibliography{nonhermitian}

\providecommand{\href}[2]{#2}\begingroup\raggedright\begin{thebibliography}{100}

\bibitem{Ashida:2020dkc}
Y.~Ashida, Z.~Gong, and M.~Ueda, {\slshape {Non-Hermitian physics},}
  \href{http://dx.doi.org/10.1080/00018732.2021.1876991}{{\em Adv. Phys.}
  {\bfseries 69} (2021) 249--435}, \href{http://arxiv.org/abs/2006.01837}{{
  arXiv:2006.01837~[cond-mat.mes-hall]}}.

\bibitem{Okuma:2022bnb}
N.~Okuma and M.~Sato, {\slshape {Non-Hermitian Topological Phenomena: A
  Review},}
  \href{http://dx.doi.org/10.1146/annurev-conmatphys-040521-033133}{{\em Ann.
  Rev. Condensed Matter Phys.} {\bfseries 14} (2023) 83--107},
  \href{http://arxiv.org/abs/2205.10379}{{
  arXiv:2205.10379~[cond-mat.mes-hall]}}.

\bibitem{Manzano:2020yyw}
D.~Manzano, {\slshape {A short introduction to the Lindblad master equation},}
  \href{http://dx.doi.org/10.1063/1.5115323}{{\em AIP Adv.} {\bfseries 10}
  (2020) 025106}, \href{http://arxiv.org/abs/1906.04478}{{
  arXiv:1906.04478~[quant-ph]}}.

\bibitem{Bender:2007nj}
C.~M. Bender, {\slshape {Making sense of non-Hermitian Hamiltonians},}
  \href{http://dx.doi.org/10.1088/0034-4885/70/6/R03}{{\em Rept. Prog. Phys.}
  {\bfseries 70} (2007) 947}, \href{http://arxiv.org/abs/hep-th/0703096}{{
  arXiv:hep-th/0703096}}.

\bibitem{Dorey:2007zx}
P.~Dorey, C.~Dunning, and R.~Tateo, {\slshape {The ODE/IM Correspondence},}
  \href{http://dx.doi.org/10.1088/1751-8113/40/32/R01}{{\em J. Phys. A}
  {\bfseries 40} (2007) R205}, \href{http://arxiv.org/abs/hep-th/0703066}{{
  arXiv:hep-th/0703066}}.

\bibitem{Strominger:2001pn}
A.~Strominger, {\slshape {The dS / CFT correspondence},}
  \href{http://dx.doi.org/10.1088/1126-6708/2001/10/034}{{\em JHEP} {\bfseries
  10} (2001) 034}, \href{http://arxiv.org/abs/hep-th/0106113}{{
  arXiv:hep-th/0106113}}.

\bibitem{Klemm:2001ea}
D.~Klemm, {\slshape {Some aspects of the de Sitter / CFT correspondence},}
  \href{http://dx.doi.org/10.1016/S0550-3213(02)00007-X}{{\em Nucl. Phys. B}
  {\bfseries 625} (2002) 295--311},
  \href{http://arxiv.org/abs/hep-th/0106247}{{ arXiv:hep-th/0106247}}.

\bibitem{Kawamoto:2025oko}
T.~Kawamoto, R.~Maeda, N.~Nakamura, and T.~Takayanagi, {\slshape {Traversable
  AdS wormhole via non-local double trace or Janus deformation},}
  \href{http://dx.doi.org/10.1007/JHEP04(2025)086}{{\em JHEP} {\bfseries 04}
  (2025) 086}, \href{http://arxiv.org/abs/2502.03531}{{
  arXiv:2502.03531~[hep-th]}}.

\bibitem{Nakanishi:1989cv}
T.~Nakanishi, {\slshape {Nonunitary Minimal Models and Rsos Models},}
  \href{http://dx.doi.org/10.1016/0550-3213(90)90320-D}{{\em Nucl. Phys. B}
  {\bfseries 334} (1990) 745--766}.

\bibitem{Schellekens:1989uf}
A.~N. Schellekens and S.~Yankielowicz, {\slshape {Field Identification Fixed
  Points in the Coset Construction},}
  \href{http://dx.doi.org/10.1016/0550-3213(90)90657-Y}{{\em Nucl. Phys. B}
  {\bfseries 334} (1990) 67--102}.

\bibitem{Mathieu:1991fz}
P.~Mathieu, D.~Senechal, and M.~Walton, {\slshape {Field identification in
  nonunitary diagonal cosets},}
  \href{http://dx.doi.org/10.1142/S0217751X92004002}{{\em Int. J. Mod. Phys. A}
  {\bfseries 7S1B} (1992) 731--764},
  \href{http://arxiv.org/abs/hep-th/9110003}{{ arXiv:hep-th/9110003}}.

\bibitem{Pasquier:1989kd}
V.~Pasquier and H.~Saleur, {\slshape {Common Structures Between Finite Systems
  and Conformal Field Theories Through Quantum Groups},}
  \href{http://dx.doi.org/10.1016/0550-3213(90)90122-T}{{\em Nucl. Phys. B}
  {\bfseries 330} (1990) 523--556}.

\bibitem{Hatano_1996}
N.~Hatano and D.~R. Nelson, {\slshape Localization transitions in non-hermitian
  quantum mechanics,} \href{http://dx.doi.org/10.1103/PhysRevLett.77.570}{{\em
  Physical Review Letters} {\bfseries 77} (July, 1996) 570–573}.

\bibitem{Mostafazadeh:2001jk}
A.~Mostafazadeh, {\slshape {PseudoHermiticity versus PT symmetry. The necessary
  condition for the reality of the spectrum},}
  \href{http://dx.doi.org/10.1063/1.1418246}{{\em J. Math. Phys.} {\bfseries
  43} (2002) 205--214}, \href{http://arxiv.org/abs/math-ph/0107001}{{
  arXiv:math-ph/0107001}}.

\bibitem{Mostafazadeh:2001nr}
A.~Mostafazadeh, {\slshape {PseudoHermiticity versus PT symmetry 2. A Complete
  characterization of nonHermitian Hamiltonians with a real spectrum},}
  \href{http://dx.doi.org/10.1063/1.1461427}{{\em J. Math. Phys.} {\bfseries
  43} (2002) 2814--2816}, \href{http://arxiv.org/abs/math-ph/0110016}{{
  arXiv:math-ph/0110016}}.

\bibitem{Mostafazadeh:2002id}
A.~Mostafazadeh, {\slshape {PseudoHermiticity versus PT symmetry 3: Equivalence
  of pseudoHermiticity and the presence of antilinear symmetries},}
  \href{http://dx.doi.org/10.1063/1.1489072}{{\em J. Math. Phys.} {\bfseries
  43} (2002) 3944--3951}, \href{http://arxiv.org/abs/math-ph/0203005}{{
  arXiv:math-ph/0203005}}.

\bibitem{Bender:1998ke}
C.~M. Bender and S.~Boettcher, {\slshape {Real spectra in nonHermitian
  Hamiltonians having PT symmetry},}
  \href{http://dx.doi.org/10.1103/PhysRevLett.80.5243}{{\em Phys. Rev. Lett.}
  {\bfseries 80} (1998) 5243--5246},
  \href{http://arxiv.org/abs/physics/9712001}{{ arXiv:physics/9712001}}.

\bibitem{Zamolodchikov:1986gt}
A.~B. Zamolodchikov, {\slshape {Irreversibility of the Flux of the
  Renormalization Group in a 2D Field Theory},} {\em JETP Lett.} {\bfseries 43}
  (1986) 730--732.

\bibitem{Castro-Alvaredo:2017udm}
O.~A. Castro-Alvaredo, B.~Doyon, and F.~Ravanini, {\slshape {Irreversibility of
  the renormalization group flow in non-unitary quantum field theory},}
  \href{http://dx.doi.org/10.1088/1751-8121/aa8a10}{{\em J. Phys. A} {\bfseries
  50} (2017) 424002}, \href{http://arxiv.org/abs/1706.01871}{{
  arXiv:1706.01871~[hep-th]}}.

\bibitem{Fernandez:2015aqe}
F.~M. Fern{\'a}ndez, {\slshape {Non-Hermitian Hamiltonians and similarity
  transformations},} \href{http://arxiv.org/abs/1502.02694}{{
  arXiv:1502.02694~[quant-ph]}}.

\bibitem{Fukusumi:2025fir}
Y.~Fukusumi and T.~Kawamoto, {\slshape {Generalizing fusion rules by shuffle:
  Symmetry-based classifications of nonlocal systems constructed from
  similarity transformations},} \href{http://arxiv.org/abs/2512.02139}{{
  arXiv:2512.02139~[hep-th]}}.

\bibitem{Guruswamy:1996rk}
S.~Guruswamy and A.~W.~W. Ludwig, {\slshape {Relating c \ensuremath{<} 0 and c
  \ensuremath{>} 0 conformal field theories},}
  \href{http://dx.doi.org/10.1016/S0550-3213(98)00059-5}{{\em Nucl. Phys. B}
  {\bfseries 519} (1998) 661--681},
  \href{http://arxiv.org/abs/hep-th/9612172}{{ arXiv:hep-th/9612172}}.

\bibitem{LeClair:2007iy}
A.~LeClair and M.~Neubert, {\slshape {Semi-Lorentz invariance, unitarity, and
  critical exponents of symplectic fermion models},}
  \href{http://dx.doi.org/10.1088/1126-6708/2007/10/027}{{\em JHEP} {\bfseries
  10} (2007) 027}, \href{http://arxiv.org/abs/0705.4657}{{
  arXiv:0705.4657~[hep-th]}}.

\bibitem{Hsieh:2022hgi}
C.-T. Hsieh and P.-Y. Chang, {\slshape {Relating non-Hermitian and Hermitian
  quantum systems at criticality},}
  \href{http://dx.doi.org/10.21468/SciPostPhysCore.6.3.062}{{\em SciPost Phys.
  Core} {\bfseries 6} (2023) 062}, \href{http://arxiv.org/abs/2211.12525}{{
  arXiv:2211.12525~[cond-mat.str-el]}}.

\bibitem{Kuniba:1990im}
A.~Kuniba and T.~Nakanishi, {\slshape {LEVEL RANK DUALITY IN FUSION RSOS
  MODELS},} in {\em {International Colloquium on Modern Quantum Field Theory}}.
\newblock 1, 1990.

\bibitem{Kuniba:1990zh}
A.~Kuniba, T.~Nakanishi, and J.~Suzuki, {\slshape {Ferromagnetizations and
  antiferromagnetizations in RSOS models},}
  \href{http://dx.doi.org/10.1016/0550-3213(91)90385-B}{{\em Nucl. Phys. B}
  {\bfseries 356} (1991) 750--774}.

\bibitem{Nakanishi:1990hj}
T.~Nakanishi and A.~Tsuchiya, {\slshape {Level rank duality of WZW models in
  conformal field theory},} \href{http://dx.doi.org/10.1007/BF02101097}{{\em
  Commun. Math. Phys.} {\bfseries 144} (1992) 351--372}.

\bibitem{Altschuler:1990th}
D.~Altschuler, M.~Bauer, and H.~Saleur, {\slshape {Level rank duality in
  nonunitary coset theories},} {\em J. Phys. A} {\bfseries 23} (1990)
  L789--L794.

\bibitem{Buican:2017rya}
M.~Buican and Z.~Laczko, {\slshape {Nonunitary Lagrangians and unitary
  non-Lagrangian conformal field theories},}
  \href{http://dx.doi.org/10.1103/PhysRevLett.120.081601}{{\em Phys. Rev.
  Lett.} {\bfseries 120} (2018) 081601},
  \href{http://arxiv.org/abs/1711.09949}{{ arXiv:1711.09949~[hep-th]}}.

\bibitem{Ferrari:2023fez}
A.~E.~V. Ferrari, N.~Garner, and H.~Kim, {\slshape {Boundary vertex algebras
  for 3d $\mathcal{N}=4$ rank-0 SCFTs},}
  \href{http://dx.doi.org/10.21468/SciPostPhys.17.2.057}{{\em SciPost Phys.}
  {\bfseries 17} (2024) 057}, \href{http://arxiv.org/abs/2311.05087}{{
  arXiv:2311.05087~[hep-th]}}.

\bibitem{Creutzig:2024ljv}
T.~Creutzig, N.~Garner, and H.~Kim, {\slshape {Mirror Symmetry and Level-rank
  Duality for 3d $\mathcal{N} = 4$ Rank 0 SCFTs},}
  \href{http://arxiv.org/abs/2406.00138}{{ arXiv:2406.00138~[hep-th]}}.

\bibitem{Chung:2014qpa}
H.-J. Chung, T.~Dimofte, S.~Gukov, and P.~Su{\l}kowski, {\slshape {3d-3d
  Correspondence Revisited},}
  \href{http://dx.doi.org/10.1007/JHEP04(2016)140}{{\em JHEP} {\bfseries 04}
  (2016) 140}, \href{http://arxiv.org/abs/1405.3663}{{
  arXiv:1405.3663~[hep-th]}}.

\bibitem{Chun:2019mal}
S.~Chun, S.~Gukov, S.~Park, and N.~Sopenko, {\slshape {3d-3d correspondence for
  mapping tori},} \href{http://dx.doi.org/10.1007/JHEP09(2020)152}{{\em JHEP}
  {\bfseries 09} (2020) 152}, \href{http://arxiv.org/abs/1911.08456}{{
  arXiv:1911.08456~[hep-th]}}.

\bibitem{Cobanera:2011wn}
E.~Cobanera, G.~Ortiz, and Z.~Nussinov, {\slshape {The Bond-Algebraic Approach
  to Dualities},} \href{http://dx.doi.org/10.1080/00018732.2011.619814}{{\em
  Adv. Phys.} {\bfseries 60} (2011) 679--798},
  \href{http://arxiv.org/abs/1103.2776}{{
  arXiv:1103.2776~[cond-mat.stat-mech]}}.

\bibitem{Cobanera:2012dc}
E.~Cobanera, G.~Ortiz, and Z.~Nussinov, {\slshape {Holographic symmetries and
  generalized order parameters for topological matter},}
  \href{http://dx.doi.org/10.1103/PhysRevB.87.041105}{{\em Phys. Rev. B}
  {\bfseries 87} (2013) 041105}, \href{http://arxiv.org/abs/1211.0564}{{
  arXiv:1211.0564~[cond-mat.stat-mech]}}.

\bibitem{Gaiotto:2014kfa}
D.~Gaiotto, A.~Kapustin, N.~Seiberg, and B.~Willett, {\slshape {Generalized
  Global Symmetries},} \href{http://dx.doi.org/10.1007/JHEP02(2015)172}{{\em
  JHEP} {\bfseries 02} (2015) 172}, \href{http://arxiv.org/abs/1412.5148}{{
  arXiv:1412.5148~[hep-th]}}.

\bibitem{Brunner:2007ur}
I.~Brunner and D.~Roggenkamp, {\slshape {Defects and bulk perturbations of
  boundary Landau-Ginzburg orbifolds},}
  \href{http://dx.doi.org/10.1088/1126-6708/2008/04/001}{{\em JHEP} {\bfseries
  04} (2008) 001}, \href{http://arxiv.org/abs/0712.0188}{{
  arXiv:0712.0188~[hep-th]}}.

\bibitem{Gaiotto:2012np}
D.~Gaiotto, {\slshape {Domain Walls for Two-Dimensional Renormalization Group
  Flows},} \href{http://dx.doi.org/10.1007/JHEP12(2012)103}{{\em JHEP}
  {\bfseries 12} (2012) 103}, \href{http://arxiv.org/abs/1201.0767}{{
  arXiv:1201.0767~[hep-th]}}.

\bibitem{Lan:2014uaa}
T.~Lan, J.~C. Wang, and X.-G. Wen, {\slshape {Gapped Domain Walls, Gapped
  Boundaries and Topological Degeneracy},}
  \href{http://dx.doi.org/10.1103/PhysRevLett.114.076402}{{\em Phys. Rev.
  Lett.} {\bfseries 114} (2015) 076402},
  \href{http://arxiv.org/abs/1408.6514}{{ arXiv:1408.6514~[cond-mat.str-el]}}.

\bibitem{Petkova:2000ip}
V.~B. Petkova and J.~B. Zuber, {\slshape {Generalized Twisted Partition
  Functions},} \href{http://dx.doi.org/10.1016/S0370-2693(01)00276-3}{{\em
  Phys. Lett. B} {\bfseries 504} (2001) 157--164},
  \href{http://arxiv.org/abs/hep-th/0011021}{{ arXiv:hep-th/0011021}}.

\bibitem{article}
A.~Ocneanu, {\slshape Paths on coxeter diagrams: From platonic solids and
  singularities to minimal models and subfactors,} {\em Lectures on Operator
  Theory} (01, 2000) .

\bibitem{Bockenhauer:1998ef}
J.~Bockenhauer and D.~E. Evans, {\slshape {Modular invariants, graphs and alpha
  induction for nets of subfactors. 3.},}
  \href{http://dx.doi.org/10.1007/s002200050673}{{\em Commun. Math. Phys.}
  {\bfseries 205} (1999) 183--228},
  \href{http://arxiv.org/abs/hep-th/9812110}{{ arXiv:hep-th/9812110}}.

\bibitem{Bockenhauer:1999ae}
J.~Bockenhauer and D.~E. Evans, {\slshape {On alpha induction, chiral
  generators and modular invariants for subfactors},}
  \href{http://dx.doi.org/10.1007/s002200050765}{{\em Commun. Math. Phys.}
  {\bfseries 208} (1999) 429--487}, \href{http://arxiv.org/abs/math/9904109}{{
  arXiv:math/9904109}}.

\bibitem{Bockenhauer:1999wt}
J.~Bockenhauer, D.~E. Evans, and Y.~Kawahigashi, {\slshape {Chiral structure of
  modular invariants for subfactors},}
  \href{http://dx.doi.org/10.1007/s002200050798}{{\em Commun. Math. Phys.}
  {\bfseries 210} (2000) 733--784}, \href{http://arxiv.org/abs/math/9907149}{{
  arXiv:math/9907149}}.

\bibitem{Thorngren:2019iar}
R.~Thorngren and Y.~Wang, {\slshape {Fusion Category Symmetry I: Anomaly
  In-Flow and Gapped Phases},} \href{http://arxiv.org/abs/1912.02817}{{
  arXiv:1912.02817~[hep-th]}}.

\bibitem{Thorngren:2021yso}
R.~Thorngren and Y.~Wang, {\slshape {Fusion category symmetry. Part II.
  Categoriosities at c = 1 and beyond},}
  \href{http://dx.doi.org/10.1007/JHEP07(2024)051}{{\em JHEP} {\bfseries 07}
  (2024) 051}, \href{http://arxiv.org/abs/2106.12577}{{
  arXiv:2106.12577~[hep-th]}}.

\bibitem{Note1}
In this sense, a part of category theories appearing in physics is less
  abstract and less general than linear algebra.

\bibitem{Zhao:2023wtg}
Y.~Zhao, H.~Wang, Y.~Hu, and Y.~Wan, {\slshape {Symmetry fractionalized
  (irrationalized) fusion rules and two domain-wall Verlinde formulae},}
  \href{http://dx.doi.org/10.1007/JHEP04(2024)115}{{\em JHEP} {\bfseries 04}
  (2024) 115}, \href{http://arxiv.org/abs/2304.08475}{{
  arXiv:2304.08475~[cond-mat.str-el]}}.

\bibitem{Fukusumi:2025clr}
Y.~Fukusumi and Y.~Furuta, {\slshape {Homomorphism, substructure and ideal:
  Elementary but rigorous aspects of renormalization group or hierarchical
  structure of topological orders},} \href{http://arxiv.org/abs/2506.23155}{{
  arXiv:2506.23155~[hep-th]}}.

\bibitem{Witten:1988hf}
E.~Witten, {\slshape {Quantum Field Theory and the Jones Polynomial},}
  \href{http://dx.doi.org/10.1007/BF01217730}{{\em Commun. Math. Phys.}
  {\bfseries 121} (1989) 351--399}.

\bibitem{Note2}
We thank Yuji Tachikawa and Jurgen Fuchs for the related discussions and many
  useful comments.

\bibitem{Dorey:2004xk}
P.~Dorey, D.~Fioravanti, C.~Rim, and R.~Tateo, {\slshape {Integrable quantum
  field theory with boundaries: The Exact g function},}
  \href{http://dx.doi.org/10.1016/j.nuclphysb.2004.06.045}{{\em Nucl. Phys. B}
  {\bfseries 696} (2004) 445--467},
  \href{http://arxiv.org/abs/hep-th/0404014}{{ arXiv:hep-th/0404014}}.

\bibitem{Dorey:2005ak}
P.~Dorey, A.~Lishman, C.~Rim, and R.~Tateo, {\slshape {Reflection factors and
  exact g-functions for purely elastic scattering theories},}
  \href{http://dx.doi.org/10.1016/j.nuclphysb.2006.02.043}{{\em Nucl. Phys. B}
  {\bfseries 744} (2006) 239--276},
  \href{http://arxiv.org/abs/hep-th/0512337}{{ arXiv:hep-th/0512337}}.

\bibitem{Fredenhagen:2009tn}
S.~Fredenhagen, M.~R. Gaberdiel, and C.~Schmidt-Colinet, {\slshape {Bulk flows
  in Virasoro minimal models with boundaries},}
  \href{http://dx.doi.org/10.1088/1751-8113/42/49/495403}{{\em J. Phys. A}
  {\bfseries 42} (2009) 495403}, \href{http://arxiv.org/abs/0907.2560}{{
  arXiv:0907.2560~[hep-th]}}.

\bibitem{Dorey:2009vg}
P.~Dorey, C.~Rim, and R.~Tateo, {\slshape {Exact g-function flow between
  conformal field theories},}
  \href{http://dx.doi.org/10.1016/j.nuclphysb.2010.03.010}{{\em Nucl. Phys. B}
  {\bfseries 834} (2010) 485--501}, \href{http://arxiv.org/abs/0911.4969}{{
  arXiv:0911.4969~[hep-th]}}.

\bibitem{Ambrosino:2025myh}
F.~Ambrosino, I.~Runkel, and G.~M.~T. Watts, {\slshape {Non-local charges from
  perturbed defects via SymTFT in 2d CFT},}
  \href{http://dx.doi.org/10.1088/1751-8121/ae0b10}{{\em J. Phys. A} {\bfseries
  58} (2025) 425401}, \href{http://arxiv.org/abs/2504.05277}{{
  arXiv:2504.05277~[hep-th]}}.

\bibitem{Ambrosino:2025pjj}
F.~Ambrosino, I.~Runkel, and G.~M.~T. Watts, {\slshape {Translation invariant
  defects as an extension of topological symmetries},}
\newblock 11, 2025.
\newblock \href{http://arxiv.org/abs/2511.02007}{{ arXiv:2511.02007~[hep-th]}}.

\bibitem{Zamolodchikov:1987ti}
A.~B. Zamolodchikov, {\slshape {Renormalization Group and Perturbation Theory
  Near Fixed Points in Two-Dimensional Field Theory},} {\em Sov. J. Nucl.
  Phys.} {\bfseries 46} (1987) 1090.

\bibitem{Zamolodchikov:1987jf}
A.~B. Zamolodchikov, {\slshape {Higher Order Integrals of Motion in
  Two-Dimensional Models of the Field Theory with a Broken Conformal
  Symmetry},} {\em JETP Lett.} {\bfseries 46} (1987) 160--164.

\bibitem{Zamolodchikov:1989hfa}
A.~B. Zamolodchikov, {\slshape {Integrable field theory from conformal field
  theory},} {\em Adv. Stud. Pure Math.} {\bfseries 19} (1989) 641--674.

\bibitem{Wan:2016php}
Y.~Wan and C.~Wang, {\slshape {Fermion Condensation and Gapped Domain Walls in
  Topological Orders},} \href{http://dx.doi.org/10.1007/JHEP03(2017)172}{{\em
  JHEP} {\bfseries 03} (2017) 172}, \href{http://arxiv.org/abs/1607.01388}{{
  arXiv:1607.01388~[cond-mat.str-el]}}.

\bibitem{Klos:2019axh}
F.~Klos and D.~Roggenkamp, {\slshape {Realizing IR theories by projections in
  the UV},} \href{http://dx.doi.org/10.1007/JHEP01(2020)097}{{\em JHEP}
  {\bfseries 01} (2020) 097}, \href{http://arxiv.org/abs/1907.12339}{{
  arXiv:1907.12339~[hep-th]}}.

\bibitem{Klos:2021gab}
F.~Klos, \href{http://dx.doi.org/10.11588/heidok.00030265}{{\em {Embedding
  topological quantum field theories functorially in the UV.}}}
\newblock PhD thesis, Heidelberg U., 2021.

\bibitem{Cardy:2017ufe}
J.~Cardy, {\slshape {Bulk Renormalization Group Flows and Boundary States in
  Conformal Field Theories},}
  \href{http://dx.doi.org/10.21468/SciPostPhys.3.2.011}{{\em SciPost Phys.}
  {\bfseries 3} (2017) 011}, \href{http://arxiv.org/abs/1706.01568}{{
  arXiv:1706.01568~[hep-th]}}.

\bibitem{Lencses:2018paa}
M.~Lencses, J.~Viti, and G.~Takacs, {\slshape {Chiral entanglement in massive
  quantum field theories in 1+1 dimensions},}
  \href{http://dx.doi.org/10.1007/JHEP01(2019)177}{{\em JHEP} {\bfseries 01}
  (2019) 177}, \href{http://arxiv.org/abs/1811.06500}{{
  arXiv:1811.06500~[hep-th]}}.

\bibitem{Ares:2020uwy}
F.~Ares, M.~A. Rajabpour, and J.~Viti, {\slshape {Scaling of the Formation
  Probabilities and Universal Boundary Entropies in the Quantum XY Spin
  Chain},} \href{http://dx.doi.org/10.1088/1742-5468/aba9d4}{{\em J. Stat.
  Mech.} {\bfseries 2008} (2020) 083111},
  \href{http://arxiv.org/abs/2004.10606}{{
  arXiv:2004.10606~[cond-mat.stat-mech]}}.

\bibitem{Kikuchi:2021qxz}
K.~Kikuchi, {\slshape {Symmetry enhancement in RCFT},}
  \href{http://arxiv.org/abs/2109.02672}{{ arXiv:2109.02672~[hep-th]}}.

\bibitem{Kikuchi:2022gfi}
K.~Kikuchi, {\slshape {Symmetry enhancement in RCFT II},}
  \href{http://arxiv.org/abs/2207.06433}{{ arXiv:2207.06433~[hep-th]}}.

\bibitem{Fukusumi:2024ejk}
Y.~Fukusumi, {\slshape {Gauging or extending bulk and boundary conformal field
  theories: Application to bulk and domain wall problem in topological matter
  and their descriptions by (mock) modular covariant},}
  \href{http://arxiv.org/abs/2412.19577}{{ arXiv:2412.19577~[hep-th]}}.

\bibitem{Choi:2025ebk}
Y.~Choi, H.~Ha, D.~Kim, Y.~Kusuki, S.~Ohyama, and S.~Ryu, {\slshape {Higher
  Structures on Boundary Conformal Manifolds: Higher Berry Phase and Boundary
  Conformal Field Theory},} \href{http://arxiv.org/abs/2507.12525}{{
  arXiv:2507.12525~[hep-th]}}.

\bibitem{Wen:2025xka}
X.~Wen, {\slshape {Space of conformal boundary conditions from the view of
  higher Berry phase: Flow of Berry curvature in parametrized BCFTs},}
  \href{http://arxiv.org/abs/2507.12546}{{ arXiv:2507.12546~[hep-th]}}.

\bibitem{Date:1987zz}
E.~Date, M.~Jimbo, T.~Miwa, and M.~Okado, {\slshape {Automorphic properties of
  local height probabilities for integrable solid-on-solid models},}
  \href{http://dx.doi.org/10.1103/PhysRevB.35.2105}{{\em Phys. Rev. B}
  {\bfseries 35} (1987) 2105--2107}.

\bibitem{Saleur:1988zx}
H.~Saleur and M.~Bauer, {\slshape {On Some Relations Between Local Height
  Probabilities and Conformal Invariance},}
  \href{http://dx.doi.org/10.1016/0550-3213(89)90014-X}{{\em Nucl. Phys. B}
  {\bfseries 320} (1989) 591--624}.

\bibitem{Foda:2017vog}
O.~Foda, {\slshape {Off-critical local height probabilities on a plane and
  critical partition functions on a cylinder},}
  \href{http://arxiv.org/abs/1711.03337}{{ arXiv:1711.03337~[hep-th]}}.

\bibitem{Calabrese:2006rx}
P.~Calabrese and J.~L. Cardy, {\slshape {Time-dependence of correlation
  functions following a quantum quench},}
  \href{http://dx.doi.org/10.1103/PhysRevLett.96.136801}{{\em Phys. Rev. Lett.}
  {\bfseries 96} (2006) 136801}, \href{http://arxiv.org/abs/cond-mat/0601225}{{
  arXiv:cond-mat/0601225}}.

\bibitem{Qi_2012}
X.-L. Qi, H.~Katsura, and A.~W.~W. Ludwig, {\slshape General relationship
  between the entanglement spectrum and the edge state spectrum of topological
  quantum states,} \href{https://doi.org/10.1103%2Fphysrevlett.108.196402}{{\em
  Physical Review Letters} {\bfseries 108} (May, 2012) }.

\bibitem{Das:2015oha}
D.~Das and S.~Datta, {\slshape {Universal features of left-right entanglement
  entropy},} \href{http://dx.doi.org/10.1103/PhysRevLett.115.131602}{{\em Phys.
  Rev. Lett.} {\bfseries 115} (2015) 131602},
  \href{http://arxiv.org/abs/1504.02475}{{ arXiv:1504.02475~[hep-th]}}.

\bibitem{Moore:1988qv}
G.~W. Moore and N.~Seiberg, {\slshape {Classical and Quantum Conformal Field
  Theory},} \href{http://dx.doi.org/10.1007/BF01238857}{{\em Commun. Math.
  Phys.} {\bfseries 123} (1989) 177}.

\bibitem{Moore:1988ss}
G.~W. Moore and N.~Seiberg, {\slshape {Naturality in Conformal Field Theory},}
  \href{http://dx.doi.org/10.1016/0550-3213(89)90511-7}{{\em Nucl. Phys. B}
  {\bfseries 313} (1989) 16--40}.

\bibitem{Moore:1989vd}
G.~W. Moore and N.~Seiberg, {\slshape {Lectures on RCFT},} in {\em {Strings
  '89, Proceedings of the Trieste Spring School on Superstrings.}}
\newblock World Scientific, 1990.
\newblock \url{{http://www.physics.rutgers.edu/~gmoore/LecturesRCFT.pdf}}.

\bibitem{Li_2008}
H.~Li and F.~D.~M. Haldane, {\slshape Entanglement spectrum as a generalization
  of entanglement entropy: Identification of topological order in non-abelian
  fractional quantum hall effect states,}
  \href{https://doi.org/10.1103%2Fphysrevlett.101.010504}{{\em Physical Review
  Letters} {\bfseries 101} (Jul, 2008) }.

\bibitem{Kitaev:2005dm}
A.~Kitaev and J.~Preskill, {\slshape {Topological entanglement entropy},}
  \href{http://dx.doi.org/10.1103/PhysRevLett.96.110404}{{\em Phys. Rev. Lett.}
  {\bfseries 96} (2006) 110404}, \href{http://arxiv.org/abs/hep-th/0510092}{{
  arXiv:hep-th/0510092}}.

\bibitem{Levin:2006zz}
M.~Levin and X.-G. Wen, {\slshape {Detecting Topological Order in a Ground
  State Wave Function},}
  \href{http://dx.doi.org/10.1103/PhysRevLett.96.110405}{{\em Phys. Rev. Lett.}
  {\bfseries 96} (2006) 110405}, \href{http://arxiv.org/abs/cond-mat/0510613}{{
  arXiv:cond-mat/0510613}}.

\bibitem{Holzhey:1994we}
C.~Holzhey, F.~Larsen, and F.~Wilczek, {\slshape {Geometric and renormalized
  entropy in conformal field theory},}
  \href{http://dx.doi.org/10.1016/0550-3213(94)90402-2}{{\em Nucl. Phys. B}
  {\bfseries 424} (1994) 443--467},
  \href{http://arxiv.org/abs/hep-th/9403108}{{ arXiv:hep-th/9403108}}.

\bibitem{Nakayama:2024msv}
Y.~Nakayama and T.~Tanaka, {\slshape {Infinitely many new renormalization group
  flows between Virasoro minimal models from non-invertible symmetries},}
  \href{http://dx.doi.org/10.1007/JHEP11(2024)137}{{\em JHEP} {\bfseries 11}
  (2024) 137}, \href{http://arxiv.org/abs/2407.21353}{{
  arXiv:2407.21353~[hep-th]}}.

\bibitem{Kikuchi:2024cjd}
K.~Kikuchi, {\slshape {Rational RG flow, extension, and Witt class},}
  \href{http://arxiv.org/abs/2412.08935}{{ arXiv:2412.08935~[hep-th]}}.

\bibitem{Chen:2025qub}
J.~Chen, Z.~Duan, Q.~Jia, and S.~Lee, {\slshape {Fermionic Non-invertible
  Symmetry Behind Supersymmetric ADE Solitons},}
  \href{http://arxiv.org/abs/2511.22129}{{ arXiv:2511.22129~[hep-th]}}.

\bibitem{Gaberdiel:2026sfg}
M.~R. Gaberdiel and L.~Merkens, {\slshape {Defects in N=1 minimal models and RG
  flows},} \href{http://arxiv.org/abs/2601.03879}{{
  arXiv:2601.03879~[hep-th]}}.

\bibitem{Ambrosino:2026umb}
F.~Ambrosino and T.~Proch{\'a}zka, {\slshape {RG flows of minimal $\mathcal
  W$-algebra CFTs via non-invertible symmetries},}
  \href{http://arxiv.org/abs/2601.18667}{{ arXiv:2601.18667~[hep-th]}}.

\bibitem{Benedetti:2026drn}
V.~Benedetti, P.~Fendley, and J.~M. Magan, {\slshape {Non-invertible symmetries
  and selection rules for RG flows of coset models},}
  \href{http://arxiv.org/abs/2603.09591}{{ arXiv:2603.09591~[hep-th]}}.

\bibitem{Higgs:1964pj}
P.~W. Higgs, {\slshape {Broken Symmetries and the Masses of Gauge Bosons},}
  \href{http://dx.doi.org/10.1103/PhysRevLett.13.508}{{\em Phys. Rev. Lett.}
  {\bfseries 13} (1964) 508--509}.

\bibitem{Cordova:2025eim}
C.~Cordova, D.~Garc\'\i{}a-Sep\'ulveda, and K.~Ohmori, {\slshape {Higgsing
  Transitions from Topological Field Theory \& Non-Invertible Symmetry in
  Chern-Simons Matter Theories},} \href{http://arxiv.org/abs/2504.03614}{{
  arXiv:2504.03614~[hep-th]}}.

\bibitem{LeClair:2001yp}
A.~LeClair, {\slshape {Chiral stabilization of the renormalization group for
  flavor and color anisotropic current interactions},}
  \href{http://dx.doi.org/10.1016/S0370-2693(01)01089-9}{{\em Phys. Lett. B}
  {\bfseries 519} (2001) 183--187},
  \href{http://arxiv.org/abs/hep-th/0105092}{{ arXiv:hep-th/0105092}}.

\bibitem{Sfetsos:2017sep}
K.~Sfetsos and K.~Siampos, {\slshape {Integrable deformations of the $G_{k_1}
  \times G_{k_2}/G_{k_1+k_2}$ coset CFTs},}
  \href{http://dx.doi.org/10.1016/j.nuclphysb.2017.12.011}{{\em Nucl. Phys. B}
  {\bfseries 927} (2018) 124--139}, \href{http://arxiv.org/abs/1710.02515}{{
  arXiv:1710.02515~[hep-th]}}.

\bibitem{Georgiou:2018gpe}
G.~Georgiou and K.~Sfetsos, {\slshape {The most general $\lambda$-deformation
  of CFTs and integrability},}
  \href{http://dx.doi.org/10.1007/JHEP03(2019)094}{{\em JHEP} {\bfseries 03}
  (2019) 094}, \href{http://arxiv.org/abs/1812.04033}{{
  arXiv:1812.04033~[hep-th]}}.

\bibitem{PhysRevB.34.6372}
H.~J. Schulz, {\slshape Phase diagrams and correlation exponents for quantum
  spin chains of arbitrary spin quantum number,}
  \href{https://link.aps.org/doi/10.1103/PhysRevB.34.6372}{{\em Phys. Rev. B}
  {\bfseries 34} (Nov, 1986) 6372--6385}.

\bibitem{PhysRevLett.88.036401}
C.~L. Kane, R.~Mukhopadhyay, and T.~C. Lubensky, {\slshape Fractional quantum
  hall effect in an array of quantum wires,}
  \href{https://link.aps.org/doi/10.1103/PhysRevLett.88.036401}{{\em Phys. Rev.
  Lett.} {\bfseries 88} (Jan, 2002) 036401}.

\bibitem{Teo:2011hq}
J.~C.~Y. Teo and C.~L. Kane, {\slshape {From Luttinger liquid to non-Abelian
  quantum Hall states},}
  \href{http://dx.doi.org/10.1103/PhysRevB.89.085101}{{\em Phys. Rev. B}
  {\bfseries 89} (2014) 085101}, \href{http://arxiv.org/abs/1111.2617}{{
  arXiv:1111.2617~[cond-mat.mes-hall]}}.

\bibitem{Kimura:2014hva}
T.~Kimura and M.~Murata, {\slshape {Current Reflection and Transmission at
  Conformal Defects: Applying BCFT to Transport Process},}
  \href{http://dx.doi.org/10.1016/j.nuclphysb.2014.05.026}{{\em Nucl. Phys. B}
  {\bfseries 885} (2014) 266--279}, \href{http://arxiv.org/abs/1402.6705}{{
  arXiv:1402.6705~[hep-th]}}.

\bibitem{Lecheminant:2015iga}
P.~Lecheminant, {\slshape {Massless renormalization group flow in SU(N)$_k$
  perturbed conformal field theory},}
  \href{http://dx.doi.org/10.1016/j.nuclphysb.2015.11.004}{{\em Nucl. Phys. B}
  {\bfseries 901} (2015) 510--525}, \href{http://arxiv.org/abs/1509.01680}{{
  arXiv:1509.01680~[cond-mat.str-el]}}.

\bibitem{Kimura:2015nka}
T.~Kimura and M.~Murata, {\slshape {Transport Process in Multi-Junctions of
  Quantum Systems},} \href{http://dx.doi.org/10.1007/JHEP07(2015)072}{{\em
  JHEP} {\bfseries 07} (2015) 072}, \href{http://arxiv.org/abs/1505.05275}{{
  arXiv:1505.05275~[hep-th]}}.

\bibitem{Fuji_2017}
Y.~Fuji and P.~Lecheminant
  \href{https://doi.org/10.1103%2Fphysrevb.95.125130}{{\em Physical Review B}
  {\bfseries 95} (Mar, 2017) }.

\bibitem{Quella:2019los}
T.~Quella and A.~Roy, {\slshape {Conformal field theory and the non-abelian
  $SU(2)_k$ chiral spin liquid},}
  \href{http://dx.doi.org/10.1088/1742-5468/ab7c62}{{\em J. Stat. Mech.}
  {\bfseries 2005} (2020) 053107}, \href{http://arxiv.org/abs/1911.01505}{{
  arXiv:1911.01505~[cond-mat.str-el]}}.

\bibitem{Cheng:2026qax}
M.~Cheng and N.~Seiberg, {\slshape {Proliferation transitions from a
  topological phase in $2+1$ dimensions},}
  \href{http://arxiv.org/abs/2603.00245}{{
  arXiv:2603.00245~[cond-mat.str-el]}}.

\bibitem{Graham:2003nc}
K.~Graham and G.~M.~T. Watts, {\slshape {Defect lines and boundary flows},}
  \href{http://dx.doi.org/10.1088/1126-6708/2004/04/019}{{\em JHEP} {\bfseries
  04} (2004) 019}, \href{http://arxiv.org/abs/hep-th/0306167}{{
  arXiv:hep-th/0306167}}.

\bibitem{Furuya:2015coa}
S.~C. Furuya and M.~Oshikawa, {\slshape {Symmetry Protection of Critical Phases
  and a Global Anomaly in $1+1$ Dimensions},}
  \href{http://dx.doi.org/10.1103/PhysRevLett.118.021601}{{\em Phys. Rev.
  Lett.} {\bfseries 118} (2017) 021601},
  \href{http://arxiv.org/abs/1503.07292}{{
  arXiv:1503.07292~[cond-mat.stat-mech]}}.

\bibitem{Numasawa:2017crf}
T.~Numasawa and S.~Yamaguch, {\slshape {Mixed Global Anomalies and Boundary
  Conformal Field Theories},}
  \href{http://dx.doi.org/10.1007/JHEP11(2018)202}{{\em JHEP} {\bfseries 11}
  (2018) 202}, \href{http://arxiv.org/abs/1712.09361}{{
  arXiv:1712.09361~[hep-th]}}.

\bibitem{Fukusumi_2022_c}
Y.~Fukusumi, {\slshape {Composing parafermions: a construction of $Z_{N}$
  fractional quantum Hall systems and a modern understanding of confinement and
  duality},} \href{http://arxiv.org/abs/2212.12999}{{
  arXiv:2212.12999~[cond-mat.str-el]}}.

\bibitem{Kikuchi:2022ipr}
K.~Kikuchi, {\slshape {RG flows from WZW models},}
  \href{http://arxiv.org/abs/2212.13851}{{ arXiv:2212.13851~[hep-th]}}.

\bibitem{Schellekens:1990ys}
A.~N. Schellekens, {\slshape {Fusion rule automorphisms from integer spin
  simple currents},} \href{http://dx.doi.org/10.1016/0370-2693(90)90065-E}{{\em
  Phys. Lett. B} {\bfseries 244} (1990) 255--260}.

\bibitem{Gato-Rivera:1990lxi}
B.~Gato-Rivera and A.~N. Schellekens, {\slshape {Complete classification of
  simple current automorphisms},}
  \href{http://dx.doi.org/10.1016/0550-3213(91)90346-Y}{{\em Nucl. Phys. B}
  {\bfseries 353} (1991) 519--537}.

\bibitem{Gato-Rivera:1991bqv}
B.~Gato-Rivera and A.~N. Schellekens, {\slshape {Complete classification of
  simple current modular invariants for (Z(p))**k},}
  \href{http://dx.doi.org/10.1007/BF02099282}{{\em Commun. Math. Phys.}
  {\bfseries 145} (1992) 85--122}.

\bibitem{Kreuzer:1993tf}
M.~Kreuzer and A.~N. Schellekens, {\slshape {Simple currents versus orbifolds
  with discrete torsion: A Complete classification},}
  \href{http://dx.doi.org/10.1016/0550-3213(94)90055-8}{{\em Nucl. Phys. B}
  {\bfseries 411} (1994) 97--121}, \href{http://arxiv.org/abs/hep-th/9306145}{{
  arXiv:hep-th/9306145}}.

\bibitem{Delmastro:2025ksn}
D.~Delmastro, A.~Sharon, and Y.~Zheng, {\slshape {Non-Local Conserved Currents
  and Continuous Non-Invertible Symmetries},}
  \href{http://arxiv.org/abs/2507.22976}{{ arXiv:2507.22976~[hep-th]}}.

\bibitem{Leng:2025qlu}
R.~Leng, C.-Y. Lee, and S.~Zhou, {\slshape {Pseudo-Hermitian QFT: relativistic
  scattering and symmetry structure},} \href{http://arxiv.org/abs/2510.27404}{{
  arXiv:2510.27404~[hep-th]}}.

\bibitem{Bender:1998gh}
C.~M. Bender, S.~Boettcher, and P.~Meisinger, {\slshape {PT symmetric quantum
  mechanics},} \href{http://dx.doi.org/10.1063/1.532860}{{\em J. Math. Phys.}
  {\bfseries 40} (1999) 2201--2229},
  \href{http://arxiv.org/abs/quant-ph/9809072}{{ arXiv:quant-ph/9809072}}.

\bibitem{Note3}
We thank Naomichi Hatano for clarifying these points.

\bibitem{Note4}
Because the nonunitary CFT we study in this section has real eigenenergy, it
  might be more precise to call them pseudo-unitary CFT or pseudo-Hermitian
  CFTs.

\bibitem{DiFrancesco:1997nk}
P.~Di~Francesco, P.~Mathieu, and D.~Senechal,
  \href{http://dx.doi.org/10.1007/978-1-4612-2256-9}{{\em {Conformal Field
  Theory}}}.
\newblock Graduate Texts in Contemporary Physics. Springer-Verlag, New York,
  1997.

\bibitem{Ginsparg:1988ui}
P.~H. Ginsparg, {\slshape {APPLIED CONFORMAL FIELD THEORY},} in {\em {Les
  Houches Summer School in Theoretical Physics: Fields, Strings, Critical
  Phenomena}}.
\newblock 9, 1988.
\newblock \href{http://arxiv.org/abs/hep-th/9108028}{{ arXiv:hep-th/9108028}}.

\bibitem{Recknagel:2013uja}
A.~Recknagel and V.~Schomerus,
  \href{http://dx.doi.org/10.1017/CBO9780511806476}{{\em {Boundary Conformal
  Field Theory and the Worldsheet Approach to D-Branes}}}.
\newblock Cambridge Monographs on Mathematical Physics. Cambridge University
  Press, 11, 2013.

\bibitem{Northe:2024tnm}
C.~Northe, {\slshape {Young Researchers School 2024 Maynooth: Lectures on CFT,
  BCFT and DCFT},} \href{http://arxiv.org/abs/2411.03381}{{
  arXiv:2411.03381~[hep-th]}}.

\bibitem{Note5}
We have used the symbol $t$ which is less common than $\tau $ to represent the
  modular parameter for a notational reason. In the successive discussion, we
  use $\tau $ for the Fibonacci anyon, satisfying the fusion rule $\tau \times
  \tau =I+\tau $.

\bibitem{Cardy:1989ir}
J.~L. Cardy, {\slshape {Boundary Conditions, Fusion Rules and the Verlinde
  Formula},} \href{http://dx.doi.org/10.1016/0550-3213(89)90521-X}{{\em Nucl.
  Phys. B} {\bfseries 324} (1989) 581--596}.

\bibitem{Belavin:1984vu}
A.~A. Belavin, A.~M. Polyakov, and A.~B. Zamolodchikov, {\slshape {Infinite
  Conformal Symmetry in Two-Dimensional Quantum Field Theory},}
  \href{http://dx.doi.org/10.1016/0550-3213(84)90052-X}{{\em Nucl. Phys. B}
  {\bfseries 241} (1984) 333--380}.

\bibitem{Zwiebach:1992ie}
B.~Zwiebach, {\slshape {Closed string field theory: Quantum action and the B-V
  master equation},} \href{http://dx.doi.org/10.1016/0550-3213(93)90388-6}{{\em
  Nucl. Phys. B} {\bfseries 390} (1993) 33--152},
  \href{http://arxiv.org/abs/hep-th/9206084}{{ arXiv:hep-th/9206084}}.

\bibitem{Gaberdiel:1997ia}
M.~R. Gaberdiel and B.~Zwiebach, {\slshape {Tensor constructions of open string
  theories. 1: Foundations},}
  \href{http://dx.doi.org/10.1016/S0550-3213(97)00580-4}{{\em Nucl. Phys. B}
  {\bfseries 505} (1997) 569--624},
  \href{http://arxiv.org/abs/hep-th/9705038}{{ arXiv:hep-th/9705038}}.

\bibitem{Sen:2016bwe}
A.~Sen, {\slshape {Reality of Superstring Field Theory Action},}
  \href{http://dx.doi.org/10.1007/JHEP11(2016)014}{{\em JHEP} {\bfseries 11}
  (2016) 014}, \href{http://arxiv.org/abs/1606.03455}{{
  arXiv:1606.03455~[hep-th]}}.

\bibitem{Erbin:2021smf}
H.~Erbin, \href{http://dx.doi.org/10.1007/978-3-030-65321-7}{{\em {String Field
  Theory: A Modern Introduction}}}, vol.~980 of {\em Lecture Notes in Physics}.
\newblock 3, 2021.
\newblock \href{http://arxiv.org/abs/2301.01686}{{ arXiv:2301.01686~[hep-th]}}.

\bibitem{Sen:2024nfd}
A.~Sen and B.~Zwiebach, {\slshape {String Field Theory: A Review},}
  \href{http://arxiv.org/abs/2405.19421}{{ arXiv:2405.19421~[hep-th]}}.

\bibitem{Verlinde:1988sn}
E.~P. Verlinde, {\slshape {Fusion Rules and Modular Transformations in 2D
  Conformal Field Theory},}
  \href{http://dx.doi.org/10.1016/0550-3213(88)90603-7}{{\em Nucl. Phys. B}
  {\bfseries 300} (1988) 360--376}.

\bibitem{Apruzzi:2021nmk}
F.~Apruzzi, F.~Bonetti, I.~Garc{\'\i}a~Etxebarria, S.~S. Hosseini, and
  S.~Schafer-Nameki, {\slshape {Symmetry TFTs from String Theory},}
  \href{http://dx.doi.org/10.1007/s00220-023-04737-2}{{\em Commun. Math. Phys.}
  {\bfseries 402} (2023) 895--949}, \href{http://arxiv.org/abs/2112.02092}{{
  arXiv:2112.02092~[hep-th]}}.

\bibitem{Kaidi:2022cpf}
J.~Kaidi, K.~Ohmori, and Y.~Zheng, {\slshape {Symmetry TFTs for Non-invertible
  Defects},} \href{http://dx.doi.org/10.1007/s00220-023-04859-7}{{\em Commun.
  Math. Phys.} {\bfseries 404} (2023) 1021--1124},
  \href{http://arxiv.org/abs/2209.11062}{{ arXiv:2209.11062~[hep-th]}}.

\bibitem{Kaidi:2023maf}
J.~Kaidi, E.~Nardoni, G.~Zafrir, and Y.~Zheng, {\slshape {Symmetry TFTs and
  anomalies of non-invertible symmetries},}
  \href{http://dx.doi.org/10.1007/JHEP10(2023)053}{{\em JHEP} {\bfseries 10}
  (2023) 053}, \href{http://arxiv.org/abs/2301.07112}{{
  arXiv:2301.07112~[hep-th]}}.

\bibitem{Fuchs:2002cm}
J.~Fuchs, I.~Runkel, and C.~Schweigert, {\slshape {TFT construction of RCFT
  correlators 1. Partition functions},}
  \href{http://dx.doi.org/10.1016/S0550-3213(02)00744-7}{{\em Nucl. Phys. B}
  {\bfseries 646} (2002) 353--497},
  \href{http://arxiv.org/abs/hep-th/0204148}{{ arXiv:hep-th/0204148}}.

\bibitem{kong2015boundarybulkrelationtopologicalorders}
L.~Kong, X.-G. Wen, and H.~Zheng, {\slshape Boundary-bulk relation for
  topological orders as the functor mapping higher categories to their
  centers,} 2015.
\newblock \url{https://arxiv.org/abs/1502.01690}.

\bibitem{Kong:2017etd}
L.~Kong and H.~Zheng, {\slshape {Gapless edges of 2d topological orders and
  enriched monoidal categories},}
  \href{http://dx.doi.org/10.1016/j.nuclphysb.2017.12.007}{{\em Nucl. Phys. B}
  {\bfseries 927} (2018) 140--165}, \href{http://arxiv.org/abs/1705.01087}{{
  arXiv:1705.01087~[cond-mat.str-el]}}.

\bibitem{Kong:2020cie}
L.~Kong, T.~Lan, X.-G. Wen, Z.-H. Zhang, and H.~Zheng, {\slshape {Algebraic
  higher symmetry and categorical symmetry -- a holographic and entanglement
  view of symmetry},}
  \href{http://dx.doi.org/10.1103/PhysRevResearch.2.043086}{{\em Phys. Rev.
  Res.} {\bfseries 2} (2020) 043086}, \href{http://arxiv.org/abs/2005.14178}{{
  arXiv:2005.14178~[cond-mat.str-el]}}.

\bibitem{Moradi:2022lqp}
H.~Moradi, S.~F. Moosavian, and A.~Tiwari, {\slshape {Topological holography:
  Towards a unification of Landau and beyond-Landau physics},}
  \href{http://dx.doi.org/10.21468/SciPostPhysCore.6.4.066}{{\em SciPost Phys.
  Core} {\bfseries 6} (2023) 066}, \href{http://arxiv.org/abs/2207.10712}{{
  arXiv:2207.10712~[cond-mat.str-el]}}.

\bibitem{Bhardwaj:2023bbf}
L.~Bhardwaj, L.~E. Bottini, D.~Pajer, and S.~Schafer-Nameki, {\slshape {The
  Club Sandwich: Gapless Phases and Phase Transitions with Non-Invertible
  Symmetries},} \href{http://arxiv.org/abs/2312.17322}{{
  arXiv:2312.17322~[hep-th]}}.

\bibitem{Huang:2023pyk}
S.-J. Huang and M.~Cheng, {\slshape {Topological holography, quantum
  criticality, and boundary states},} \href{http://arxiv.org/abs/2310.16878}{{
  arXiv:2310.16878~[cond-mat.str-el]}}.

\bibitem{Wen:2024udn}
R.~Wen, W.~Ye, and A.~C. Potter, {\slshape {Topological holography for
  fermions},} \href{http://arxiv.org/abs/2404.19004}{{
  arXiv:2404.19004~[cond-mat.str-el]}}.

\bibitem{Fukusumi:2024cnl}
Y.~Fukusumi, {\slshape {Fusion rule in conformal field theories and topological
  orders: A unified view of correspondence and (fractional) supersymmetry and
  their relation to topological holography},}
  \href{http://arxiv.org/abs/2405.05178}{{ arXiv:2405.05178~[hep-th]}}.

\bibitem{Huang:2024ror}
S.-J. Huang, {\slshape {Fermionic quantum criticality through the lens of
  topological holography},} \href{http://arxiv.org/abs/2405.09611}{{
  arXiv:2405.09611~[cond-mat.str-el]}}.

\bibitem{Bhardwaj:2024ydc}
L.~Bhardwaj, K.~Inamura, and A.~Tiwari, {\slshape {Fermionic Non-Invertible
  Symmetries in (1+1)d: Gapped and Gapless Phases, Transitions, and Symmetry
  TFTs},} \href{http://arxiv.org/abs/2405.09754}{{ arXiv:2405.09754~[hep-th]}}.

\bibitem{Fukusumi:2025ljx}
Y.~Fukusumi and S.~Yahagi, {\slshape {Extending fusion rules with finite
  subgroups: For a general understanding of quotient or gauging},}
  \href{http://arxiv.org/abs/2508.08639}{{ arXiv:2508.08639~[hep-th]}}.

\bibitem{Motamarri:2023abx}
V.~Motamarri, C.~McLauchlan, and B.~Beri, {\slshape {SymTFT out of equilibrium:
  from time crystals to braided drives and Floquet codes},}
  \href{http://arxiv.org/abs/2312.17176}{{
  arXiv:2312.17176~[cond-mat.str-el]}}.

\bibitem{McLauchlan:2025rnw}
C.~McLauchlan, V.~Motamarri, and B.~B{\'e}ri, {\slshape {Classifying
  one-dimensional Floquet phases through two-dimensional topological order},}
  \href{http://arxiv.org/abs/2512.15868}{{
  arXiv:2512.15868~[cond-mat.str-el]}}.

\bibitem{Grimm:2001dr}
U.~Grimm, {\slshape {Spectrum of a duality twisted Ising quantum chain},}
  \href{http://dx.doi.org/10.1088/0305-4470/35/3/101}{{\em J. Phys. A}
  {\bfseries 35} (2002) L25--L30}, \href{http://arxiv.org/abs/hep-th/0111157}{{
  arXiv:hep-th/0111157}}.

\bibitem{Belletete:2018eua}
J.~Bellet\^ete, A.~M. Gainutdinov, J.~L. Jacobsen, H.~Saleur, and T.~S.
  Tavares, {\slshape {Topological Defects in Lattice Models and Affine
  Temperley\textendash{}Lieb Algebra},}
  \href{http://dx.doi.org/10.1007/s00220-022-04618-0}{{\em Commun. Math. Phys.}
  {\bfseries 400} (2023) 1203--1254}, \href{http://arxiv.org/abs/1811.02551}{{
  arXiv:1811.02551~[hep-th]}}.

\bibitem{Belletete:2020gst}
J.~Bellet\^ete, A.~M. Gainutdinov, J.~L. Jacobsen, H.~Saleur, and T.~S.
  Tavares, {\slshape {Topological defects in periodic RSOS models and anyonic
  chains},} \href{http://arxiv.org/abs/2003.11293}{{
  arXiv:2003.11293~[math-ph]}}.

\bibitem{Seiberg:2023cdc}
N.~Seiberg and S.-H. Shao, {\slshape {Majorana chain and Ising model --
  (non-invertible) translations, anomalies, and emanant symmetries},}
  \href{http://arxiv.org/abs/2307.02534}{{
  arXiv:2307.02534~[cond-mat.str-el]}}.

\bibitem{Seiberg:2024gek}
N.~Seiberg, S.~Seifnashri, and S.-H. Shao, {\slshape {Non-invertible symmetries
  and LSM-type constraints on a tensor product Hilbert space},}
  \href{http://arxiv.org/abs/2401.12281}{{
  arXiv:2401.12281~[cond-mat.str-el]}}.

\bibitem{Sinha:2025jhh}
M.~Sinha, T.~S. Tavares, A.~Roy, and H.~Saleur, {\slshape {Integrability and
  lattice discretizations of all Topological Defect Lines in minimal CFTs},}
  \href{http://arxiv.org/abs/2509.04257}{{ arXiv:2509.04257~[hep-th]}}.

\bibitem{Note6}
For example, the Fibonacci anyon $\tau $ with the fusion rule $\tau \times \tau
  =I+\tau $ generates invertible symmetry $\protect \mathcal {Q}_{\tau }$ under
  the relation $\protect \mathcal {Q}_{\tau }(\protect \mathcal {Q}_{\tau
  }-\protect \mathcal {Q}_{I})=\protect \mathcal {Q}_{I}$.

\bibitem{Note7}
A $\protect \mathbb {C}$-linear category will correspond to the linear algebra,
  but it is less common in the context of fusion rules.

\bibitem{onn2006quantumdimensionsnonarchimedeandegenerations}
U.~Onn and J.~Stokman, {\slshape Quantum dimensions and their non-archimedean
  degenerations,} 2006.
\newblock \url{https://arxiv.org/abs/math/0606222}.

\bibitem{Gannon:2003de}
T.~Gannon, {\slshape {Comments on nonunitary conformal field theories},}
  \href{http://dx.doi.org/10.1016/j.nuclphysb.2003.07.030}{{\em Nucl. Phys. B}
  {\bfseries 670} (2003) 335--358},
  \href{http://arxiv.org/abs/hep-th/0305070}{{ arXiv:hep-th/0305070}}.

\bibitem{Beltaos:2010ka}
E.~Beltaos and T.~Gannon, {\slshape {The $W_{N}$ minimal model
  classification},} \href{http://dx.doi.org/10.1007/s00220-012-1473-4}{{\em
  Commun. Math. Phys.} {\bfseries 312} (2012) 337--360},
  \href{http://arxiv.org/abs/1004.1205}{{ arXiv:1004.1205~[hep-th]}}.

\bibitem{Milovanovic:1996nj}
M.~Milovanovic and N.~Read, {\slshape {Edge excitations of paired fractional
  quantum Hall states},}
  \href{http://dx.doi.org/10.1103/PhysRevB.53.13559}{{\em Phys. Rev. B}
  {\bfseries 53} (1996) 13559}, \href{http://arxiv.org/abs/cond-mat/9602113}{{
  arXiv:cond-mat/9602113}}.

\bibitem{Ino:1998by}
K.~Ino, {\slshape {Modular invariants in the fractional quantum Hall effect},}
  \href{http://dx.doi.org/10.1016/S0550-3213(98)00598-7}{{\em Nucl. Phys. B}
  {\bfseries 532} (1998) 783--806},
  \href{http://arxiv.org/abs/cond-mat/9804198}{{ arXiv:cond-mat/9804198}}.

\bibitem{Fukusumi_2022}
Y.~Fukusumi, {\slshape {Gaplessness protected by bulk-edge correspondence},}
  \href{http://arxiv.org/abs/2212.12996}{{
  arXiv:2212.12996~[cond-mat.str-el]}}.

\bibitem{Fuchs:1991ci}
J.~Fuchs, {\slshape {Quantum dimensions},}.

\bibitem{etingof2015tensor}
P.~Etingof, S.~Gelaki, D.~Nikshych, and V.~Ostrik, {\em Tensor categories},
  vol.~205.
\newblock American Mathematical Soc., 2015.

\bibitem{Kong:2019cuu}
L.~Kong and H.~Zheng, {\slshape {A mathematical theory of gapless edges of 2d
  topological orders. Part II},}
  \href{http://dx.doi.org/10.1016/j.nuclphysb.2021.115384}{{\em Nucl. Phys. B}
  {\bfseries 966} (2021) 115384}, \href{http://arxiv.org/abs/1912.01760}{{
  arXiv:1912.01760~[cond-mat.str-el]}}.

\bibitem{SCHWARZ1982141}
A.~Schwarz, {\slshape Field theories with no local conservation of the electric
  charge,}
  \href{https://www.sciencedirect.com/science/article/pii/0550321382901900}{{\em
  Nuclear Physics B} {\bfseries 208} (1982) 141--158}.

\bibitem{SCHWARZ1982427}
A.~Schwarz and Y.~Tyupkin, {\slshape Grand unification and mirror particles,}
  \href{https://www.sciencedirect.com/science/article/pii/0550321382902656}{{\em
  Nuclear Physics B} {\bfseries 209} (1982) 427--432}.

\bibitem{Fredenhagen:1988fj}
K.~Fredenhagen, K.-H. Rehren, and B.~Schroer, {\slshape {Superselection Sectors
  with Braid Group Statistics and Exchange Algebras. 1. General Theory},}
  \href{http://dx.doi.org/10.1007/BF01217906}{{\em Commun. Math. Phys.}
  {\bfseries 125} (1989) 201}.

\bibitem{Fuchs:1989rv}
J.~Fuchs and P.~van Driel, {\slshape {Some Symmetries of Quantum Dimensions},}
  \href{http://dx.doi.org/10.1063/1.528673}{{\em J. Math. Phys.} {\bfseries 31}
  (1990) 1770--1775}.

\bibitem{Fredenhagen:1992yz}
K.~Fredenhagen, K.-H. Rehren, and B.~Schroer, {\slshape {Superselection sectors
  with braid group statistics and exchange algebras. 2. Geometric aspects and
  conformal covariance},}
  \href{http://dx.doi.org/10.1142/S0129055X92000170}{{\em Rev. Math. Phys.}
  {\bfseries 4} (1992) 113--157}.

\bibitem{Cardy:1986gw}
J.~L. Cardy, {\slshape {Effect of Boundary Conditions on the Operator Content
  of Two-Dimensional Conformally Invariant Theories},}
  \href{http://dx.doi.org/10.1016/0550-3213(86)90596-1}{{\em Nucl. Phys. B}
  {\bfseries 275} (1986) 200--218}.

\bibitem{Ishibashi:1988kg}
N.~Ishibashi, {\slshape {The Boundary and Crosscap States in Conformal Field
  Theories},} \href{http://dx.doi.org/10.1142/S0217732389000320}{{\em Mod.
  Phys. Lett. A} {\bfseries 4} (1989) 251}.

\bibitem{elDeeb:2015jgf}
O.~el~Deeb, {\slshape {The critical boundary RSOS $\mathcal M$(3,5) model},}
  \href{http://dx.doi.org/10.1134/S0040577917120078}{{\em Theor. Math. Phys.}
  {\bfseries 193} (2017) 1811--1825}, \href{http://arxiv.org/abs/1512.02185}{{
  arXiv:1512.02185~[hep-th]}}.

\bibitem{Feiguin:2006ydp}
A.~Feiguin, S.~Trebst, A.~W.~W. Ludwig, M.~Troyer, A.~Kitaev, Z.~Wang, and
  M.~H. Freedman, {\slshape {Interacting anyons in topological quantum liquids:
  The golden chain},}
  \href{http://dx.doi.org/10.1103/PhysRevLett.98.160409}{{\em Phys. Rev. Lett.}
  {\bfseries 98} (2007) 160409}, \href{http://arxiv.org/abs/cond-mat/0612341}{{
  arXiv:cond-mat/0612341}}.

\bibitem{Ardonne:2011wxx}
E.~Ardonne, J.~Gukelberger, A.~W.~W. Ludwig, S.~Trebst, and M.~Troyer,
  {\slshape {Microscopic models of interacting Yang{\textendash}Lee anyons},}
  \href{http://dx.doi.org/10.1088/1367-2630/13/4/045006}{{\em New J. Phys.}
  {\bfseries 13} (2011) 045006}.

\bibitem{Buican:2017rxc}
M.~Buican and A.~Gromov, {\slshape {Anyonic Chains, Topological Defects, and
  Conformal Field Theory},}
  \href{http://dx.doi.org/10.1007/s00220-017-2995-6}{{\em Commun. Math. Phys.}
  {\bfseries 356} (2017) 1017--1056}, \href{http://arxiv.org/abs/1701.02800}{{
  arXiv:1701.02800~[hep-th]}}.

\bibitem{Affleck:1991tk}
I.~Affleck and A.~W.~W. Ludwig, {\slshape {Universal noninteger 'ground state
  degeneracy' in critical quantum systems},}
  \href{http://dx.doi.org/10.1103/PhysRevLett.67.161}{{\em Phys. Rev. Lett.}
  {\bfseries 67} (1991) 161--164}.

\bibitem{Behrend:1999bn}
R.~E. Behrend, P.~A. Pearce, V.~B. Petkova, and J.-B. Zuber, {\slshape
  {Boundary conditions in rational conformal field theories},}
  \href{http://dx.doi.org/10.1016/S0550-3213(99)00592-1}{{\em Nucl. Phys. B}
  {\bfseries 570} (2000) 525--589},
  \href{http://arxiv.org/abs/hep-th/9908036}{{ arXiv:hep-th/9908036}}.

\bibitem{Fukusumi:2020irh}
Y.~Fukusumi and S.~Iino, {\slshape {Open spin chain realization of a
  topological defect in a one-dimensional Ising model: Boundary and bulk
  symmetry},} \href{http://dx.doi.org/10.1103/PhysRevB.104.125418}{{\em Phys.
  Rev. B} {\bfseries 104} (2021) 125418},
  \href{http://arxiv.org/abs/2004.04415}{{ arXiv:2004.04415~[hep-th]}}.

\bibitem{Okada:2024qmk}
M.~Okada and Y.~Tachikawa, {\slshape {Non-invertible symmetries act locally by
  quantum operations},} \href{http://arxiv.org/abs/2403.20062}{{
  arXiv:2403.20062~[hep-th]}}.

\bibitem{Smith:2021luc}
P.~Boyle~Smith, {\slshape {Boundary States and Anomalous Symmetries of
  Fermionic Minimal Models},} \href{http://arxiv.org/abs/2102.02203}{{
  arXiv:2102.02203~[hep-th]}}.

\bibitem{Weizmann}
H.~Ebisu and M.~Watanabe, {\slshape {Fermionization of conformal boundary
  states},} \href{http://dx.doi.org/10.1103/PhysRevB.104.195124}{{\em Phys.
  Rev. B} {\bfseries 104} (2021) 195124},
  \href{http://arxiv.org/abs/2103.01101}{{ arXiv:2103.01101~[hep-th]}}.

\bibitem{Fukusumi:2021zme}
Y.~Fukusumi, Y.~Tachikawa, and Y.~Zheng, {\slshape {Fermionization and boundary
  states in 1+1 dimensions},}
  \href{http://dx.doi.org/10.21468/SciPostPhys.11.4.082}{{\em SciPost Phys.}
  {\bfseries 11} (2021) 082}, \href{http://arxiv.org/abs/2103.00746}{{
  arXiv:2103.00746~[hep-th]}}.

\bibitem{Pollmann:2009ryx}
F.~Pollmann, A.~M. Turner, E.~Berg, and M.~Oshikawa, {\slshape {Entanglement
  spectrum of a topological phase in one dimension},}
  \href{http://dx.doi.org/10.1103/PhysRevB.81.064439}{{\em Phys. Rev. B}
  {\bfseries 81} (2010) 064439}, \href{http://arxiv.org/abs/0910.1811}{{
  arXiv:0910.1811~[cond-mat.str-el]}}.

\bibitem{Pollmann:2009mhk}
F.~Pollmann, E.~Berg, A.~M. Turner, and M.~Oshikawa, {\slshape {Symmetry
  protection of topological phases in one-dimensional quantum spin systems},}
  \href{http://dx.doi.org/10.1103/PhysRevB.85.075125}{{\em Phys. Rev. B}
  {\bfseries 85} (2012) 075125}, \href{http://arxiv.org/abs/0909.4059}{{
  arXiv:0909.4059~[cond-mat.str-el]}}.

\bibitem{Fukusumi:2023vjm}
Y.~Fukusumi, {\slshape {Protected edge modes based on the bulk and boundary
  renormalization group: A relationship between duality and generalized
  symmetry},} \href{http://arxiv.org/abs/2312.12887}{{
  arXiv:2312.12887~[hep-th]}}.

\bibitem{Wu:2023ezm}
Y.-H. Wu, Y.~Zhang, H.-H. Tu, and M.~Cheng, {\slshape {Impurity screening by
  defects in (1+1)$d$ quantum critical systems},}
  \href{http://arxiv.org/abs/2307.09519}{{
  arXiv:2307.09519~[cond-mat.str-el]}}.

\bibitem{Kikuchi:2023gpj}
K.~Kikuchi, {\slshape {Ground state degeneracy and module category},}
  \href{http://arxiv.org/abs/2311.00746}{{ arXiv:2311.00746~[hep-th]}}.

\bibitem{Note8}
However, to study a larger fusion ring symmetry $\protect \mathbf {C}''(\supset
  \{ I,\tau \})$, one needs to do the same analysis in the respective
  situation.

\bibitem{Affleck:1998nq}
I.~Affleck, M.~Oshikawa, and H.~Saleur, {\slshape {Boundary critical phenomena
  in the three state Potts model},}
  \href{http://dx.doi.org/10.1088/0305-4470/31/28/003}{{\em J. Phys. A}
  {\bfseries 31} (1998) 5827}, \href{http://arxiv.org/abs/cond-mat/9804117}{{
  arXiv:cond-mat/9804117}}.

\bibitem{Fuchs:1999zi}
J.~Fuchs and C.~Schweigert, {\slshape {Symmetry breaking boundaries. 1. General
  theory},} \href{http://dx.doi.org/10.1016/S0550-3213(99)00406-X}{{\em Nucl.
  Phys. B} {\bfseries 558} (1999) 419--483},
  \href{http://arxiv.org/abs/hep-th/9902132}{{ arXiv:hep-th/9902132}}.

\bibitem{Quella:2002ct}
T.~Quella and V.~Schomerus, {\slshape {Symmetry breaking boundary states and
  defect lines},} \href{http://dx.doi.org/10.1088/1126-6708/2002/06/028}{{\em
  JHEP} {\bfseries 06} (2002) 028},
  \href{http://arxiv.org/abs/hep-th/0203161}{{ arXiv:hep-th/0203161}}.

\bibitem{Fuchs:1993et}
J.~Fuchs, {\slshape {Fusion rules in conformal field theory},}
  \href{http://dx.doi.org/10.1002/prop.2190420102}{{\em Fortsch. Phys.}
  {\bfseries 42} (1994) 1--48}, \href{http://arxiv.org/abs/hep-th/9306162}{{
  arXiv:hep-th/9306162}}.

\bibitem{Ishikawa:2002wx}
H.~Ishikawa and T.~Tani, {\slshape {Novel construction of boundary states in
  coset conformal field theories},}
  \href{http://dx.doi.org/10.1016/S0550-3213(02)01011-8}{{\em Nucl. Phys. B}
  {\bfseries 649} (2003) 205--242},
  \href{http://arxiv.org/abs/hep-th/0207177}{{ arXiv:hep-th/0207177}}.

\bibitem{Ishikawa:2005ea}
H.~Ishikawa and T.~Tani, {\slshape {Twisted Boundary States and Representation
  of Generalized Fusion Algebra},}
  \href{http://dx.doi.org/10.1016/j.nuclphysb.2006.01.031}{{\em Nucl. Phys. B}
  {\bfseries 739} (2006) 328--388},
  \href{http://arxiv.org/abs/hep-th/0510242}{{ arXiv:hep-th/0510242}}.

\bibitem{Aksoy:2025rmg}
O.~M. Aksoy and X.-G. Wen, {\slshape {Phases with non-invertible symmetries in
  1+1D symmetry protected topological orders as duality automorphisms},}
  \href{http://arxiv.org/abs/2503.21764}{{
  arXiv:2503.21764~[cond-mat.str-el]}}.

\bibitem{Kikuchi:2023cgg}
K.~Kikuchi, {\slshape {Classification of connected \'etale algebras in
  pre-modular fusion categories up to rank three},}
  \href{http://arxiv.org/abs/2311.15631}{{ arXiv:2311.15631~[math.QA]}}.

\bibitem{Kikuchi:2023eor}
K.~Kikuchi, {\slshape {Classification of connected \'etale algebras in modular
  fusion categories up to rank five},} \href{http://arxiv.org/abs/2312.13353}{{
  arXiv:2312.13353~[math.QA]}}.

\bibitem{Kikuchi:2024hwf}
K.~Kikuchi, K.-S. Kam, and F.-H. Huang, {\slshape {Classification of connected
  \'etale algebras in multiplicity-free modular fusion categories at rank
  six},} \href{http://arxiv.org/abs/2402.00403}{{ arXiv:2402.00403~[math.QA]}}.

\bibitem{Kikuchi:2024pex}
K.~Kikuchi, {\slshape {Classification of connected \'etale algebras in
  multiplicity-free modular fusion categories up to rank nine},}
  \href{http://arxiv.org/abs/2404.16125}{{ arXiv:2404.16125~[math.QA]}}.

\bibitem{Inamura_2021}
K.~Inamura, {\slshape Topological field theories and symmetry protected
  topological phases with fusion category symmetries,}
  \href{http://dx.doi.org/10.1007/JHEP05(2021)204}{{\em Journal of High Energy
  Physics} {\bfseries 2021} (May, 2021) }.

\bibitem{Inamura:2021szw}
K.~Inamura, {\slshape {On lattice models of gapped phases with fusion category
  symmetries},} \href{http://dx.doi.org/10.1007/JHEP03(2022)036}{{\em JHEP}
  {\bfseries 03} (2022) 036}, \href{http://arxiv.org/abs/2110.12882}{{
  arXiv:2110.12882~[cond-mat.str-el]}}.

\bibitem{Fechisin:2023odt}
C.~Fechisin, N.~Tantivasadakarn, and V.~V. Albert, {\slshape {Noninvertible
  Symmetry-Protected Topological Order in a Group-Based Cluster State},}
  \href{http://dx.doi.org/10.1103/PhysRevX.15.011058}{{\em Phys. Rev. X}
  {\bfseries 15} (2025) 011058}, \href{http://arxiv.org/abs/2312.09272}{{
  arXiv:2312.09272~[cond-mat.str-el]}}.

\bibitem{Pace:2024acq}
S.~D. Pace, H.~T. Lam, and {\"O}.~M. Aksoy, {\slshape {(SPT-)LSM theorems from
  projective non-invertible symmetries},}
  \href{http://dx.doi.org/10.21468/SciPostPhys.18.1.028}{{\em SciPost Phys.}
  {\bfseries 18} (2025) 028}, \href{http://arxiv.org/abs/2409.18113}{{
  arXiv:2409.18113~[cond-mat.str-el]}}.

\bibitem{Seifnashri:2024dsd}
S.~Seifnashri and S.-H. Shao, {\slshape {Cluster state as a non-invertible
  symmetry protected topological phase},}
  \href{http://arxiv.org/abs/2404.01369}{{
  arXiv:2404.01369~[cond-mat.str-el]}}.

\bibitem{Cao:2024qjj}
W.~Cao, L.~Li, and M.~Yamazaki, {\slshape {Generating lattice non-invertible
  symmetries},} \href{http://dx.doi.org/10.21468/SciPostPhys.17.4.104}{{\em
  SciPost Phys.} {\bfseries 17} (2024) 104},
  \href{http://arxiv.org/abs/2406.05454}{{
  arXiv:2406.05454~[cond-mat.str-el]}}.

\bibitem{Jia:2024bng}
Z.~Jia, {\slshape {Generalized cluster states from Hopf algebras:
  non-invertible symmetry and Hopf tensor network representation},}
  \href{http://dx.doi.org/10.1007/JHEP09(2024)147}{{\em JHEP} {\bfseries 09}
  (2024) 147}, \href{http://arxiv.org/abs/2405.09277}{{
  arXiv:2405.09277~[quant-ph]}}.

\bibitem{Jia:2024zdp}
Z.~Jia, {\slshape {Weak Hopf non-invertible symmetry-protected topological spin
  liquid and lattice realization of (1+1)D symmetry topological field theory},}
  \href{http://arxiv.org/abs/2412.15336}{{ arXiv:2412.15336~[hep-th]}}.

\bibitem{Cao:2025qhg}
W.~Cao, M.~Yamazaki, and L.~Li, {\slshape {Duality viewpoint of noninvertible
  symmetry protected topological phases},}
  \href{http://arxiv.org/abs/2502.20435}{{
  arXiv:2502.20435~[cond-mat.str-el]}}.

\bibitem{Chung:2025ulc}
K.~T.~K. Chung, U.~Borla, A.~H. Nevidomskyy, and S.~Moroz, {\slshape
  {Spontaneously Broken Non-Invertible Symmetries in Transverse-Field Ising
  Qudit Chains},} \href{http://arxiv.org/abs/2508.11003}{{
  arXiv:2508.11003~[cond-mat.str-el]}}.

\bibitem{atiyah1969introduction}
M.~Atiyah and I.~Macdonald, {\em Introduction to Commutative Algebra}.
\newblock Addison-Wesley series in mathematics. Addison-Wesley Publishing
  Company, 1969.
\newblock \url{https://books.google.com.tw/books?id=senlelzZPBsC}.

\bibitem{Kong:2024ykr}
L.~Kong, Z.-H. Zhang, J.~Zhao, and H.~Zheng, {\slshape {Higher condensation
  theory},} \href{http://arxiv.org/abs/2403.07813}{{
  arXiv:2403.07813~[cond-mat.str-el]}}.

\bibitem{Benedetti:2024utz}
V.~Benedetti, H.~Casini, and J.~M. Magan, {\slshape {Selection rules for RG
  flows of minimal models},} \href{http://arxiv.org/abs/2412.16587}{{
  arXiv:2412.16587~[hep-th]}}.

\bibitem{Buican:2025zpm}
M.~Buican, R.~Geiko, M.~Moses, and B.~Shi, {\slshape {An Algebraic Theory of
  Gapped Domain Wall Partons},} \href{http://arxiv.org/abs/2506.22544}{{
  arXiv:2506.22544~[cond-mat.str-el]}}.

\bibitem{Antinucci:2025fjp}
A.~Antinucci, C.~Copetti, Y.~Gai, and S.~Schafer-Nameki, {\slshape {Categorical
  Anomaly Matching},} \href{http://arxiv.org/abs/2508.00982}{{
  arXiv:2508.00982~[hep-th]}}.

\bibitem{Stanishkov:2016pvi}
M.~Stanishkov, {\slshape {RG domain wall for the general $ \widehat{su}(2) $
  coset models},} \href{http://dx.doi.org/10.1007/JHEP08(2016)096}{{\em JHEP}
  {\bfseries 08} (2016) 096}, \href{http://arxiv.org/abs/1606.03605}{{
  arXiv:1606.03605~[hep-th]}}.

\bibitem{Stanishkov:2016rgv}
M.~Stanishkov, {\slshape {Second order RG flow in general $
  \widehat{\mathrm{su}}(2) $ coset models},}
  \href{http://dx.doi.org/10.1007/JHEP09(2016)040}{{\em JHEP} {\bfseries 09}
  (2016) 040}, \href{http://arxiv.org/abs/1606.04328}{{
  arXiv:1606.04328~[hep-th]}}.

\bibitem{Poghosyan:2022ecv}
H.~Poghosyan and R.~Poghossian, {\slshape {RG flows between $W_3$ minimal
  models},} \href{http://dx.doi.org/10.22323/1.412.0039}{{\em PoS} {\bfseries
  Regio2021} (2022) 039}.

\bibitem{Poghosyan:2023brb}
A.~Poghosyan and H.~Poghosyan, {\slshape {A note on RG domain wall between
  successive $ {A}_2^{(p)} $ minimal models},}
  \href{http://dx.doi.org/10.1007/JHEP08(2023)072}{{\em JHEP} {\bfseries 08}
  (2023) 072}, \href{http://arxiv.org/abs/2305.05997}{{
  arXiv:2305.05997~[hep-th]}}.

\bibitem{Cogburn:2023xzw}
C.~V. Cogburn, A.~L. Fitzpatrick, and H.~Geng, {\slshape {CFT and lattice
  correlators near an RG domain wall between minimal models},}
  \href{http://dx.doi.org/10.21468/SciPostPhysCore.7.2.021}{{\em SciPost Phys.
  Core} {\bfseries 7} (2024) 021}, \href{http://arxiv.org/abs/2308.00737}{{
  arXiv:2308.00737~[hep-th]}}.

\bibitem{Note9}
The literature on conformal interface provides the corresponding analytical
  data, but they often require extensive analytical calculations.

\bibitem{Gepner:1990gr}
D.~Gepner, {\slshape {Fusion rings and geometry},}
  \href{http://dx.doi.org/10.1007/BF02101511}{{\em Commun. Math. Phys.}
  {\bfseries 141} (1991) 381--411}.

\bibitem{Note10}
However, it should be kept in mind that this pioneering work appeared much
  earlier, before the appearance of the fusion category symmetry.

\bibitem{Kaidi:2021gbs}
J.~Kaidi, Z.~Komargodski, K.~Ohmori, S.~Seifnashri, and S.-H. Shao, {\slshape
  {Higher central charges and topological boundaries in 2+1-dimensional
  TQFTs},} \href{http://arxiv.org/abs/2107.13091}{{
  arXiv:2107.13091~[hep-th]}}.

\bibitem{Gukov:2015qea}
S.~Gukov, {\slshape {Counting RG flows},}
  \href{http://dx.doi.org/10.1007/JHEP01(2016)020}{{\em JHEP} {\bfseries 01}
  (2016) 020}, \href{http://arxiv.org/abs/1503.01474}{{
  arXiv:1503.01474~[hep-th]}}.

\bibitem{Konechny:2023xvo}
A.~Konechny, {\slshape {RG boundaries and Cardy\textquoteright{}s variational
  ansatz for multiple perturbations},}
  \href{http://dx.doi.org/10.1007/JHEP11(2023)004}{{\em JHEP} {\bfseries 11}
  (2023) 004}, \href{http://arxiv.org/abs/2306.13719}{{
  arXiv:2306.13719~[hep-th]}}.

\bibitem{PhysRevB.98.205136}
Y.~Li, X.~Chen, and M.~P.~A. Fisher, {\slshape Quantum zeno effect and the
  many-body entanglement transition,}
  \href{https://link.aps.org/doi/10.1103/PhysRevB.98.205136}{{\em Phys. Rev. B}
  {\bfseries 98} (Nov, 2018) 205136}.

\bibitem{PhysRevB.100.134306}
Y.~Li, X.~Chen, and M.~P.~A. Fisher, {\slshape Measurement-driven entanglement
  transition in hybrid quantum circuits,}
  \href{https://link.aps.org/doi/10.1103/PhysRevB.100.134306}{{\em Phys. Rev.
  B} {\bfseries 100} (Oct, 2019) 134306}.

\bibitem{Skinner:2018tjl}
B.~Skinner, J.~Ruhman, and A.~Nahum, {\slshape {Measurement-Induced Phase
  Transitions in the Dynamics of Entanglement},}
  \href{http://dx.doi.org/10.1103/PhysRevX.9.031009}{{\em Phys. Rev. X}
  {\bfseries 9} (2019) 031009}, \href{http://arxiv.org/abs/1808.05953}{{
  arXiv:1808.05953~[cond-mat.stat-mech]}}.

\bibitem{Hung:2015hfa}
L.-Y. Hung and Y.~Wan, {\slshape {Generalized ADE classification of topological
  boundaries and anyon condensation},}
  \href{http://dx.doi.org/10.1007/JHEP07(2015)120}{{\em JHEP} {\bfseries 07}
  (2015) 120}, \href{http://arxiv.org/abs/1502.02026}{{
  arXiv:1502.02026~[cond-mat.str-el]}}.

\bibitem{Note11}
A fiber functor should be distinguished from a tensor functor corresponding to
  a massless RG.

\bibitem{Frohlich:2009gb}
J.~Frohlich, J.~Fuchs, I.~Runkel, and C.~Schweigert,
  \href{http://dx.doi.org/10.1142/9789814304634_0056}{{\slshape {Defect Lines,
  Dualities and Generalised Orbifolds},}} in {\em {16th International Congress
  on Mathematical Physics}}, pp.~608--613.
\newblock 2010.
\newblock \href{http://arxiv.org/abs/0909.5013}{{ arXiv:0909.5013~[math-ph]}}.

\bibitem{Lu:2022ver}
D.-C. Lu and Z.~Sun, {\slshape {On triality defects in 2d CFT},}
  \href{http://dx.doi.org/10.1007/JHEP02(2023)173}{{\em JHEP} {\bfseries 02}
  (2023) 173}, \href{http://arxiv.org/abs/2208.06077}{{
  arXiv:2208.06077~[hep-th]}}.

\bibitem{Perez-Lona:2023djo}
A.~Perez-Lona, D.~Robbins, E.~Sharpe, T.~Vandermeulen, and X.~Yu, {\slshape
  {Notes on gauging noninvertible symmetries. Part I. Multiplicity-free
  cases},} \href{http://dx.doi.org/10.1007/JHEP02(2024)154}{{\em JHEP}
  {\bfseries 02} (2024) 154}, \href{http://arxiv.org/abs/2311.16230}{{
  arXiv:2311.16230~[hep-th]}}.

\bibitem{Choi:2023vgk}
Y.~Choi, D.-C. Lu, and Z.~Sun, {\slshape {Self-duality under gauging a
  non-invertible symmetry},}
  \href{http://dx.doi.org/10.1007/JHEP01(2024)142}{{\em JHEP} {\bfseries 01}
  (2024) 142}, \href{http://arxiv.org/abs/2310.19867}{{
  arXiv:2310.19867~[hep-th]}}.

\bibitem{Diatlyk:2023fwf}
O.~Diatlyk, C.~Luo, Y.~Wang, and Q.~Weller, {\slshape {Gauging non-invertible
  symmetries: topological interfaces and generalized orbifold groupoid in 2d
  QFT},} \href{http://dx.doi.org/10.1007/JHEP03(2024)127}{{\em JHEP} {\bfseries
  03} (2024) 127}, \href{http://arxiv.org/abs/2311.17044}{{
  arXiv:2311.17044~[hep-th]}}.

\bibitem{Perez-Lona:2024sds}
A.~Perez-Lona, D.~Robbins, E.~Sharpe, T.~Vandermeulen, and X.~Yu, {\slshape
  {Notes on gauging noninvertible symmetries. Part II. Higher multiplicity
  cases},} \href{http://dx.doi.org/10.1007/JHEP05(2025)066}{{\em JHEP}
  {\bfseries 05} (2025) 066}, \href{http://arxiv.org/abs/2408.16811}{{
  arXiv:2408.16811~[hep-th]}}.

\bibitem{Kaplan:1983fs}
D.~B. Kaplan and H.~Georgi, {\slshape {SU(2) x U(1) Breaking by Vacuum
  Misalignment},} \href{http://dx.doi.org/10.1016/0370-2693(84)91177-8}{{\em
  Phys. Lett. B} {\bfseries 136} (1984) 183--186}.

\bibitem{Kaplan:1983sm}
D.~B. Kaplan, H.~Georgi, and S.~Dimopoulos, {\slshape {Composite Higgs
  Scalars},} \href{http://dx.doi.org/10.1016/0370-2693(84)91178-X}{{\em Phys.
  Lett. B} {\bfseries 136} (1984) 187--190}.

\bibitem{Georgi:1984af}
H.~Georgi and D.~B. Kaplan, {\slshape {Composite Higgs and Custodial SU(2)},}
  \href{http://dx.doi.org/10.1016/0370-2693(84)90341-1}{{\em Phys. Lett. B}
  {\bfseries 145} (1984) 216--220}.

\bibitem{Dugan:1984hq}
M.~J. Dugan, H.~Georgi, and D.~B. Kaplan, {\slshape {Anatomy of a Composite
  Higgs Model},} \href{http://dx.doi.org/10.1016/0550-3213(85)90221-4}{{\em
  Nucl. Phys. B} {\bfseries 254} (1985) 299--326}.

\bibitem{Panico:2015jxa}
G.~Panico and A.~Wulzer,
  \href{http://dx.doi.org/10.1007/978-3-319-22617-0}{{\em {The Composite
  Nambu-Goldstone Higgs}}}, vol.~913.
\newblock Springer, 2016.
\newblock \href{http://arxiv.org/abs/1506.01961}{{ arXiv:1506.01961~[hep-ph]}}.

\bibitem{Watanabe:2019xul}
H.~Watanabe, {\slshape {Counting Rules of Nambu{\textendash}Goldstone Modes},}
  \href{http://dx.doi.org/10.1146/annurev-conmatphys-031119-050644}{{\em Ann.
  Rev. Condensed Matter Phys.} {\bfseries 11} (2020) 169--187},
  \href{http://arxiv.org/abs/1904.00569}{{ arXiv:1904.00569~[cond-mat.other]}}.

\bibitem{Brauner:2024juy}
T.~Brauner, {\slshape {Effective Field Theory for Spontaneously Broken
  Symmetry},} \href{http://dx.doi.org/10.1007/978-3-031-48378-3}{{\em Lect.
  Notes Phys.} {\bfseries 1023} (2024) pp.},
  \href{http://arxiv.org/abs/2404.14518}{{ arXiv:2404.14518~[hep-th]}}.

\bibitem{Fendley:1999gb}
P.~Fendley, {\slshape {Sigma models as perturbed conformal field theories},}
  \href{http://dx.doi.org/10.1103/PhysRevLett.83.4468}{{\em Phys. Rev. Lett.}
  {\bfseries 83} (1999) 4468--4471},
  \href{http://arxiv.org/abs/hep-th/9906036}{{ arXiv:hep-th/9906036}}.

\bibitem{Zhao:2025zsb}
Y.~Zhao and Y.~Wan, {\slshape {Landau-Ginzburg Paradigm of Topological
  Phases},} \href{http://arxiv.org/abs/2506.05319}{{
  arXiv:2506.05319~[cond-mat.str-el]}}.

\bibitem{Apruzzi:2025hvs}
F.~Apruzzi, N.~Dondi, I.~Garc{\'\i}a~Etxebarria, H.~T. Lam, and
  S.~Schafer-Nameki, {\slshape {Symmetry TFTs for Continuous Spacetime
  Symmetries},} \href{http://arxiv.org/abs/2509.07965}{{
  arXiv:2509.07965~[hep-th]}}.

\bibitem{Wess:1971yu}
J.~Wess and B.~Zumino, {\slshape {Consequences of anomalous Ward identities},}
  \href{http://dx.doi.org/10.1016/0370-2693(71)90582-X}{{\em Phys. Lett. B}
  {\bfseries 37} (1971) 95--97}.

\bibitem{Witten:1983ar}
E.~Witten, {\slshape {Nonabelian Bosonization in Two-Dimensions},}
  \href{http://dx.doi.org/10.1007/BF01215276}{{\em Commun. Math. Phys.}
  {\bfseries 92} (1984) 455--472}.

\bibitem{Witten:1983tw}
E.~Witten, {\slshape {Global Aspects of Current Algebra},}
  \href{http://dx.doi.org/10.1016/0550-3213(83)90063-9}{{\em Nucl. Phys. B}
  {\bfseries 223} (1983) 422--432}.

\bibitem{Goddard:1984vk}
P.~Goddard, A.~Kent, and D.~I. Olive, {\slshape {Virasoro Algebras and Coset
  Space Models},} \href{http://dx.doi.org/10.1016/0370-2693(85)91145-1}{{\em
  Phys. Lett. B} {\bfseries 152} (1985) 88--92}.

\bibitem{Goddard:1984hg}
P.~Goddard and D.~I. Olive, {\slshape {Kac-Moody Algebras, Conformal Symmetry
  and Critical Exponents},}
  \href{http://dx.doi.org/10.1016/0550-3213(85)90344-X}{{\em Nucl. Phys. B}
  {\bfseries 257} (1985) 226--252}.

\bibitem{Goddard:1986ee}
P.~Goddard, A.~Kent, and D.~I. Olive, {\slshape {Unitary Representations of the
  Virasoro and Supervirasoro Algebras},}
  \href{http://dx.doi.org/10.1007/BF01464283}{{\em Commun. Math. Phys.}
  {\bfseries 103} (1986) 105--119}.

\bibitem{Borsato:2023dis}
R.~Borsato, {\slshape {Lecture notes on current\textendash{}current
  deformations},} \href{http://dx.doi.org/10.1140/epjc/s10052-024-12966-5}{{\em
  Eur. Phys. J. C} {\bfseries 84} (2024) 648},
  \href{http://arxiv.org/abs/2312.13847}{{ arXiv:2312.13847~[hep-th]}}.

\bibitem{Cordova:2023jip}
C.~Cordova and D.~Garc\'\i{}a-Sep\'ulveda, {\slshape {Non-Invertible Anyon
  Condensation and Level-Rank Dualities},}
  \href{http://arxiv.org/abs/2312.16317}{{ arXiv:2312.16317~[hep-th]}}.

\bibitem{Frohlich:2003hm}
J.~Frohlich, J.~Fuchs, I.~Runkel, and C.~Schweigert, {\slshape {Correspondences
  of ribbon categories},}
  \href{http://dx.doi.org/10.1016/j.aim.2005.04.007}{{\em Adv. Math.}
  {\bfseries 199} (2006) 192--329}, \href{http://arxiv.org/abs/math/0309465}{{
  arXiv:math/0309465}}.

\bibitem{Bais:2008ni}
F.~A. Bais and J.~K. Slingerland, {\slshape {Condensate induced transitions
  between topologically ordered phases},}
  \href{http://dx.doi.org/10.1103/PhysRevB.79.045316}{{\em Phys. Rev. B}
  {\bfseries 79} (2009) 045316}, \href{http://arxiv.org/abs/0808.0627}{{
  arXiv:0808.0627~[cond-mat.mes-hall]}}.

\bibitem{Berkovich:1997ht}
A.~Berkovich, B.~M. McCoy, A.~Schilling, and S.~O. Warnaar, {\slshape {Bailey
  flows and Bose-Fermi identities for the conformal coset models (A(1)(1))N x
  (A(1)(1))N-prime / (A(1)(1))N+N-prime},}
  \href{http://dx.doi.org/10.1016/S0550-3213(97)82955-0}{{\em Nucl. Phys. B}
  {\bfseries 499} (1997) 621--649},
  \href{http://arxiv.org/abs/hep-th/9702026}{{ arXiv:hep-th/9702026}}.

\bibitem{Dorey:2000zb}
P.~Dorey, C.~Dunning, and R.~Tateo, {\slshape {New families of flows between
  two-dimensional conformal field theories},}
  \href{http://dx.doi.org/10.1016/S0550-3213(00)00185-1}{{\em Nucl. Phys. B}
  {\bfseries 578} (2000) 699--727},
  \href{http://arxiv.org/abs/hep-th/0001185}{{ arXiv:hep-th/0001185}}.

\bibitem{Dunning:2002cu}
C.~Dunning, {\slshape {Massless flows between minimal W models},}
  \href{http://dx.doi.org/10.1016/S0370-2693(02)01938-X}{{\em Phys. Lett. B}
  {\bfseries 537} (2002) 297--305},
  \href{http://arxiv.org/abs/hep-th/0204090}{{ arXiv:hep-th/0204090}}.

\bibitem{Blondeau-Fournier:2017otv}
O.~Blondeau-Fournier, P.~Mathieu, and T.~A. Welsh, {\slshape {A quartet of
  fermionic expressions for $M(k,2k\pm1)$ Virasoro characters via half-lattice
  paths},} \href{http://dx.doi.org/10.1016/j.nuclphysb.2017.09.023}{{\em Nucl.
  Phys. B} {\bfseries 924} (2017) 643--683},
  \href{http://arxiv.org/abs/1705.06775}{{ arXiv:1705.06775~[math-ph]}}.

\bibitem{PhysRevLett.60.956}
F.~D.~M. Haldane and E.~H. Rezayi, {\slshape Spin-singlet wave function for the
  half-integral quantum hall effect,}
  \href{https://link.aps.org/doi/10.1103/PhysRevLett.60.956}{{\em Phys. Rev.
  Lett.} {\bfseries 60} (Mar, 1988) 956--959}.

\bibitem{PhysRevB.75.075317}
S.~H. Simon, E.~H. Rezayi, N.~R. Cooper, and I.~Berdnikov, {\slshape
  Construction of a paired wave function for spinless electrons at filling
  fraction $\nu =2/5$,}
  \href{https://link.aps.org/doi/10.1103/PhysRevB.75.075317}{{\em Phys. Rev. B}
  {\bfseries 75} (Feb, 2007) 075317}.

\bibitem{Davenport:2012fcs}
S.~C. Davenport, E.~Ardonne, N.~Regnault, and S.~H. Simon, {\slshape
  {Spin-singlet Gaffnian wave function for fractional quantum Hall systems},}
  \href{http://dx.doi.org/10.1103/PhysRevB.87.045310}{{\em Phys. Rev. B}
  {\bfseries 87} (2013) 045310}, \href{http://arxiv.org/abs/1210.8143}{{
  arXiv:1210.8143~[cond-mat.str-el]}}.

\bibitem{Bernevig_2008}
B.~A. Bernevig and F.~D.~M. Haldane, {\slshape Model fractional quantum hall
  states and jack polynomials,}
  \href{https://doi.org/10.1103%2Fphysrevlett.100.246802}{{\em Physical Review
  Letters} {\bfseries 100} (Jun, 2008) }.

\bibitem{Yuzhu_2023}
W.~Yuzhu and Y.~Bo, {\slshape Geometric fluctuation of conformal hilbert spaces
  and multiple graviton modes in fractional quantum hall effect,}
  \href{https://doi.org/10.1038%2Fs41467-023-38036-0}{{\em Nature
  Communications} {\bfseries 14} (Apr, 2023) }.

\bibitem{Bourgine:2024ycr}
J.-E. Bourgine and Y.~Matsuo, {\slshape {Calogero model for the non-Abelian
  quantum Hall effect},}
  \href{http://dx.doi.org/10.1103/PhysRevB.109.155158}{{\em Phys. Rev. B}
  {\bfseries 109} (2024) 155158}, \href{http://arxiv.org/abs/2401.03087}{{
  arXiv:2401.03087~[hep-th]}}.

\bibitem{Yurov:1989yu}
V.~P. Yurov and A.~B. Zamolodchikov, {\slshape {TRUNCATED CONFORMAL SPACE
  APPROACH TO SCALING LEE-YANG MODEL},}
  \href{http://dx.doi.org/10.1142/S0217751X9000218X}{{\em Int. J. Mod. Phys. A}
  {\bfseries 5} (1990) 3221--3246}.

\bibitem{Yurov:1991my}
V.~P. Yurov and A.~B. Zamolodchikov, {\slshape {Truncated fermionic space
  approach to the critical 2-D Ising model with magnetic field},}
  \href{http://dx.doi.org/10.1142/S0217751X91002161}{{\em Int. J. Mod. Phys. A}
  {\bfseries 6} (1991) 4557--4578}.

\bibitem{Quella:2006de}
T.~Quella, I.~Runkel, and G.~M.~T. Watts, {\slshape {Reflection and
  transmission for conformal defects},}
  \href{http://dx.doi.org/10.1088/1126-6708/2007/04/095}{{\em JHEP} {\bfseries
  04} (2007) 095}, \href{http://arxiv.org/abs/hep-th/0611296}{{
  arXiv:hep-th/0611296}}.

\bibitem{Note12}
The term, accidental, is different from that used to mean the small (or
  irrelevant) breaking of symmetry.

\bibitem{Lencses:2023evr}
M.~Lencs{\'e}s, A.~Miscioscia, G.~Mussardo, and G.~Tak{\'a}cs, {\slshape {$
  \mathcal{PT} $ breaking and RG flows between multicritical Yang-Lee fixed
  points},} \href{http://dx.doi.org/10.1007/JHEP09(2023)052}{{\em JHEP}
  {\bfseries 09} (2023) 052}, \href{http://arxiv.org/abs/2304.08522}{{
  arXiv:2304.08522~[cond-mat.stat-mech]}}.

\bibitem{Lencses:2024wib}
M.~Lencs{\'e}s, A.~Miscioscia, G.~Mussardo, and G.~Tak{\'a}cs, {\slshape
  {Ginzburg-Landau description for multicritical Yang-Lee models},}
  \href{http://dx.doi.org/10.1007/JHEP08(2024)224}{{\em JHEP} {\bfseries 08}
  (2024) 224}, \href{http://arxiv.org/abs/2404.06100}{{
  arXiv:2404.06100~[cond-mat.stat-mech]}}.

\bibitem{Katsevich:2025ojk}
A.~Katsevich, I.~R. Klebanov, Z.~Sun, and G.~Tarnopolsky, {\slshape {Towards a
  Quintic Ginzburg-Landau Description of the $(2,7)$ Minimal Model},}
  \href{http://arxiv.org/abs/2510.19085}{{ arXiv:2510.19085~[hep-th]}}.

\bibitem{Note13}
We thank Andrei Katsevich for notifying us the subtelties of $M(2,7)\rightarrow
  M(2,5)$ in the literature.

\bibitem{Smirnov:1990vm}
F.~A. Smirnov, {\slshape {Reductions of the sine-Gordon model as a perturbation
  of minimal models of conformal field theory},}
  \href{http://dx.doi.org/10.1016/0550-3213(90)90255-C}{{\em Nucl. Phys. B}
  {\bfseries 337} (1990) 156--180}.

\bibitem{tHooft:1979rat}
G.~'t~Hooft, {\slshape {Naturalness, chiral symmetry, and spontaneous chiral
  symmetry breaking},}
  \href{http://dx.doi.org/10.1007/978-1-4684-7571-5_9}{{\em NATO Sci. Ser. B}
  {\bfseries 59} (1980) 135--157}.

\bibitem{Schultz:1964fv}
T.~D. Schultz, D.~C. Mattis, and E.~H. Lieb, {\slshape {Two-Dimensional Ising
  Model as a Soluble Problem of Many Fermions},}
  \href{http://dx.doi.org/10.1103/RevModPhys.36.856}{{\em Rev. Mod. Phys.}
  {\bfseries 36} (1964) 856--871}.

\bibitem{Cho:2017fgz}
G.~Y. Cho, S.~Ryu, and C.-T. Hsieh, {\slshape {Anomaly Manifestation of
  Lieb-Schultz-Mattis Theorem and Topological Phases},}
  \href{http://dx.doi.org/10.1103/PhysRevB.96.195105}{{\em Phys. Rev. B}
  {\bfseries 96} (2017) 195105}, \href{http://arxiv.org/abs/1705.03892}{{
  arXiv:1705.03892~[cond-mat.str-el]}}.

\bibitem{Lieb:1961fr}
E.~H. Lieb, T.~Schultz, and D.~Mattis, {\slshape {Two soluble models of an
  antiferromagnetic chain},}
  \href{http://dx.doi.org/10.1016/0003-4916(61)90115-4}{{\em Annals Phys.}
  {\bfseries 16} (1961) 407--466}.

\bibitem{Haldane:1981zza}
F.~D.~M. Haldane, {\slshape {Luttinger liquid theory of one-dimensional quantum
  fluids. I. Properties of the Luttinger model and their extension to the
  general 1D interacting spinless Fermi gas},}
  \href{http://dx.doi.org/10.1088/0022-3719/14/19/010}{{\em J. Phys. C}
  {\bfseries 14} (1981) 2585--2609}.

\bibitem{Haldane:1982rj}
F.~D.~M. Haldane, {\slshape {Continuum dynamics of the 1-D Heisenberg
  antiferromagnetic identification with the O(3) nonlinear sigma model},}
  \href{http://dx.doi.org/10.1016/0375-9601(83)90631-X}{{\em Phys. Lett. A}
  {\bfseries 93} (1983) 464--468}.

\bibitem{Haldane:1983ru}
F.~D.~M. Haldane, {\slshape {Nonlinear field theory of large spin Heisenberg
  antiferromagnets. Semiclassically quantized solitons of the one-dimensional
  easy Axis Neel state},}
  \href{http://dx.doi.org/10.1103/PhysRevLett.50.1153}{{\em Phys. Rev. Lett.}
  {\bfseries 50} (1983) 1153--1156}.

\bibitem{haldane2016groundstatepropertiesantiferromagnetic}
F.~D.~M. Haldane, {\slshape Ground state properties of antiferromagnetic chains
  with unrestricted spin: Integer spin chains as realisations of the o(3)
  non-linear sigma model,} 2016.
\newblock \url{https://arxiv.org/abs/1612.00076}.

\bibitem{Wamer:2019oge}
K.~Wamer, M.~Lajk\'o, F.~Mila, and I.~Affleck, {\slshape {Generalization of the
  Haldane conjecture to SU($n$) chains},}
  \href{http://dx.doi.org/10.1016/j.nuclphysb.2020.114932}{{\em Nucl. Phys. B}
  {\bfseries 952} (2020) 114932}, \href{http://arxiv.org/abs/1910.08196}{{
  arXiv:1910.08196~[cond-mat.str-el]}}.

\bibitem{moore_nonabelions_1991}
G.~Moore and N.~Read, {\slshape Nonabelions in the fractional quantum hall
  effect,}
  \href{http://www.sciencedirect.com/science/article/pii/055032139190407O}{{\em
  Nuclear Physics B} {\bfseries 360} (Aug., 1991) 362--396}.

\bibitem{Cappelli:1996np}
A.~Cappelli and G.~R. Zemba, {\slshape {Modular invariant partition functions
  in the quantum Hall effect},}
  \href{http://dx.doi.org/10.1016/S0550-3213(97)00110-7}{{\em Nucl. Phys. B}
  {\bfseries 490} (1997) 595--632},
  \href{http://arxiv.org/abs/hep-th/9605127}{{ arXiv:hep-th/9605127}}.

\bibitem{Frohlich:2000qs}
J.~Frohlich, B.~Pedrini, C.~Schweigert, and J.~Walcher, {\slshape {Universality
  in quantum Hall systems: Coset construction of incompressible states},}
  \href{http://dx.doi.org/10.1023/A:1010389232079}{{\em J. Stat. Phys.}
  {\bfseries 103} (2001) 527--567},
  \href{http://arxiv.org/abs/cond-mat/0002330}{{ arXiv:cond-mat/0002330}}.

\bibitem{Schoutens:2015uia}
K.~Schoutens and X.-G. Wen, {\slshape {Simple-current algebra constructions of
  2+1-dimensional topological orders},}
  \href{http://dx.doi.org/10.1103/PhysRevB.93.045109}{{\em Phys. Rev. B}
  {\bfseries 93} (2016) 045109}, \href{http://arxiv.org/abs/1508.01111}{{
  arXiv:1508.01111~[cond-mat.str-el]}}.

\bibitem{Fukusumi:2022xxe}
Y.~Fukusumi and B.~Yang, {\slshape {Fermionic fractional quantum Hall states: A
  modern approach to systems with bulk-edge correspondence},}
  \href{http://dx.doi.org/10.1103/PhysRevB.108.085123}{{\em Phys. Rev. B}
  {\bfseries 108} (2023) 085123}, \href{http://arxiv.org/abs/2212.12993}{{
  arXiv:2212.12993~[cond-mat.str-el]}}.

\bibitem{Vafa:1986wx}
C.~Vafa, {\slshape {Modular Invariance and Discrete Torsion on Orbifolds},}
  \href{http://dx.doi.org/10.1016/0550-3213(86)90379-2}{{\em Nucl. Phys. B}
  {\bfseries 273} (1986) 592--606}.

\bibitem{Dixon:1986jc}
L.~J. Dixon, J.~A. Harvey, C.~Vafa, and E.~Witten, {\slshape {Strings on
  Orbifolds. 2.},} \href{http://dx.doi.org/10.1016/0550-3213(86)90287-7}{{\em
  Nucl. Phys. B} {\bfseries 274} (1986) 285--314}.

\bibitem{Hamidi:1986vh}
S.~Hamidi and C.~Vafa, {\slshape {Interactions on Orbifolds},}
  \href{http://dx.doi.org/10.1016/0550-3213(87)90006-X}{{\em Nucl. Phys. B}
  {\bfseries 279} (1987) 465--513}.

\bibitem{Dunbar:1993hr}
D.~C. Dunbar and K.~G. Joshi, {\slshape {Maverick examples of coset conformal
  field theories},} \href{http://dx.doi.org/10.1142/S0217732393003196}{{\em
  Mod. Phys. Lett. A} {\bfseries 8} (1993) 2803--2814},
  \href{http://arxiv.org/abs/hep-th/9309093}{{ arXiv:hep-th/9309093}}.

\bibitem{Gannon:1992np}
T.~Gannon, {\slshape {Partition functions for heterotic WZW conformal field
  theories},} \href{http://dx.doi.org/10.1016/0550-3213(93)90127-B}{{\em Nucl.
  Phys. B} {\bfseries 402} (1993) 729--753},
  \href{http://arxiv.org/abs/hep-th/9209042}{{ arXiv:hep-th/9209042}}.

\bibitem{Gannon:1998rw}
T.~Gannon and M.~A. Walton, {\slshape {Heterotic modular invariants and level
  rank duality},} \href{http://dx.doi.org/10.1016/S0550-3213(98)00385-X}{{\em
  Nucl. Phys. B} {\bfseries 536} (1998) 553--574},
  \href{http://arxiv.org/abs/hep-th/9804040}{{ arXiv:hep-th/9804040}}.

\bibitem{Pedrini:1999iy}
B.~Pedrini, C.~Schweigert, and J.~Walcher, {\slshape {New maverick coset
  theories},} \href{http://dx.doi.org/10.1016/S0370-2693(99)01115-6}{{\em Phys.
  Lett. B} {\bfseries 466} (1999) 206--210},
  \href{http://arxiv.org/abs/hep-th/9908185}{{ arXiv:hep-th/9908185}}.

\bibitem{Note14}
Corresponding to this anomaly-free condition of the non-group like objects,
  there will exist intrinsically nonabelian anyonic TOs, or $\protect \mathbf
  {A}_{\protect \text {ub}}$-symmetry enriched TOs where the fundamental
  ``electron operators'' constructing the wavefunctions are nonabelian anyon
  belonging to $\protect \mathbf {A}_{\protect \text {ub}}\boxtimes \protect
  \mathbf {A'}_{\protect \text {ub}}$ or $\protect \overline {\protect \mathbf
  {A}_{\protect \text {ub}}}\boxtimes \protect \mathbf {A'}_{\protect \text
  {ub}}$.

\bibitem{Johnson-Freyd:2020usu}
T.~Johnson-Freyd, {\slshape {On the Classification of Topological Orders},}
  \href{http://dx.doi.org/10.1007/s00220-022-04380-3}{{\em Commun. Math. Phys.}
  {\bfseries 393} (2022) 989--1033}, \href{http://arxiv.org/abs/2003.06663}{{
  arXiv:2003.06663~[math.CT]}}.

\bibitem{Zeev:2022cnv}
R.~B. Zeev, B.~Ergun, E.~Milan, and S.~S. Razamat, {\slshape {Categorical
  structure of the set of all CFTs},}
  \href{http://dx.doi.org/10.1103/PhysRevD.110.025019}{{\em Phys. Rev. D}
  {\bfseries 110} (2024) 025019}, \href{http://arxiv.org/abs/2212.11022}{{
  arXiv:2212.11022~[hep-th]}}.

\bibitem{Nivesvivat:2025odb}
R.~Nivesvivat and S.~Ribault, {\slshape {Fusion rules and structure constants
  of E-series minimal models},}
  \href{http://dx.doi.org/10.21468/SciPostPhys.18.5.163}{{\em SciPost Phys.}
  {\bfseries 18} (2025) 163}, \href{http://arxiv.org/abs/2502.14295}{{
  arXiv:2502.14295~[hep-th]}}.

\bibitem{Note15}
More precisely, linear magma satisfying the pentagon relation.

\bibitem{liebmann2019nonassociativealgebrasquantumphysics}
M.~Liebmann, H.~Rühaak, and B.~Henschenmacher, {\slshape Non-associative
  algebras and quantum physics -- a historical perspective,} 2019.
\newblock \url{https://arxiv.org/abs/1909.04027}.

\bibitem{Lauchli:2013jga}
A.~M. L{\"a}uchli, {\slshape {Operator content of real-space entanglement
  spectra at conformal critical points},}
  \href{http://arxiv.org/abs/1303.0741}{{
  arXiv:1303.0741~[cond-mat.stat-mech]}}.

\bibitem{Ohmori:2014eia}
K.~Ohmori and Y.~Tachikawa, {\slshape {Physics at the entangling surface},}
  \href{http://dx.doi.org/10.1088/1742-5468/2015/04/P04010}{{\em J. Stat.
  Mech.} {\bfseries 1504} (2015) P04010},
  \href{http://arxiv.org/abs/1406.4167}{{ arXiv:1406.4167~[hep-th]}}.

\bibitem{Fossati:2023zyz}
M.~Fossati, F.~Ares, and P.~Calabrese, {\slshape {Symmetry-resolved
  entanglement in critical non-Hermitian systems},}
  \href{http://dx.doi.org/10.1103/PhysRevB.107.205153}{{\em Phys. Rev. B}
  {\bfseries 107} (2023) 205153}, \href{http://arxiv.org/abs/2303.05232}{{
  arXiv:2303.05232~[cond-mat.stat-mech]}}.

\bibitem{Rottoli:2024tvr}
F.~Rottoli, M.~Fossati, and P.~Calabrese, {\slshape {Entanglement Hamiltonian
  in the non-Hermitian SSH model},}
  \href{http://dx.doi.org/10.1088/1742-5468/ad4860}{{\em J. Stat. Mech.}
  {\bfseries 2024} (2024) 063102}, \href{http://arxiv.org/abs/2402.04776}{{
  arXiv:2402.04776~[quant-ph]}}.

\bibitem{Shimizu:2025kse}
H.~Shimizu and K.~Kawabata, {\slshape {Complex entanglement entropy for complex
  conformal field theory},} \href{http://dx.doi.org/10.1103/n578-ljd5}{{\em
  Phys. Rev. B} {\bfseries 112} (2025) 085112},
  \href{http://arxiv.org/abs/2502.02001}{{
  arXiv:2502.02001~[cond-mat.stat-mech]}}.

\bibitem{Chou:2025awd}
K.-H. Chou, X.-J. Yu, and P.-Y. Chang, {\slshape {PT symmetry-enriched
  non-unitary criticality},} \href{http://arxiv.org/abs/2509.09587}{{
  arXiv:2509.09587~[quant-ph]}}.

\bibitem{Bianchini:2014uta}
D.~Bianchini, O.~A. Castro-Alvaredo, B.~Doyon, E.~Levi, and F.~Ravanini,
  {\slshape {Entanglement Entropy of Non Unitary Conformal Field Theory},}
  \href{http://dx.doi.org/10.1088/1751-8113/48/4/04FT01}{{\em J. Phys. A}
  {\bfseries 48} (2015) 04FT01}, \href{http://arxiv.org/abs/1405.2804}{{
  arXiv:1405.2804~[hep-th]}}.

\bibitem{Couvreur:2016mbr}
R.~Couvreur, J.~L. Jacobsen, and H.~Saleur, {\slshape {Entanglement in
  nonunitary quantum critical spin chains},}
  \href{http://dx.doi.org/10.1103/PhysRevLett.119.040601}{{\em Phys. Rev.
  Lett.} {\bfseries 119} (2017) 040601},
  \href{http://arxiv.org/abs/1611.08506}{{
  arXiv:1611.08506~[cond-mat.stat-mech]}}.

\bibitem{Knizhnik:1987xp}
V.~G. Knizhnik, {\slshape {Analytic Fields on Riemann Surfaces. 2},}
  \href{http://dx.doi.org/10.1007/BF01225373}{{\em Commun. Math. Phys.}
  {\bfseries 112} (1987) 567--590}.

\bibitem{Calabrese:2004eu}
P.~Calabrese and J.~L. Cardy, {\slshape {Entanglement entropy and quantum field
  theory},} \href{http://dx.doi.org/10.1088/1742-5468/2004/06/P06002}{{\em J.
  Stat. Mech.} {\bfseries 0406} (2004) P06002},
  \href{http://arxiv.org/abs/hep-th/0405152}{{ arXiv:hep-th/0405152}}.

\bibitem{Calabrese:2009qy}
P.~Calabrese and J.~Cardy, {\slshape {Entanglement entropy and conformal field
  theory},} \href{http://dx.doi.org/10.1088/1751-8113/42/50/504005}{{\em J.
  Phys. A} {\bfseries 42} (2009) 504005},
  \href{http://arxiv.org/abs/0905.4013}{{
  arXiv:0905.4013~[cond-mat.stat-mech]}}.

\bibitem{Lu:2025myv}
H.-H. Lu and P.-Y. Chang, {\slshape {Biorthogonal quench dynamics of
  entanglement and quantum geometry in PT-symmetric non-Hermitian systems},}
  other thesis, 7, 2025.

\bibitem{Tu:2021xje}
Y.-T. Tu, Y.-C. Tzeng, and P.-Y. Chang, {\slshape {R{\'e}nyi entropies and
  negative central charges in non-Hermitian quantum systems},}
  \href{http://dx.doi.org/10.21468/SciPostPhys.12.6.194}{{\em SciPost Phys.}
  {\bfseries 12} (2022) 194}, \href{http://arxiv.org/abs/2107.13006}{{
  arXiv:2107.13006~[cond-mat.str-el]}}.

\bibitem{Yang:2024ebm}
P.-Y. Yang and Y.-C. Tzeng, {\slshape {Entanglement Hamiltonian and effective
  temperature of non-Hermitian quantum spin ladders},}
  \href{http://dx.doi.org/10.21468/SciPostPhysCore.7.4.074}{{\em SciPost Phys.
  Core} {\bfseries 7} (2024) 074}, \href{http://arxiv.org/abs/2409.17062}{{
  arXiv:2409.17062~[quant-ph]}}.

\bibitem{Goldstein:2017bua}
M.~Goldstein and E.~Sela, {\slshape {Symmetry-resolved entanglement in
  many-body systems},}
  \href{http://dx.doi.org/10.1103/PhysRevLett.120.200602}{{\em Phys. Rev.
  Lett.} {\bfseries 120} (2018) 200602},
  \href{http://arxiv.org/abs/1711.09418}{{
  arXiv:1711.09418~[cond-mat.stat-mech]}}.

\bibitem{Xavier:2018kqb}
J.~C. Xavier, F.~C. Alcaraz, and G.~Sierra, {\slshape {Equipartition of the
  entanglement entropy},}
  \href{http://dx.doi.org/10.1103/PhysRevB.98.041106}{{\em Phys. Rev. B}
  {\bfseries 98} (2018) 041106}, \href{http://arxiv.org/abs/1804.06357}{{
  arXiv:1804.06357~[cond-mat.stat-mech]}}.

\bibitem{Kusuki:2023bsp}
Y.~Kusuki, S.~Murciano, H.~Ooguri, and S.~Pal, {\slshape {Symmetry-resolved
  entanglement entropy, spectra {\&} boundary conformal field theory},}
  \href{http://dx.doi.org/10.1007/JHEP11(2023)216}{{\em JHEP} {\bfseries 11}
  (2023) 216}, \href{http://arxiv.org/abs/2309.03287}{{
  arXiv:2309.03287~[hep-th]}}.

\bibitem{Saura-Bastida:2024yye}
P.~Saura-Bastida, A.~Das, G.~Sierra, and J.~Molina-Vilaplana, {\slshape
  {Categorical-symmetry resolved entanglement in conformal field theory},}
  \href{http://dx.doi.org/10.1103/PhysRevD.109.105026}{{\em Phys. Rev. D}
  {\bfseries 109} (2024) 105026}, \href{http://arxiv.org/abs/2402.06322}{{
  arXiv:2402.06322~[hep-th]}}.

\bibitem{Das:2024qdx}
A.~Das, J.~Molina-Vilaplana, and P.~Saura-Bastida, {\slshape {Generalized
  symmetry resolution of entanglement in conformal field theory for twisted and
  anyonic sectors},} \href{http://dx.doi.org/10.1103/PhysRevD.110.125005}{{\em
  Phys. Rev. D} {\bfseries 110} (2024) 125005},
  \href{http://arxiv.org/abs/2409.02162}{{ arXiv:2409.02162~[hep-th]}}.

\bibitem{Choi:2024tri}
Y.~Choi, B.~C. Rayhaun, and Y.~Zheng, {\slshape {Generalized Tube Algebras,
  Symmetry-Resolved Partition Functions, and Twisted Boundary States},}
  \href{http://arxiv.org/abs/2409.02159}{{ arXiv:2409.02159~[hep-th]}}.

\bibitem{Heymann:2024vvf}
J.~Heymann and T.~Quella, {\slshape {Revisiting the symmetry-resolved
  entanglement for noninvertible symmetries in 1+1d conformal field theories},}
  \href{http://dx.doi.org/10.1103/lr47-yv3j}{{\em Phys. Rev. D} {\bfseries 112}
  (2025) 025004}, \href{http://arxiv.org/abs/2409.02315}{{
  arXiv:2409.02315~[hep-th]}}.

\bibitem{Choi:2024wfm}
Y.~Choi, B.~C. Rayhaun, and Y.~Zheng, {\slshape {Noninvertible
  Symmetry-Resolved Affleck-Ludwig-Cardy Formula and Entanglement Entropy from
  the Boundary Tube Algebra},}
  \href{http://dx.doi.org/10.1103/PhysRevLett.133.251602}{{\em Phys. Rev.
  Lett.} {\bfseries 133} (2024) 251602},
  \href{http://arxiv.org/abs/2409.02806}{{ arXiv:2409.02806~[hep-th]}}.

\bibitem{Castro-Alvaredo:2024azg}
O.~A. Castro-Alvaredo and L.~Santamar{\'\i}a-Sanz, {\slshape {Symmetry-resolved
  measures in quantum field theory: A short review},}
  \href{http://dx.doi.org/10.1142/S0217984924300023}{{\em Mod. Phys. Lett. B}
  {\bfseries 39} (2025) 2430002}, \href{http://arxiv.org/abs/2403.06652}{{
  arXiv:2403.06652~[hep-th]}}.

\bibitem{Bhattacharyya:2025tmg}
A.~Bhattacharyya, S.~Ghosh, S.~Pal, and J.~Santara, {\slshape {On the
  resolution of categorical symmetries in (Non-) Unitary Rational CFTs},}
  \href{http://arxiv.org/abs/2511.16363}{{ arXiv:2511.16363~[hep-th]}}.

\bibitem{BAXTER198118}
R.~Baxter, {\slshape Corner transfer matrices,}
  \href{https://www.sciencedirect.com/science/article/pii/037843718190203X}{{\em
  Physica A: Statistical Mechanics and its Applications} {\bfseries 106} (1981)
  18--27}.

\bibitem{Baxter_2007}
R.~J. Baxter, {\slshape Corner transfer matrices in statistical mechanics,}
  \href{http://dx.doi.org/10.1088/1751-8113/40/42/S05}{{\em Journal of Physics
  A: Mathematical and Theoretical} {\bfseries 40} (Oct., 2007) 12577–12588}.

\bibitem{Chui:2001kw}
C.~H.~O. Chui, C.~Mercat, W.~P. Orrick, and P.~A. Pearce, {\slshape {Integrable
  lattice realizations of conformal twisted boundary conditions},}
  \href{http://dx.doi.org/10.1016/S0370-2693(01)00982-0}{{\em Phys. Lett. B}
  {\bfseries 517} (2001) 429--435},
  \href{http://arxiv.org/abs/hep-th/0106182}{{ arXiv:hep-th/0106182}}.

\bibitem{Chui:2002bp}
C.~H.~O. Chui, C.~Mercat, and P.~A. Pearce, {\slshape {Integrable and conformal
  twisted boundary conditions for sl(2) A-D-E lattice models},}
  \href{http://dx.doi.org/10.1088/0305-4470/36/11/301}{{\em J. Phys. A}
  {\bfseries 36} (2003) 2623--2662},
  \href{http://arxiv.org/abs/hep-th/0210301}{{ arXiv:hep-th/0210301}}.

\bibitem{Breckenridge:1996is}
J.~C. Breckenridge, R.~C. Myers, A.~W. Peet, and C.~Vafa, {\slshape {D-branes
  and spinning black holes},}
  \href{http://dx.doi.org/10.1016/S0370-2693(96)01460-8}{{\em Phys. Lett. B}
  {\bfseries 391} (1997) 93--98}, \href{http://arxiv.org/abs/hep-th/9602065}{{
  arXiv:hep-th/9602065}}.

\bibitem{Maldacena:1997de}
J.~M. Maldacena, A.~Strominger, and E.~Witten, {\slshape {Black hole entropy in
  M theory},} \href{http://dx.doi.org/10.1088/1126-6708/1997/12/002}{{\em JHEP}
  {\bfseries 12} (1997) 002}, \href{http://arxiv.org/abs/hep-th/9711053}{{
  arXiv:hep-th/9711053}}.

\bibitem{Dijkgraaf:2000fq}
R.~Dijkgraaf, J.~M. Maldacena, G.~W. Moore, and E.~P. Verlinde, {\slshape {A
  Black hole Farey tail},} \href{http://arxiv.org/abs/hep-th/0005003}{{
  arXiv:hep-th/0005003}}.

\bibitem{Hosseini:2020vgl}
S.~M. Hosseini, K.~Hristov, Y.~Tachikawa, and A.~Zaffaroni, {\slshape
  {Anomalies, Black strings and the charged Cardy formula},}
  \href{http://dx.doi.org/10.1007/JHEP09(2020)167}{{\em JHEP} {\bfseries 09}
  (2020) 167}, \href{http://arxiv.org/abs/2006.08629}{{
  arXiv:2006.08629~[hep-th]}}.

\bibitem{Petkova:2000dv}
V.~B. Petkova and J.-B. Zuber,
  \href{http://dx.doi.org/10.1142/9789812799968_0001}{{\slshape {Conformal
  boundary conditions and what they teach us},}} in {\em {Eotvos Summer School
  in Physics: Nonperturbative QFT Methods and Their Applications}}, pp.~1--35.
\newblock 8, 2000.
\newblock \href{http://arxiv.org/abs/hep-th/0103007}{{ arXiv:hep-th/0103007}}.

\bibitem{Nakata:2020luh}
Y.~Nakata, T.~Takayanagi, Y.~Taki, K.~Tamaoka, and Z.~Wei, {\slshape {New
  holographic generalization of entanglement entropy},}
  \href{http://dx.doi.org/10.1103/PhysRevD.103.026005}{{\em Phys. Rev. D}
  {\bfseries 103} (2021) 026005}, \href{http://arxiv.org/abs/2005.13801}{{
  arXiv:2005.13801~[hep-th]}}.

\bibitem{Akal:2021dqt}
I.~Akal, T.~Kawamoto, S.-M. Ruan, T.~Takayanagi, and Z.~Wei, {\slshape {Page
  curve under final state projection},}
  \href{http://dx.doi.org/10.1103/PhysRevD.105.126026}{{\em Phys. Rev. D}
  {\bfseries 105} (2022) 126026}, \href{http://arxiv.org/abs/2112.08433}{{
  arXiv:2112.08433~[hep-th]}}.

\bibitem{Caputa:2024gve}
P.~Caputa, B.~Chen, T.~Takayanagi, and T.~Tsuda, {\slshape {Thermal
  pseudo-entropy},} \href{http://dx.doi.org/10.1007/JHEP01(2025)003}{{\em JHEP}
  {\bfseries 01} (2025) 003}, \href{http://arxiv.org/abs/2411.08948}{{
  arXiv:2411.08948~[hep-th]}}.

\bibitem{Kawamoto:2023ade}
T.~Kawamoto and Y.-k. Suzuki, {\slshape {Entanglement entropy via double-cone
  regularization},} \href{http://dx.doi.org/10.1103/PhysRevD.110.046011}{{\em
  Phys. Rev. D} {\bfseries 110} (2024) 046011},
  \href{http://arxiv.org/abs/2401.00219}{{ arXiv:2401.00219~[hep-th]}}.

\bibitem{Saad:2018bqo}
P.~Saad, S.~H. Shenker, and D.~Stanford, {\slshape {A semiclassical ramp in SYK
  and in gravity},} \href{http://arxiv.org/abs/1806.06840}{{
  arXiv:1806.06840~[hep-th]}}.

\bibitem{Chen:2023hra}
Y.~Chen, V.~Ivo, and J.~Maldacena, {\slshape {Comments on the double cone
  wormhole},} \href{http://dx.doi.org/10.1007/JHEP04(2024)124}{{\em JHEP}
  {\bfseries 04} (2024) 124}, \href{http://arxiv.org/abs/2310.11617}{{
  arXiv:2310.11617~[hep-th]}}.

\bibitem{Roy:2025hew}
A.~Roy, S.~L. Lukyanov, and H.~Saleur, {\slshape {Boundary conditions for the
  entanglement cut in two-dimensional conformal field theories},}
  \href{http://dx.doi.org/10.1103/677y-pxs5}{{\em Phys. Rev. B} {\bfseries 112}
  (2025) L121406}, \href{http://arxiv.org/abs/2503.12674}{{
  arXiv:2503.12674~[quant-ph]}}.

\end{thebibliography}\endgroup

\end{document}